\documentclass[sn-basic]{sn-jnl}

\usepackage{graphicx}%
\usepackage{multirow}%
\usepackage{amsmath,amssymb,amsfonts}%
\usepackage{amsthm}%
\usepackage{mathrsfs}%
\usepackage[title]{appendix}%
\usepackage{xcolor}%
\usepackage{textcomp}%
\usepackage{manyfoot}%
\usepackage{booktabs}%
\usepackage{algorithm}%
\usepackage{algorithmicx}%
\usepackage{algpseudocode}%
\usepackage{listings}%
\usepackage{natbib}

\newcommand\sun{\odot}%
\newcommand\kms{\mbox{km~s$^{-1}$}}

\newcommand{\hh}{\mbox{H}$_2$}
\newcommand{\Lya}{Lyman-$\alpha$~} 
\newcommand{\Lyb}{Lyman-$\beta$~}
\newcommand{\OI}{\hbox{O$\,\rm \scriptstyle I\ $}}
\newcommand{\CII}{\hbox{C$\,\rm \scriptstyle II\ $}}
\newcommand{\CIV}{\hbox{C$\,\rm \scriptstyle IV$}}
\newcommand{\SiII}{\hbox{Si$\,\rm \scriptstyle II\ $}}
\newcommand{\SiIV}{\hbox{Si$\,\rm \scriptstyle IV\ $}}
\newcommand{\HI}{H\,{\sc i}}
\newcommand{\HeII}{\hbox{He$\,\rm \scriptstyle II\ $}}
\newcommand{\CaII}{\hbox{Ca$\,\rm \scriptstyle II\ $}}
\newcommand{\MgII}{\hbox{Mg$\,\rm \scriptstyle II\ $}}

\begin{document}

\title[Article Title]{Galaxy Formation and Symbiotic Evolution with the Inter-Galactic Medium in the Age of ELT-ANDES}


\author*[1,2,3]{\fnm{Valentina} \sur{D'Odorico}}\email{valentina.dodorico@inaf.it}

\author[4]{\fnm{James S.} \sur{Bolton}}

\author[5]{\fnm{Lise} \sur{Christensen}}

\author[6,7]{\fnm{Annalisa} \sur{De Cia}}

\author[8,9]{\fnm{Erik} \sur{Zackrisson}}
%
\author[8]{\fnm{Aron} \sur{Kordt}}
\author[10,5]{\fnm{Luca} \sur{Izzo}}
\author[11]{\fnm{Jiangtao} \sur{Li}}
\author[12]{\fnm{Roberto} \sur{Maiolino}}
\author[13,14]{\fnm{Alessandro} \sur{Marconi}}
\author[15,16]{\fnm{Philipp} \sur{Richter}}
\author[17]{\fnm{Andrea} \sur{Saccardi}}
\author[13,14]{\fnm{Stefania} \sur{Salvadori}}
\author[13,14]{\fnm{Irene} \sur{Vanni}}
\author[1,3]{\fnm{Chiara} \sur{Feruglio}}
\author[18]{\fnm{Michele} \sur{Fumagalli}}
\author[5]{\fnm{Johan P. U.} \sur{Fynbo}}
\author[19]{\fnm{Pasquier} \sur{Noterdaeme}}
\author[20]{\fnm{Polychronis} \sur{Papaderos}}
\author[6]{\fnm{C\'eline} \sur{P\'eroux}}
\author[21]{\fnm{Aprajita} \sur{Verma}}
\author[1]{\fnm{Paolo} \sur{Di Marcantonio}}
\author[22]{\fnm{Livia} \sur{Origlia}}
\author[23]{\fnm{Alessio} \sur{Zanutta}}


\affil*[1]{INAF - Osservatorio Astronomico di Trieste, Via Tiepolo 11, I-34143 Trieste, Italy}
\affil[2]{Scuola Normale Superiore, Piazza dei Cavalieri 7, I-56126 Pisa, Italy}
\affil[3]{IFPU - Institute for Fundamental Physics of the Universe, Via Beirut 2, 34014 Trieste, Italy}
\affil[4]{School of Physics and Astronomy, University of Nottingham, University Park, Nottingham NG7 2RD, UK}
\affil[5]{\orgdiv{Niels Bohr Institute}, \orgname{University of Copenhagen}, \orgaddress{\street{Jagtvej 128}, \postcode{DK-2200} \city{Copenhagen}, \country{Denmark}}}
\affil[6]{\orgname{European Southern Observatory}, \orgaddress{\street{Karl-Schwarzschild Str. 2}, \postcode{D-85748} \city{Garching bei M\"unchen}, \country{Germany}}}
\affil[7]{Department of Astronomy, University of Geneva, Chemin Pegasi 51, 1290 Versoix, Switzerland}
\affil[8]{\orgdiv{Department of Physics and Astronomy}, \orgname{Uppsala University}, \orgaddress{\street{Box 516}, \postcode{SE-751 20} \city{Uppsala}, \country{Sweden}}}
\affil[9]{Swedish Collegium for Advanced Study, Linneanum, Thunbergsv{\"a}gen 2, SE-752 38 Uppsala, Sweden}
\affil[10]{INAF - Osservatorio Astronomico di Capodimonte, Salita Moiariello 16, I-80131 Napoli, Italy}
\affil[11]{Purple Mountain Observatory, Chinese Academy of Sciences, 10 Yuanhua Road, Nanjing 210023, People’s Republic of China}
\affil[12]{Cavendish Laboratory, University of Cambridge, 19 J. J. Thomson Ave., Cambridge CB3 0HE, UK}
\affil[13]{Dipartimento di Fisica e Astronomia, Università degli Studi di Firenze, Via G. Sansone 1, Italy}
\affil[14]{INAF - Osservatorio Astrofisico di Arcetri, Largo E. Fermi 5, I-50125 Firenze, Italy}
\affil[15]{\orgdiv{Institut f\"ur Physik und Astronomie}, \orgname{Universit\"at Potsdam}, \orgaddress{\street{Haus 28, Karl-Liebknecht-Str.\,24/25}, \postcode{14476} \city{Potsdam}, \country{Germany}}}
\affil[16]{\orgname{Leibniz-Institut f\"ur Astrophysik (AIP)}, \orgaddress{\street{An der Sternwarte 16}, \postcode{14482} \city{Potsdam}, \country{Germany}}}
\affil[17]{GEPI, Observatoire de Paris, Universit\'e PSL, CNRS, 5 Place Jules Janssen, 92190 Meudon, France}
\affil[18]{Dipartimento di Fisica G. Occhialini, Universit\`a degli Studi di Milano-Bicocca, Piazza della Scienza 3, 20126 Milano, Italy}
\affil[19]{\orgdiv{Institut d'Astrophysique de Paris, UMR 7095, CNRS-SU}, \orgaddress{\street{98bis bd Arago}, \city{75014 Paris}, \country{France}}}
\affil[20]{Instituto de Astrof\'{i}sica e Ci\^{e}ncias do Espaço - Centro de Astrof\'isica da Universidade do Porto, Rua das Estrelas, 4150-762 Porto, Portugal}
\affil[21]{Sub-department of Astrophysics, University of Oxford, Denys Wilkinson Building, Keble Road, Oxford OX1 3RH, UK}
\affil[22]{INAF - Osservatorio di Astrofisica e Scienza dello Spazio di Bologna, Via Gobetti 93/3, I-40129 Bologna, Italy}
\affil[23]{INAF - Osservatorio Astronomico di Brera, via E. Bianchi 46, 23807 Merate, Italy}


\abstract{High-resolution absorption spectroscopy toward bright background sources has had a paramount role in understanding early galaxy formation, the evolution of the intergalactic medium and the reionisation of the Universe. 
However, these studies are now approaching the boundaries of what can be achieved at ground-based 8-10m class telescopes. The identification of primeval systems at the highest redshifts, within the reionisation epoch and even into the dark ages, and of the products of the first generation of stars and the chemical enrichment of the early Universe, requires observing very faint targets with a signal-to-noise ratio high enough to detect very weak spectral signatures. In this paper, we describe the giant leap forward that will be enabled by ANDES, the high-resolution spectrograph for the ELT, in these key science fields, together with a brief, non-exhaustive overview of other extragalactic research topics that will be pursued by this instrument, and its synergistic use with other facilities that will become available in the early 2030s.}

\keywords{Galaxy formation and evolution, Intergalactic medium, Circumgalactic medium, Interstellar medium, High-resolution spectroscopy, ANDES}



\maketitle

\section{Introduction}
\label{sec:intro}

After an initial hot, dense phase, the Universe expanded and cooled enough to enable protons and electrons to form the first hydrogen atoms. The epoch that followed is known as the “Dark Ages”. The first generation of stars and galaxies formed from gas collapse approximately $\sim100-200$ Myrs after the Big Bang and started flooding its surroundings with highly energetic ultraviolet (UV) radiation. These UV photons (with energies $E>13.6\rm\, eV$) brought the hydrogen atoms in the intergalactic medium (IGM) back into a highly ionised state. This important physical transition, which impacted almost every baryon in the Universe, is known as the Epoch of Reionisation (EoR). The same first stars enabled also the production of the first chemical elements heavier than hydrogen and helium (dubbed ``metals") beyond those produced by Big Bang nucleosynthesis, thereby opening the stage for the chemical enrichment of the early Universe.

The EoR had a profound impact in shaping the Universe as we observe it today. This is why some of the most pressing questions in modern cosmology and astrophysics are related to the timing and topology of reionisation and to the nature of the sources providing the ionising photons.

At present, the ionising photon budget for reionisation is thought to have been dominated by radiation emitted by the first generations of star-forming galaxies \citep[e.g.][]{jiang2022,matthee2022}, although it is still debated whether accreting black holes may also have provided a substantial contribution \citep[e.g.][]{grazian2022,fontanot2023,Madau2024}. The fraction of ionising photons that eventually escaped from these sources into the IGM is also uncertain.  Despite the recent discovery by JWST of unexpected excesses of both bright galaxies \citep[e.g.,][]{mcleod2024} and faint active galactic nuclei \citep[AGN, e.g.][]{harikane2023,maiolino2023c,maiolino2024} at $z\gtrsim7-10$, a complete account of all ionising sources in the early Universe remains very challenging.  

A less direct but powerful, complementary approach is to focus on the signature that the first galaxies will leave on their surrounding gaseous environments. Intergalactic gas in the early Universe represents the reservoir for the formation and sustenance of gravitationally bound luminous sources, but is also sensitive to the escape of ionising radiation and heavy elements from them, even from those that are too faint to be directly imaged. EoR sources will therefore have a measurable impact on the ionisation state, temperature and metallicity of the IGM  \citep[e.g.,][]{becker2015}. 

The investigation of this interplay between galaxies and the IGM through cosmic history is central to the understanding of galaxy formation and evolution. 
Most of the interaction between galaxies and the IGM happens on scales of the order of 1 physical Mpc or less, corresponding to  $\lesssim2$ arcmin at $z\sim2.5$ ($\lesssim 3$ arcmin at $z\sim6$), and this region has been aptly termed the circum-galactic medium (CGM).

Spectroscopic observations focused on absorption lines are the method of choice to study the diffuse gas outside galaxies but also the interstellar medium (ISM). Multiple, complementary probes are available: neutral hydrogen Lyman-series lines (in particular, \HI\ Lyman-$\alpha$ and Lyman-$\beta$), transitions of heavier elements (e.g. C, O, Si, Mg, Al and Fe, to cite the most common ones), as well as molecular transitions (due to e.g. H$_2$, HD, CO). Accurate chemical abundances and physical conditions can be derived from absorption spectroscopy up to the highest redshifts probed by bright background sources, at variance with other high-$z$ galaxy measurements derived from emission line diagnostics. 

In this paper, we highlight the outstanding questions related with galaxy formation and evolution that can be answered with the ArmazoNes high Dispersion Echelle Spectrograph (ANDES\footnote{https://andes.inaf.it/}, \citealt{Marconi2022}) for the Extremely Large Telescope (ELT), mainly through the properties of the ISM, CGM and IGM studied in absorption in the spectra of bright background sources. ANDES is a second generation instrument for the ELT. It is funded by an external consortium and will go to the telescope soon after the first light instruments, which are funded by ESO. At present, it is in the preliminary design phase. The baseline design foresees a wavelength range extending from the blue to the near-IR ($0.4-1.8$ $\mu$m) and a spectral resolution $R\sim100,000$. The U and the K band (total range $0.35 - 2.4$ $\mu$m) are considered as goals in the design, while a possible mode at lower resolution, $R\sim50,000$, is under discussion.

High-resolution spectroscopy has been key to investigate the properties of the diffuse gas in terms of spatial distribution, physical conditions and chemical abundances at intermediate redshift, $z\sim2-4$, e.g. determining the temperature of the IGM and the intensity of the UV background impinging on it \citep[e.g.][and references therein]{gaikwad2021}, assessing the existence of large amount of metals in the gas surrounding galaxies \citep[e.g.][]{turner2014}, but also their presence in the low-density intergalactic gas \citep[e.g.][]{dodorico2016}.
%
ANDES coupled with the unprecedented collecting area of the ELT, will allow us for the first time to carry out detailed studies of the topology of reionisation.  ANDES will be sensitive to very weak and narrow transmission peaks, and to larger \HI\ optical depths than allowed by present 8-10m class telescopes. Furthermore, the unmatched sensitivity to weak metal absorption lines will allow us to explore, at $z \sim2-5$, extremely metal poor ([Fe/H]$ < -4$) galactic environments, possibly retaining the signature of chemical enrichment by the first stars, as well as the metal content of the low-density IGM. In the EoR, ANDES will grant us access to critical elements, like zinc, which are  beyond the reach of current 8-10m class telescopes. 

The paper is structured as follows. Section~\ref{sec:reioniz} describes the science case which deals with the characterisation of the epoch of reionisation through the study of the \Lya\ and \Lyb\ forests in the spectra of $z\gtrsim 6$ quasars and $\gamma$-ray bursts (GRBs). Section~\ref{sec:metal} highlights how ANDES will unveil the chemical enrichment of the early Universe and the nature of the first stars. Section~\ref{sec:transient} is dedicated to the extragalactic transient phenomena whose investigation will be allowed by ANDES. In Section~\ref{sec:addit}, we briefly describe a few additional science cases hinting at the potentiality of ANDES in this field. In Sections~\ref{sec:Uband} and \ref{sec:R50}, we report the benefits of an extension to the U band and of the addition of a mode with resolving power $R\simeq 50,000$, respectively. Finally, in Section~\ref{sec:synergy}, we highlight the synergies between ANDES at the ELT and other surveys or facilities.


\section{Investigating the reionisation epoch with absorption spectroscopy}
\label{sec:reioniz}

In the forthcoming decade, the detection of the redshifted \HI\ $21$-cm emission and/or absorption from the neutral IGM at redshift $z>6$ is a key goal of radio interferometric arrays such as the Square Kilometre Array \citep[SKA, e.g.,][]{koopmans2015}.  This approach holds enormous promise for characterising the nature of the first stars and galaxies across a broad range of time that stretches from reionisation at $z\sim 6$ to the end of the cosmic dark ages at $z\sim 20$--$30$.  
However, the $21$-cm observations are subject to challenging foregrounds, and during reionisation the $21$-cm power spectrum and/or tomographic maps will be primarily sensitive to the distribution of the neutral hydrogen gas on the $>10$ comoving Mpc scales typical of ionised regions \citep[e.g.,][]{lu2024}.  Furthermore, the results of $21$-cm experiments will require independent verification by other methods.

\begin{figure}
\begin{center}
  \includegraphics[width=\textwidth]{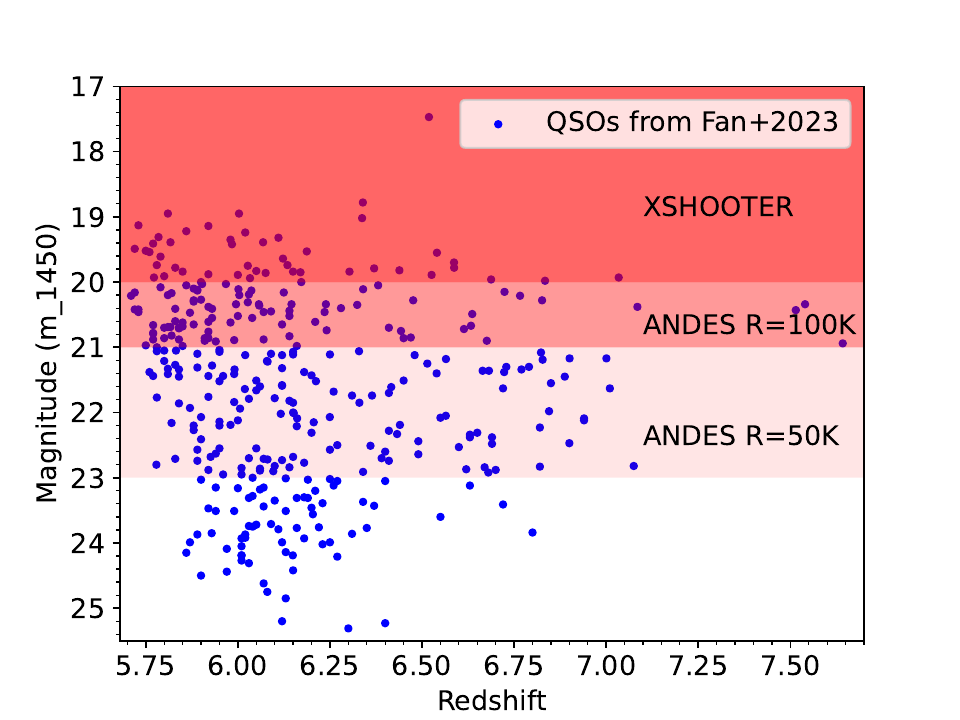}
  \vspace{-0.1cm}
  \caption{Known quasars at $z \ge 5.7$ \citep[blue dots, based on][]{fanARAA2023}, in the redshift, AB magnitude space. The shaded regions with degrading color identify the magnitude range of the quasars for which a S/N~$\sim 30$ per pixel can be reached with $\approx10$h of observations with VLT-XSHOOTER at $R\sim10,000$, ELT-ANDES for the baseline resolution of $R=100\,000$ and a possible lower resolution mode with $R=50\,000$, respectively.}  
  \label{fig:qsoplot}
\end{center}
\end{figure}

\begin{figure}
\begin{center}
  \includegraphics[width=1.00\textwidth]{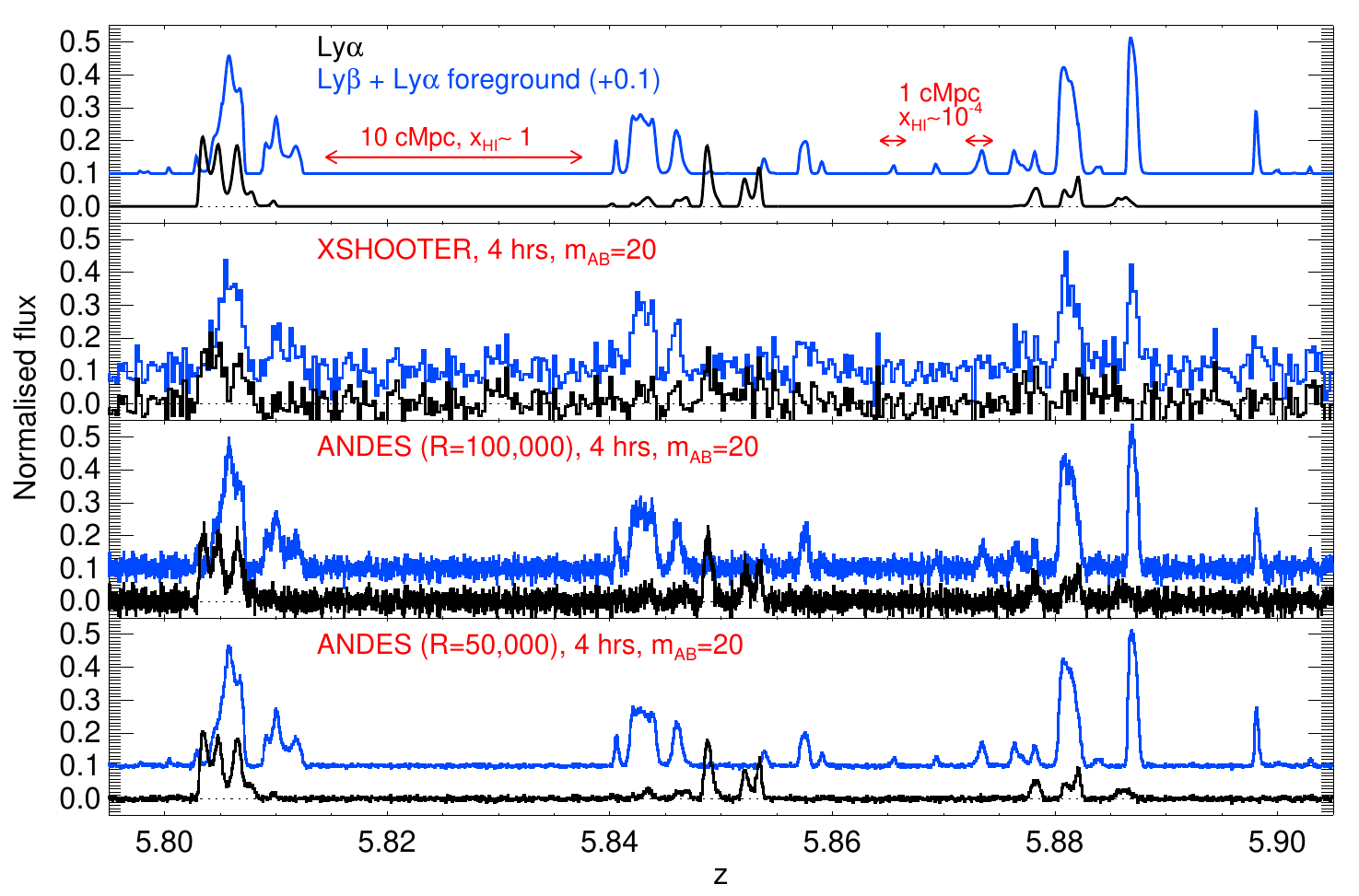}
  \vspace{-0.3cm}
  \caption{The {\it upper panel} shows a segment of the \Lya (black curve) and coeval \Lyb (blue curve) forests in the spectrum of a $z=6.1$ quasar, drawn from a cosmological hydrodynamical simulation coupled with radiative transfer \citep{puchwein2023}. Note that the \Lyb forest flux has been offset by $+0.1$ for presentation purposes, and it also includes foreground \Lya forest absorption at $z^{\prime}=\lambda_{\beta}(1+z)/\lambda_{\alpha}$, where $\lambda_{\alpha}=1215.67$ \AA\ and  $\lambda_{\beta}=1025.72$ \AA.  
  The lower three panels show simulated observations with an exposure time of 4 hours assuming a magnitude $m_{\mathrm{AB}}=20$ for the target quasar, obtained with ({\it from top to bottom}): VLT-XSHOOTER at $R\sim10,900$ \citep[see e.g.,][]{dodorico2023} with S/N$\sim12$, ELT-ANDES for the baseline resolution of $R=100\,000$, reaching S/N~$\sim60$ and for the lower resolution mode with $R=50\,000$ with S/N~$\sim100$. All signal-to-noise ratios are per 10 \kms\ spectral element.  
  }
  \label{fig:Lyseries}
\end{center}
\end{figure}

\begin{figure}[h!]
\begin{center}
  \includegraphics[width=0.85\textwidth]{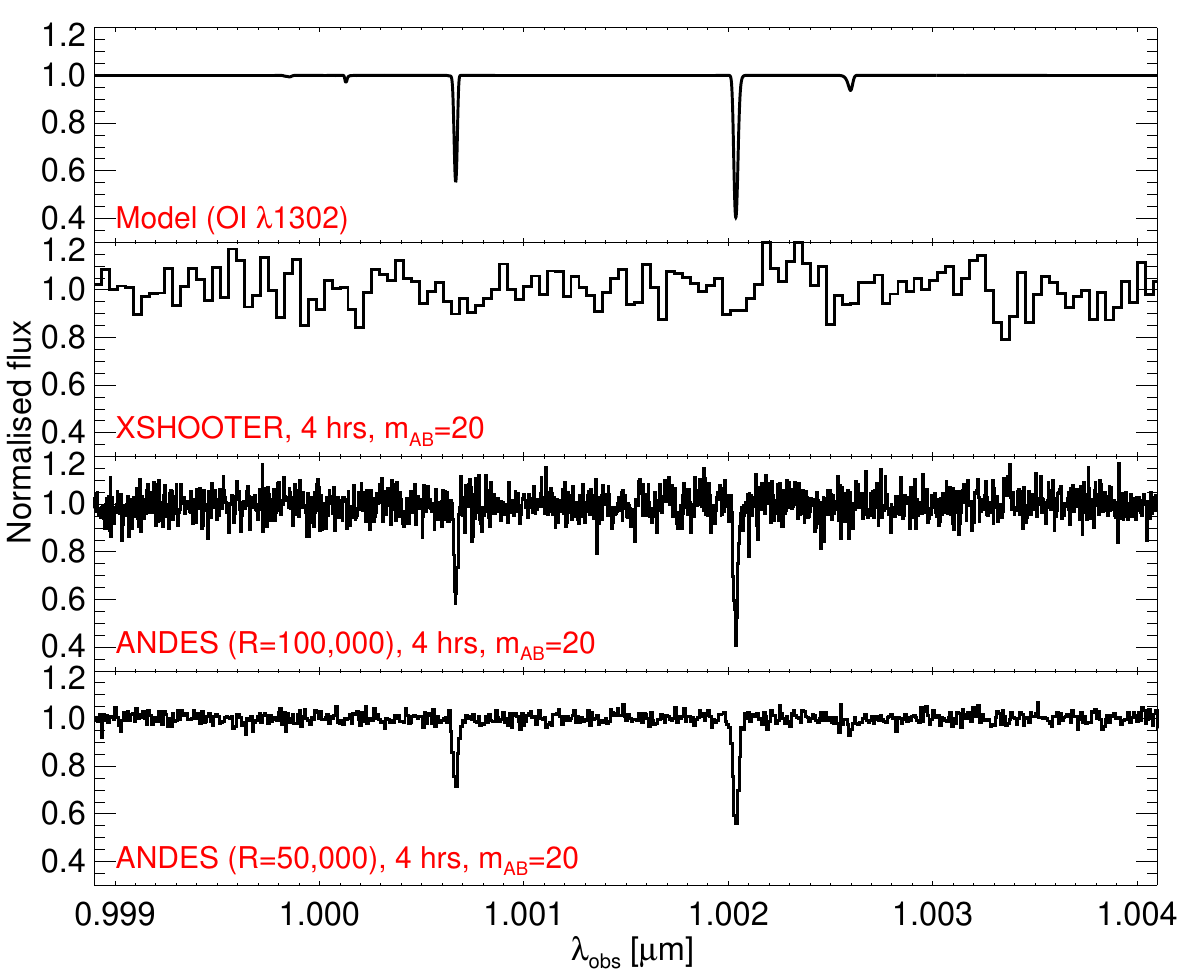}
  \vspace{-0.1cm}
  \caption{\OI absorbers at $z\sim 6.7$ in the simulated spectrum of a quasar with magnitude $m_{\rm AB}=20$. The stronger lines in the plotted spectrum have an equivalent width of $\sim0.01$\AA\, and $\sim 0.02$\AA, respectively. See text for the details of the simulation. 
  As for Figure~\ref{fig:Lyseries}, the lower three panels show simulated observations with VLT-XSHOOTER at $R\sim10,900$ and ELT-ANDES for the baseline resolution of $R=100,000$ and a possible lower resolution mode with $R=50,000$.}  
  \label{fig:OIforest}
\end{center}
\end{figure}

An alternative, well-established and highly complementary approach, that is also sensitive to gas on much smaller scales ($<1$ comoving Mpc, see Figure~\ref{fig:Lyseries}), is absorption line spectroscopy.  Indeed, the majority of our current observational understanding of the later stages of reionisation has been derived from studying the IGM using optical and near-IR spectroscopy.  For example, variations in the \Lya opacity observed in the spectra of bright quasars at $z>5.5$ are consistent with the final stages of reionisation ending as late as $z\sim5.3$ \citep{kulkarni2019,bosman2022}.   The rather small ionising photon mean free path at $z\sim6$ inferred from the Lyman continuum opacity measured in high redshift quasar spectra points to a similar conclusion \citep{becker2021,zhu2023}.  The widths of \Lya transmission spikes at $z>5$ have been used to measure the temperature of the IGM \citep{gaikwad2020}, a quantity that is sensitive to the spectral shape of the ionising radiation from the first galaxies.  Furthermore, the three-dimensional relationship between \Lya transmission spikes, metal lines, and high redshift galaxies has yielded new insights into the escape of enriched gas \citep{diaz2021}  and the escape fraction of ionising radiation into the IGM \citep{meyer2020,kashino2023}.  Dark gaps and pixels in the \Lya forest can be used to measure (or place limits on) the IGM neutral fraction at $5<z<7$ \citep{zhu2022,jin2023}.  The power spectrum of the \Lya forest transmission at $z\gtrsim 5$ has been used to place leading constraints on the thermal state of the IGM and the coldness of cold dark matter \citep{Irsic2017,Irsic2024,Villasenor2023}. Finally, \Lya damping wings in the spectra of quasars \citep{wang2020}, gamma-ray bursts \citep{lidz2021} and (now, with JWST) star forming galaxies at $z=6$--$12$ \citep{umeda2024,heintz2024} are sensitive to intergalactic and circumgalactic neutral hydrogen deep into the reionisation era.

The enormous advantage that ANDES brings to this endeavour is the dramatic improvement in spectral resolution and sensitivity that is attainable with a high resolution spectrograph on the ELT.  At present, only modest samples of the brightest ($m_{\rm AB} \lesssim 20$ mag) among the quasars at $z\gtrsim6$ can be gathered with high enough sensitivity to carry out reliable studies at the EoR and only at intermediate resolution ($R\lesssim 10,000$) \citep[e.g.][]{dodorico2023}. ANDES will allow us to at least double the current number of observed $z\gtrsim6$ quasars (moving the magnitude threshold to $m_{\rm AB} \sim 21$ mag, see Fig.~~\ref{fig:qsoplot}) and study both Lyman-series transmission features and metal absorption lines with a sensitivity and a resolution ($R\sim100,000$) never obtained before. Should a $R=50,000$ mode also be implemented in ANDES, then we could have access to the majority of the quasars at $z\gtrsim 6$ known today \citep[for an updated catalogue see][]{fanARAA2023} with $m_{\rm AB} \sim 21-23$ mag, that cannot be used today for the proposed studies, therefore outstandingly improving the statistical significance of constraints on key properties (e.g., the average H~I fraction).   This will allow ANDES to constrain the patchiness of the reionisation process, the properties of the ultraviolet background (UVB) radiation, and the thermal state and chemical enrichment of the IGM to high precision.   

Figure~\ref{fig:Lyseries} illustrates a mock line of sight through the IGM at $z\sim6$ drawn from a cosmological hydrodynamical simulation that has been coupled with radiative transfer \citep{puchwein2023}, and that assumes a late, inhomogeneous reionisation that completes at $z\sim 5.7$. The increasingly neutral IGM at $z\geq 5.5$ produces largely saturated \Lya forest absorption (i.e. the Gunn-Peterson trough, \citealt{gunnpeterson65}) blueward of the \Lya of the background source. However, underdense, highly ionised regions of the IGM may nevertheless produce narrow transmission features in this part of the spectrum.  The smaller absorption cross-section of the \Lyb transition furthermore allows transmission from ionised regions that otherwise remain saturated in \Lya absorption.  For example, several regions with saturated \Lya absorption (black curves) are highlighted with horizontal red arrows in the upper panel of Figure~\ref{fig:Lyseries}.  There is a small $\sim 10$ comoving Mpc island of fully neutral hydrogen ($x_{\rm HI}=1$) at $z\sim 5.83$ that exhibits coeval saturated \Lyb absorption (blue curves).  In contrast, the region of saturated \Lya absorption between $z=5.855$--$5.875$ is highly ionised ($x_{\rm HI}\sim 10^{-4}$), as indicated by the coeval \Lyb transmission peaks that extend over scales of $\lesssim 1$ comoving Mpc.  
The simulated observations in the lower three panels highlight the unique capabilities of ANDES for probing Lyman series absorption at the tail-end of the reionisation era.  
The narrow features in the forest are easily detected by ELT-ANDES, but are not accessible with VLT-XSHOOTER.  By detecting such features, ANDES will provide constraints on the patchiness of the reionisation, the temperature of the IGM \citep{gaikwad2020}, and the nature of the dark matter  \citep{Irsic2024} at scales that are highly complementary to those probed by SKA. Furthermore, the scatter in the tail of the transmitted flux distribution, particularly at very low transmission/high optical depth, should also be sensitive to the \emph{type} of sources responsible for driving reionisation (e.g. galaxies or AGN), rather than just the timing of reionisation \citep{Asthana2024}.  The locations of these transmission spikes may also be correlated with galaxies detected in emission with IFU observations or slitless spectroscopy \citep[e.g.][]{kakiichi2018,meyer2020,kashino2023}, yielding insight into the relationship between the IGM and the sources of ionising photons.  The latter science case will benefit from the use of the MOSAIC multi-object spectrograph for the ELT \citep{hammer2021}. 


\smallskip

Above redshift $z\sim6.5$, the \Lya and \Lyb forests become progressively less informative for what concerns the amount and the distribution on neutral hydrogen, due to the almost complete absorption of the quasar flux.  However, the remaining \Lya transmission in quasar proximity zones in high resolution spectra at $z>6$ will still contain information on the gas temperature \citep{Bolton2012}, density \citep{ChenGnedin2021,chen2022} and properties of the dark matter \citep{DaviesHennawi2023}.   High-resolution, $z\gtrsim6.5$ quasar spectra will also be useful for the study of the topology of the reionisation process thanks to the presence of metal absorption lines in the region of the spectrum not affected by \HI\ absorption. Some of these metal absorption lines could trace the mostly neutral IGM, if this is polluted to metallicities $\log (Z/Z_{\sun}) \sim[-3.5,-2.5]$. Such a minimal level of metal pollution is expected at the EoR if massive stars in galaxies are the main sources of ionising photons and metals \citep[e.g.][]{madau2001,pallottini2014,jaacks2018}.  
The \OI line at $1302$ \AA\ is particularly promising: the O and H first ionisation potentials are almost identical, and \OI should be in very tight charge exchange equilibrium with \HI. Other low-ionisation transitions like \SiII $1260$ \AA, \CII $1334$ \AA\ and \MgII   $2796$, $2803$ \AA\ might also be observable tracers of the neutral IGM \citep[e.g.][]{hennawi2021}. At high redshift, overdense regions are the first to be polluted to high metallicity but the last to remain permanently ionised, as a result of the short recombination times. Such regions should produce a fluctuating \OI forest, which, if observed, would indicate large quantities of neutral hydrogen \citep{oh2002}. Furthermore, cosmological hydrodynamical simulations with on-the-fly multi-frequency radiative transfer shows that, concurrent with the development of the hydrogen reionisation process, \OI absorbers undergo a decrease in their footprint around haloes and a decrease in the number of observed systems \citep{doughty19}. This decrease occurs preferentially for absorbers that are farther from haloes (in the IGM and CGM) and that have lower equivalent width (EW~$\lesssim 0.05$ \AA). State-of-the-art observations at intermediate resolution ($R\lesssim 10,000$) of a sample of $\sim200$ quasars covering the redshift range $3.2 < z < 6.5$ show a drop of the number density of \OI absorbers with  EW~$\geq 0.05$ \AA\ at $z\lesssim 5.7$ \citep{becker2019,sebastian2024}, evocative of the evolution suggested by simulations due to the change in the ionisation state of the CGM gas.  However, present facilities do not allow to gather statistically significant, complete samples of low equivalent width \OI absorption lines.  

Only thanks to the sensitivity and resolving power of ANDES, it will be possible to detect the weak metal lines due to, in particular, \OI $1302$, \SiII $1260$ and \CII $1334$ \AA\ up to the highest redshifts probed by quasars (and possibly GRBs, see Section~\ref{sec:transient}). An example of the capability of ANDES to reveal these tracers of the early chemical enrichment is illustrated by the simulated spectrum of a $z \sim 7$ quasar in Figure~\ref{fig:OIforest}. The model is drawn from a cosmological hydrodynamical simulation of inhomogeneous reionisation with a volume averaged neutral fraction of $\langle x_{\rm HI}\rangle=0.38$ at $z=6.7$ \citep{puchwein2023}. Following \citet{keating2014}, a density dependent metallicity of $[\rm O/H]=-2.65+1.3\log_{10}(\Delta/80)$ is assumed, where $\Delta=\rho/\langle \rho \rangle$ is the gas density relative to the mean and $(\rm O/H)_{\odot}=-3.31$ \citep{Asplund2009}. In this model, strong \OI absorbers typically arise from circumgalactic gas that remains well-shielded from UV radiation (i.e. with a neutral hydrogen fraction $x_{\rm HI}\sim 1$, see also \citealt{doughty19}). 
The figure also shows how this kind of science is challenging for current facilities (e.g. VLT-XSHOOTER).  

\section{The metal era: the onset of metal enrichment in the Early Universe}
\label{sec:metal}

Metals in the ISM and CGM of galaxies have a crucial role in shaping the formation and evolution of galaxies, stars, planets, and molecules. All chemical elements heavier than beryllium that we observe today were formed in diverse stellar processes throughout the history of the Universe. 
The first generation of (Pop~III) stars were likely more massive than those observed today (and thus shorter-lived), possibly with masses between $\approx 1$ and 1000 $M_\odot$ \citep[e.g.][]{Hirano14, Rossi2021, Pagnini2023, klessen2023}. 
Pop~III stars were born from pristine gas and polluted their surroundings with the first metals, thus starting the chemical enrichment of the early universe and marking the onset of the ``metal era". 
 
 Simulations predict that the first stars appeared around $z\sim20$--30 in the so-called \hh{}-cooling low-mass minihalos (e.g. \citealt{abel2002,yoshida2003,bromm2013,klessen2023}, but see also \citealt{naoz2006} for a model where they appear at $z\sim65$)  and that they rapidly enriched the ISM \citep[e.g.][]{greif2008,wise2012} along with the IGM \citep[e.g.][]{tornatore2007,pallottini2014,jaacks2018} with their newly produced heavy elements. Observationally, direct detection of Pop~III stars is beyond the reach of present and currently foreseen facilities\footnote{With the possible exception of gravitationally lensed, individual Pop~III stars \citep[e.g.][]{windhorst18,surace18}, which are potentially detectable in imaging, but for which it would be extremely hard, if not impossible, to establish whether they actually are Pop~III (e.g., by measuring their metallicity).}, with JWST having the potential to unveil emitting gas at low-metallicity at high-$z$ \citep{Vanzella23,maiolino2023b}. A powerful way to investigate the early chemical enrichment is instead to study the chemical signature of Pop~III stars in gas throughout the universe by comparing the gas (relative) abundances with the expected metal yields of Pop~III stars and related stellar populations (Sect.~\ref{subsec:[X/Y]}). ANDES will be transformational in unveiling the early metal enrichment by studying very-metal-poor absorbing systems tracing pristine gas at intermediate redshifts (Sect. \ref{sec pristine}) as well as gas-rich galaxies or absorbing systems at very high redshift towards quasars and GRBs (Sect. \ref{sec high z}). 
 
\begin{figure}[h]%
\centering
\includegraphics[width=0.9\textwidth]{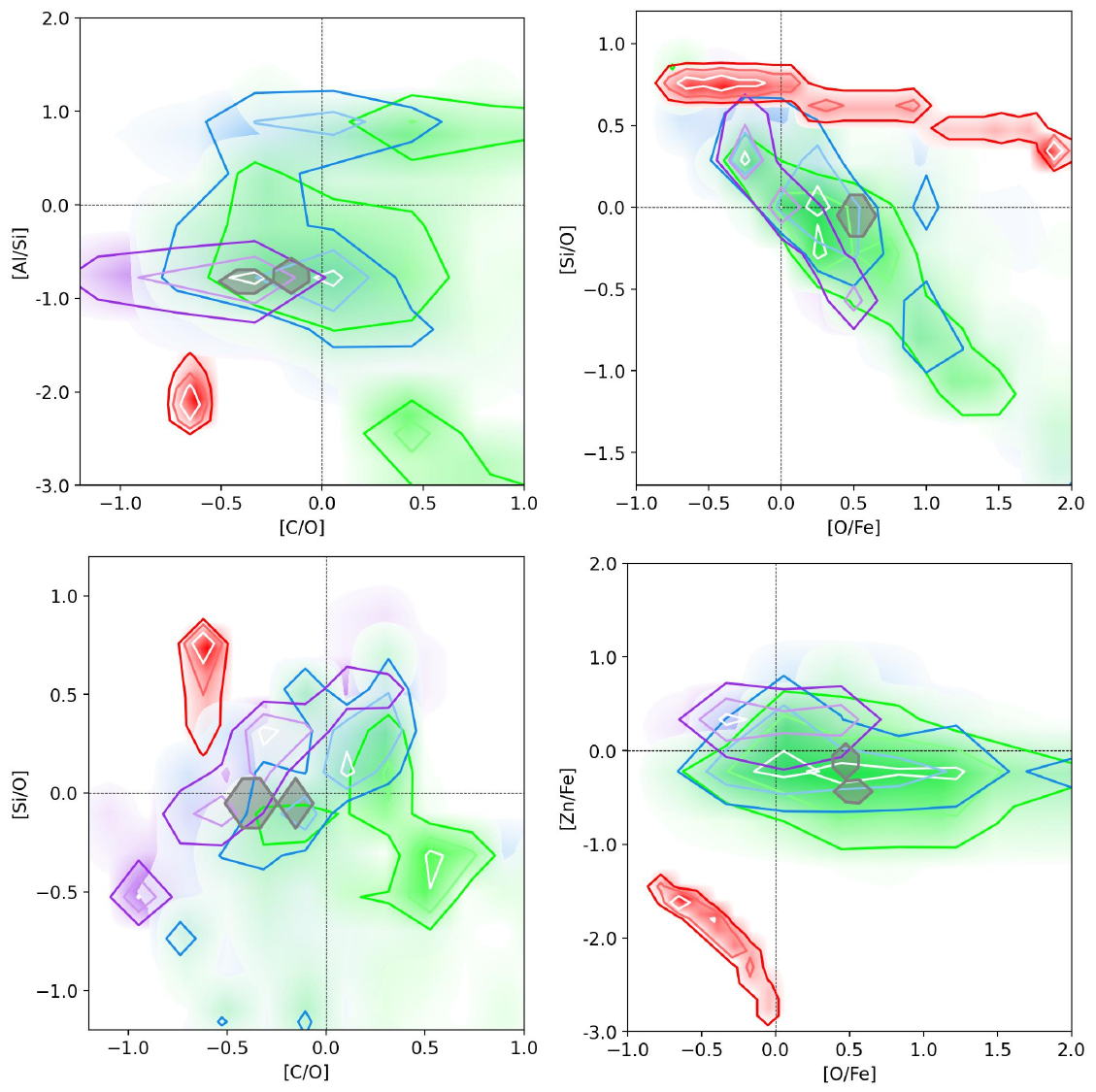}
\caption{Chemical abundance ratios predicted for a gaseous environment imprinted by a single zero-metallicity Pop III SN (faint SNe: green; core-collapse SNe: blue; hypernovae: purple; PISN: red). The contours with decreasing color intensity identify 30\%, 60\%, and 90\% probability densities, corresponding to the fraction of environments predicting those abundances for each SN type. For comparison, we also show the predicted chemical abundances of environments solely imprinted by normal Pop II stars (grey regions).}
\label{fig [X/Y]}
\end{figure}

\subsection{Expected metal yields of Pop~III stars}
\label{subsec:[X/Y]}

The first Pop~III SNe can explode with a variety of energies. Pop~III stars with $m_*=140-260\ M_{\odot}$ are predicted to end their life as Pair Instability Supernovae \citep[PISN, e.g.][]{Heger02,takahashi2018} whose explosion energy increases with the stellar mass ($E=10^{52-53}$~erg) and is able to completely destroy the progenitor star. Conversely, Pop~III stars in the mass range $m_*=10-100\ M_{\odot}$ can, independently of their mass, evolve as SNe with different energies, ${\rm E=(0.3-10)\times 10^{51}}$~erg, thus releasing in the surrounding gas different chemical elements \citep[e.g.,][]{heger10}. General theoretical models for Pop~III stars' enrichment can be used to pinpoint the chemical signature of these different Pop III SNe \citep{salvadori2019,Welsh19, vanni2023a}. 
In Fig.~\ref{fig [X/Y]}, we report the diagnostic diagrams introduced by \citet{vanni2024}, which enable us to disentangle the chemical signature left by different Pop III SNe in high-$z$ absorbers, where the \HI\ column density can not be always measured (see Sect.~\ref{sec high z}). It is clear from the figure that gaseous environments solely imprinted by PISNe dwell in very specific regions of the diagrams. Thus, we can potentially identify them by measuring the proposed abundance ratios. Conversely, environments imprinted by less massive, $m_*=10-100\ M_{\odot}$, Pop III SNe overlap for several chemical abundances and reside in regions of the diagrams where we can also find the imprint of normal Pop~II SNe \citep[see][for details]{vanni2023a}. This makes their identification more challenging. 
We should also emphasize that while the yields of PISNe are extremely solid, the predictions for less massive Pop~III (and Pop~II) SNe are more unsure \citep[e.g. see Sec. 2.2 of][]{Welsh19}. To minimize this issue in Fig.~\ref{fig [X/Y]} we adopted the most complete set of yields for Pop~III SNe with $m_*=10-100\ M_{\odot}$ by \citet{heger10}, which accounts for different possible explosion energies and stellar mixing efficiencies (see Sec. 5 and Fig. 10 by \citealt{Koutsouridou2023} for a comparison with other yields). For Pop~II SNe, furthermore, we show their IMF-integrated contribution for two different sets of yields \citep{woosley1995,LC2018}.

\subsection{Metal-poor gas at intermediate redshift}
\label{sec pristine}
Finding and studying near-pristine gas can be an effective way to trace gas enriched by the first generation of stars. Very-metal poor \citep[$Z < 10^{-2} Z_\odot$][]{Cooke11,Beers05} gas in absorbing systems such as Lyman Limit systems (LLS) and Damped Lyman-$\alpha$ systems (DLAs)\footnote{These absorption systems are defined by their \HI\ column density. LLS have $17.3 \le \log (N / \rm{cm}^{-2}) < 20.3$, while DLAs are characterised by $ \log (N / \rm{cm}^{-2}) \ge 20.3$} has been observed at $z \sim 2-4$ \citep[e.g.][]{Fumagalli11,Welsh19,Welsh20,Robert22}. The existence of such systems at intermediate redshifts implies that the mixing of metals in gas is still incomplete 10~Gyr after the Big Bang, so it is promising to search for signatures of the first generation of stars in gas even at cosmic noon. In some cases gas that had possibly been enriched by a single generation of stars was observed \citep{cooke12,Crighton16,Cooke17}. Recently, \citet{saccardi2023b} observed carbon-enhanced very metal-poor absorbing systems, which represent the high-$z$ analogue of carbon enhanced metal poor (CEMP-no) stars observed in the Local Group. These peculiar absorbers are likely the first example of high-$z$ gas clouds that have preserved the chemical yields of Pop~III stars exploding as low-energy SNe \citep[see][]{salvadori2023}. Despite the long searches, we are still missing observations of environments imprinted by the chemical products of the more massive PISNe, which are the key to constrain the unknown initial mass function of Pop~III stars \citep{Koutsouridou2024}. Recent metal-poor star \citep{xing2023} and distant absorber \citep{christensen2023} candidates were unfortunately not confirmed by subsequent, higher-resolution observations \citep{skuladottir2024b,thibodeaux2024,vanni2024}.
%
The ANDES spectrograph with its high resolution and the access to a larger number of candidates will put us in the best position to finally discover these absorbers in the distant Universe.

\begin{figure}[h]%
\centering
\includegraphics[width=0.9\textwidth]{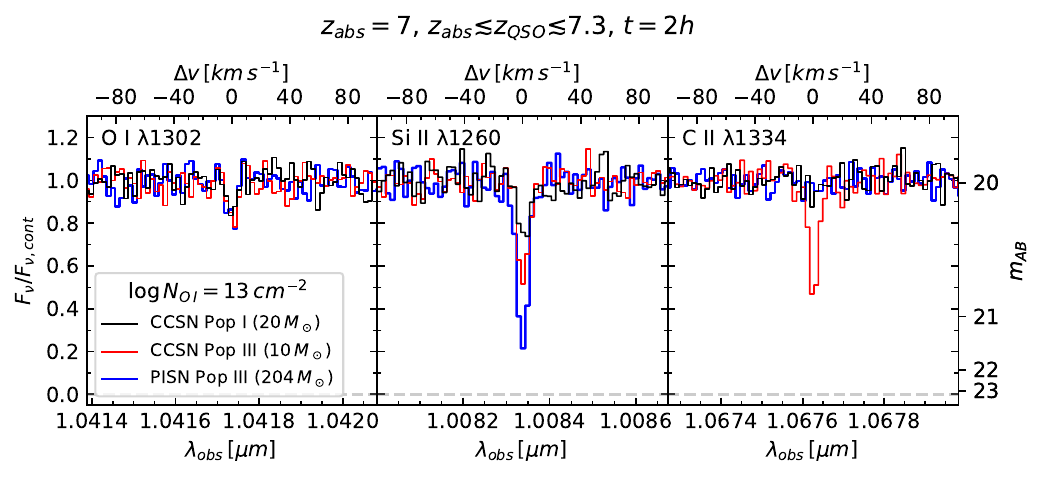}
\caption{Simulated ANDES spectrum of a metal absorption system at $z=7.0$, with  $\mathrm{N}(\mathrm{O~I})=10^{13}$ cm$^{-2}$ along the line of sight to a background $m_{\rm AB} = 20$ quasar observed for 2 hours (noise level based on the ANDES v.1.1 exposure time calculator). The panels display zoomed-in regions around the OI $\lambda$1302, SiII $\lambda$1260 and CII $\lambda$1334 absorption features in three different scenarios enriched by the ejecta from: the core collapse supernova (CCSN) of a $Z=1/3 Z_\odot$, $m_* = 20\  M_\odot$ star \citep[black line;][yields]{Sukhbold16}, the Pop III CCSN of a $m_* = 10\ M_\odot$ star \citep[red line;][yields]{heger10} and the Pop III PISN from a $m_* = 204\ M_\odot$ star \citep[blue line;][yields]{Heger02}. The three scenarios can be clearly distinguished from the ANDES spectrum. }\label{fig quasar7}
\end{figure}

\begin{figure}[h]%
\centering
\includegraphics[width=0.9\textwidth]{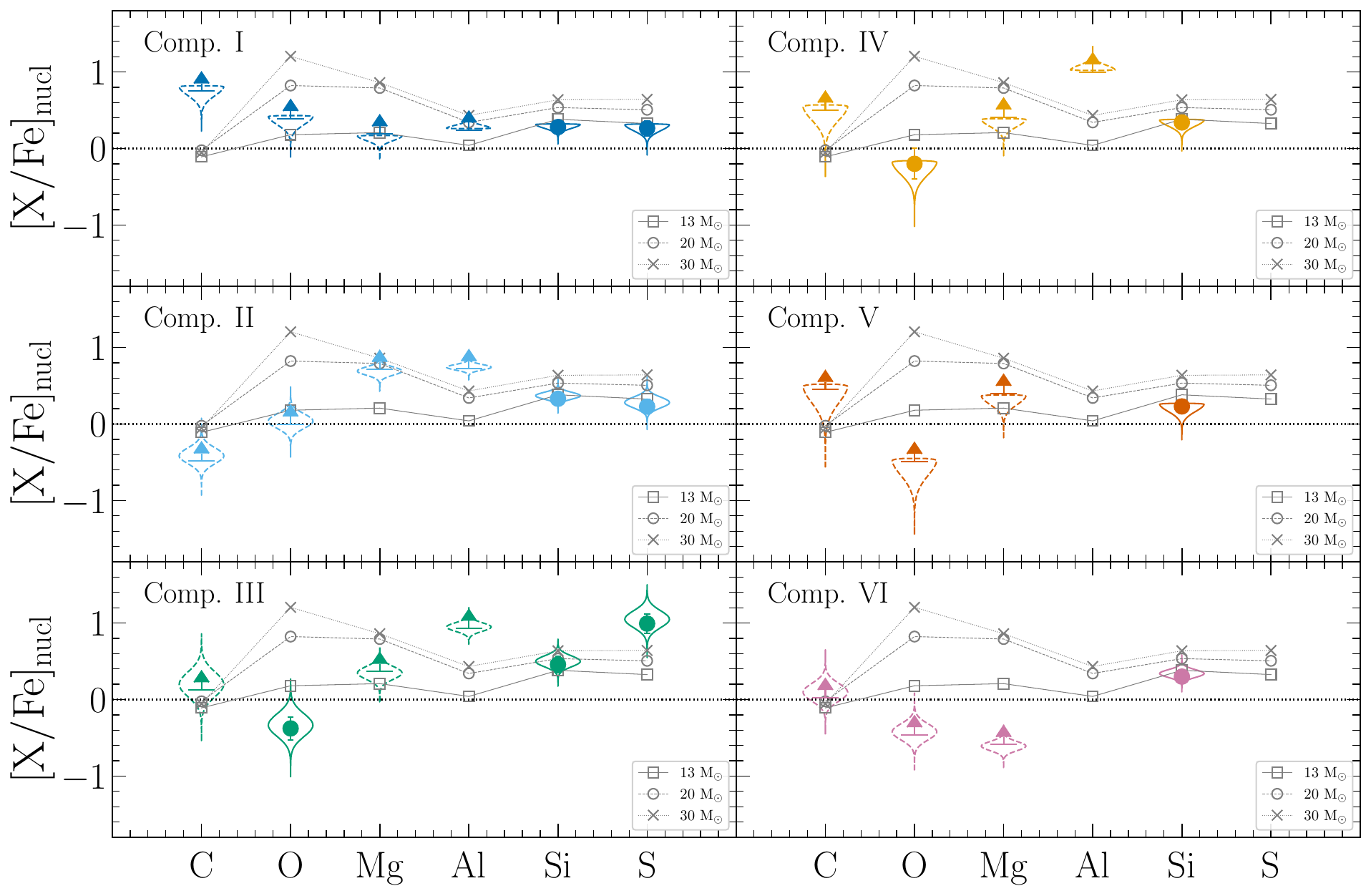}
\caption{Relative abundances corrected for dust depletion, $[X/\mbox{Fe}]_{\rm nucl}$, observed in individual gas components/clouds (one per panel) associated with a $z=6.3$ GRB, observed with VLT-XSHOOTER \citep[adapted from][]{saccardi2023a}. In grey, nucleosynthetic yields for type II SNe \citep{Kobayashi2006} are reported as a function of different progenitor masses $m_* =13, 20, 30\,M_{\odot}$ (respectively solid, dashed, and dotted lines) for a chosen metallicity $\log(Z/Z_{\odot})=-1.3$. Similar studies will be extended to numerous targets thanks to an instrument like ELT-ANDES.}\label{fig GRB}
\end{figure}

\subsection{Relative chemical abundances in gas at $z\gtrsim6$}
\label{sec high z}

At higher redshift, the metallicity of DLAs decreases \citep[e.g.,][]{decia18} and the probability of finding gas enriched by a single generation of stars increases. Recent surveys are finding an increasing number of quasars around the reionisation epoch, which can be used to study the gas in absorption \citep{fanARAA2023,dodorico2023}. 

Because of the increasing cosmic mean density and neutral gas fraction, above $z\sim6$ it is no longer possible to measure the neutral hydrogen column density of individual absorption features and therefore measure the absorbers' metallicity. Nevertheless, the study of the relative abundances of elements heavier than helium can still be used in a powerful way to characterise the metal pattern and thus the chemical enrichment of the absorbing gas (see Fig.~\ref{fig [X/Y]}). 

The state-of-the-art samples of optically thick, metal-poor absorbers at $z \gtrsim 5.5$ \citep{christensen2023,sodini2024} show an increase in the dispersion of the relative chemical abundances with respect to lower redshift samples, which has been interpreted as a signature of the contribution to enrichment by Pop~III stars \citep{kulkarni2019,vanni2024}. A few systems in those samples present peculiar chemical abundances which could be ascribed by enrichment due exclusively to Pop~III progenitors. However, only ANDES can  decipher these scenarii with higher-resolution and higher S/N spectroscopic data, which are beyond the reach of present 8-10m telescopes.

Figure \ref{fig quasar7} illustrates simulated ANDES spectra of a $z=7.0$ absorber characterized by a neutral oxygen column density $\mathrm{N}(\mathrm{O~I})=10^{13}$ cm$^{-2}$. Here, we consider three different single-SN enrichment scenarios for the absorber -- two featuring core-collapse SNe from either a metal-enriched or metal-free (Pop~III), moderate-mass ($10-20$ $M_\odot$) progenitor star, and one featuring a PISN from a very massive ($\approx 200\ M_\odot$) Pop~III progenitor. By measuring the strengths of the \OI $\lambda$1302, \SiII $\lambda$1260 and \CII $\lambda$1334 \AA\ absorption features in the intervening gas cloud, ANDES can readily distinguish the three cases and probe the existence of very massive Pop~III stars in the early Universe. 

 \smallskip

The relative abundances in gas are also sensitive tracers of the depletion of metals into dust, as well as to the nucleosynthetic signature imprinted in the gas by the previous stellar populations. These effects can be well disentangled by studying the relative abundances of different metals having different refractory and nucleosynthetic properties \citep[e.g][]{decia16}. This is important because significant amounts of dust have been observed out to the reionisation epoch \citep{Watson15,saccardi2023a}, so that dust depletion needs to be excluded or accounted for to be able to unveil the often subtle effects of nucleosynthesis of specific stellar populations. Figure \ref{fig GRB} shows the relative abundances in individual gas components/clouds within and around the host galaxy of a $z=6.3$ GRB \citep{saccardi2023a}. After correcting for dust depletion \citep[e.g.][]{saccardi2023a,decia21}, the gas relative abundances can be directly compared to stellar yields. This comparison is displayed in Fig. \ref{fig GRB}, where some of the gas clouds/components show high levels of $\alpha$-element enhancements, and a high [Al/O], which is not well reproduced by the models.\footnote{Note that the \HI\ column density of individual components and, as a consequence, their metallicity, cannot be determined observationally. Consequently, we have adopted the model yields by \citet{Kobayashi2006} for  a metallicity  $\log(Z/Z_{\odot})=-1.3$ (corresponding to an absolute metallicity $Z=0.001$) close to the overall metallicity of the host galaxy ([M/H]$_{\rm tot} = -1.72 \pm 0.13$; \citealt{saccardi2023a}).}   

\section{Extragalactic Transients}
\label{sec:transient}

Extragalactic transients such as GRBs, various types of supernovae, novae, counterparts of gravitational wave events, and tidal-disruption events unexpectedly light up superposed on the light of their host galaxy, and offer us for a limited amount of time new insights into the physical nature of these stellar- or galactic phenomena, and the galaxies that host them. This provides a complementary study of galaxies since the detection of transients does not depend on the host galaxy luminosity. The current limitation to these studies is the faintness of most targets, which ANDES will overcome with the unprecedented sensitivity of the ELT and with its near-IR optimised capabilities.

\subsection{Gamma-ray bursts}
GRBs are the most energetic events after the Big Bang. GRBs lasting longer than about 2 seconds are predominantly associated with the explosions of fast-rotating massive stars \citep{hjorth03,woosley06}. Their afterglows can be used as bright cosmic beacons even at very high redshifts to unveil the chemical and kinematical properties of the galaxies that host them based on rest-frame UV absorption-line spectroscopy \citep[e.g.][]{vreeswijk04,prochaska07,decia12,bolmer19}. Absorption line spectra typically trace the warm and cold, predominantly neutral ISM in the host galaxy at distances greater than $\sim$100 pc from the explosion sites \citep{vreeswijk07}. 

While GRB afterglows have been studied since 1997, numerous questions remain unanswered, and even to this date in 2023, the diversity of events remains a puzzle, including exceptionally long-duration or bright events, and compact object mergers \citep[e.g.][]{kann21,williams23,levan2024}.
ANDES will provide a leap forward with a rapid acquisition of transient sources in the near-IR, combined with a wide spectral coverage at a high resolution.

Searches for the afterglows conducted in the visible spectral range miss about half of them  \citep[e.g.][]{fynbo09,prochaska09,fynbo14}. We know that X-ray afterglows are often associated with bright, more metal-rich and dustier hosts than the optically detected afterglows \citep{perley17}. There hence is a large potential for spectroscopy of these dusty sight-lines with the near-IR coverage of ANDES. 

\begin{figure}[h]%
\centering
\includegraphics[width=0.48\textwidth, angle=-90, clip, trim=3.5cm 0.cm 3cm 0cm]{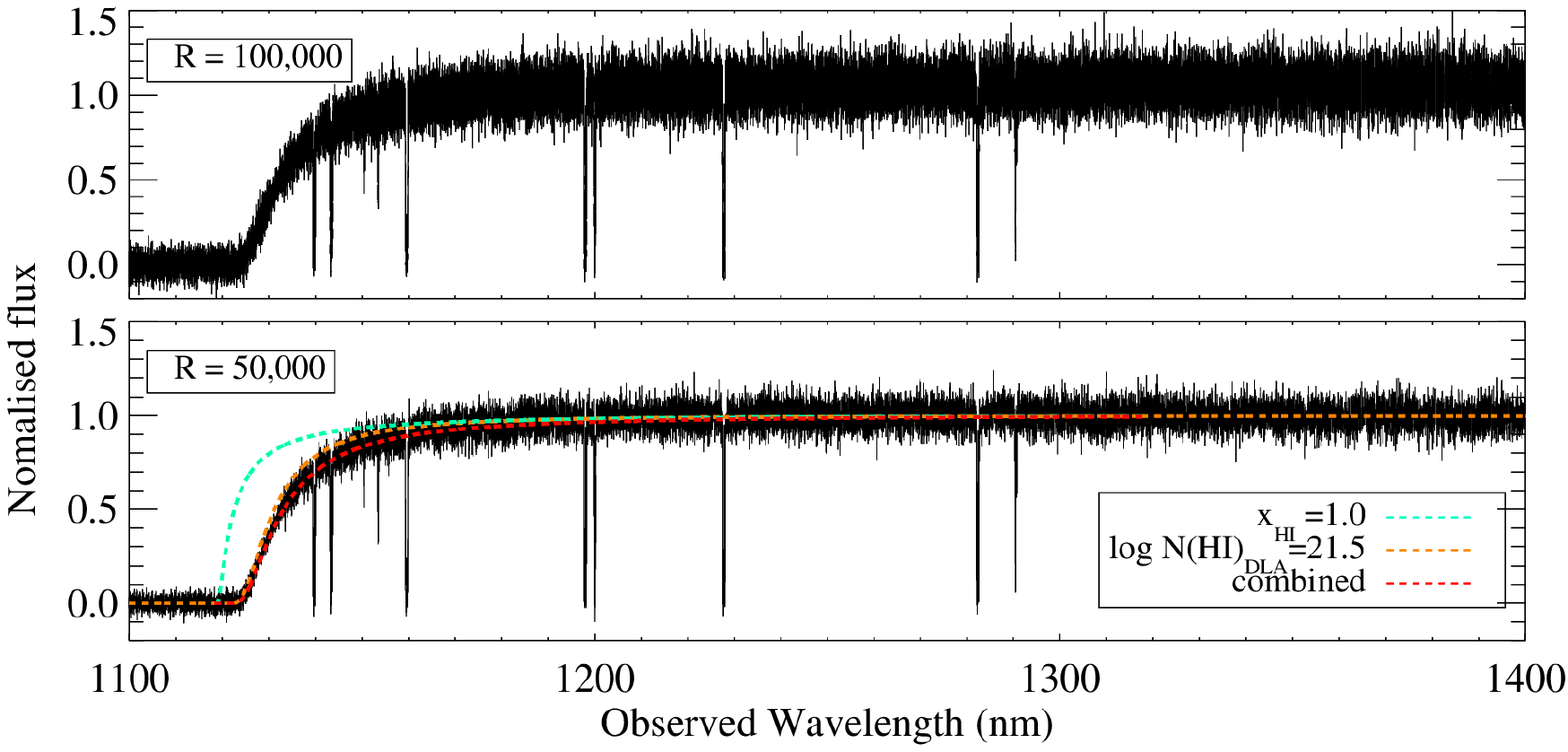}
\caption{Simulated ANDES $J$-band spectrum of a simulated GRB afterglow with a magnitude of $J$=20 at $z=8.2$ with strong damping wing from a fully neutral IGM combined with a strong intrinsic hydrogen Ly$\alpha$ damping wing and with metal absorption lines corresponding to a 1\% solar metallicity.}
\label{fig:GRB_sim}
\end{figure}

Currently, only a few GRBs at $z\gtrsim6$ have very detailed chemical information \citep{hartoog2015,saccardi2023a}. The current record for a GRB with a secured afterglow spectrum is $z=8.2$ \citep{tanvir09,salvaterra09}, with another explosion at photometric redshift $z=9.4$ \citep{cucchiara11}. Long duration GRBs explode in star-forming galaxies and hence the highest redshift explosions will identify the exact location of the earliest generation of stars. Studying the absorption lines imprinted in the spectra of GRB afterglows with ANDES will allow a crucial test on the chemical yields in the early universe as described in Section.~\ref{sec:metal}. Combining the ISM metallicity detected in absorption with emission-line based galactic  metallicities would allow us to accurately calibrate the latter diagnostics, which remains a challenge to date even at intermediate redshifts \citep{schady2024}. Furthermore, the estimate of the IGM neutral hydrogen fraction, $x_{\rm HI}$, from the measured \Lya damping wing in high-redshift afterglows is less complex compared to quasar spectra (see Sec.~\ref{sec:reioniz}) due to the simple power-law spectral distribution of afterglows,  even in the presence of the contribution of the host galaxy DLA \citep{hartoog2015}. Fig.~\ref{fig:GRB_sim} illustrates a simulated $J$-band spectrum of a $z=8.2$ GRB with a 1\% solar metallicity and a strong damping wing from an intrinsic DLA in a fully neutral IGM. The simulated spectrum shows the signal that ANDES will obtain in a hour integration on a source with a magnitude $J_{\mathrm{AB}}$=20 in the nominal $R$=100,000 resolution, and an $R$= 50,000 mode.
Simulations of ELT-ANDES spectra with varying neutral gas fractions and DLA column densities have demonstrated that the two components can reliably be disentangled \citep{tanvir21b}. For DLA column densities below $\log N($\HI$)=23$ the neutral gas fraction can be recovered with a 10\% accuracy, while the DLA column density from the GRB host galaxy is a secure measurement at the high spectral resolution.

With VLT-UVES, astronomers made a breakthrough in discovering variable fine-structure lines powered by the photo-excitation and ionising effect of the GRB afterglows on the surrounding gas, that allowed the determination of distances of the absorbing clouds out to a kpc from the GRB \citep[e.g.][]{vreeswijk07,delia10,decia12,vreeswijk13}. Such studies are challenging because the afterglow fades quickly, and it is impossible to detect the lines at phases later than one day after the burst. With a higher sensitivity, ANDES will overcome this main limitation and detect the line-variability as the GRB afterglows fade in time. This provides a unique opportunity to probe the sizes and internal ISM structure of high redshift galaxies, including that in very faint galaxies.

\subsection{Superluminous Supernovae}
Superluminous Supernovae (SLSNe) are a class of extremely energetic SNe 
\citep{quimby11,galyam12,galyam19}, which are at least 10 and often 100 times more luminous than normal SNe \citep[e.g.][]{nicholl13,inserra17,decia18}.
Because of their brightness, they can be observed out to high redshifts \citep[currently $z\sim4$,][]{cooke12}. They are particularly intriguing because their explosion processes and progenitor stars are highly debated, and may be associated with very massive stars, with potential links or analogies to pair-instability SNe 
and Pop III stars \citep{galyam09,heger10}. The host galaxies of SLSNe have low metallicities and extreme properties \citep{neill11,vreeswijk14,leloudas15,lunnan15,perley16,schulze18}. Spectral observations of SLSNe with 8-10m telescopes are already photon-starved and limited to mostly low-resolution studies, with a few exceptions that unveil host-galaxy ISM environments that are very different from any other galaxies we know of \citep{vreeswijk14,yan18}. In addition, SLSNe have a potential use for cosmology out to much higher $z$ than currently done with SNe Type Ia \citep{inserra13,scovacricchi16}. SLSNe spectra are diverse and fundamental to assess the properties of the explosions \citep{quimby18}, as well as their host galaxies in absorption \citep{vreeswijk14,yan18}. Current time-domain survey facilities such as the Zwicky Transient Facility \citep[ZTF,][]{graham19} discover about 30 SLSNe per year \citep{chen2023b,chen2023a}. 

In the future,  ANDES will be ideal to follow up selected SLSNe  out to high $z$, and will bring new insights into chemical properties of the extreme host galaxies. In addition, the observations will also bring important information on the physics and diversity of these energetic explosions, the stellar evolution of most massive stars and an intriguing perspective for SLSN high-$z$ cosmology. 

\begin{figure}[h]%
\centering
\includegraphics[width=\textwidth]{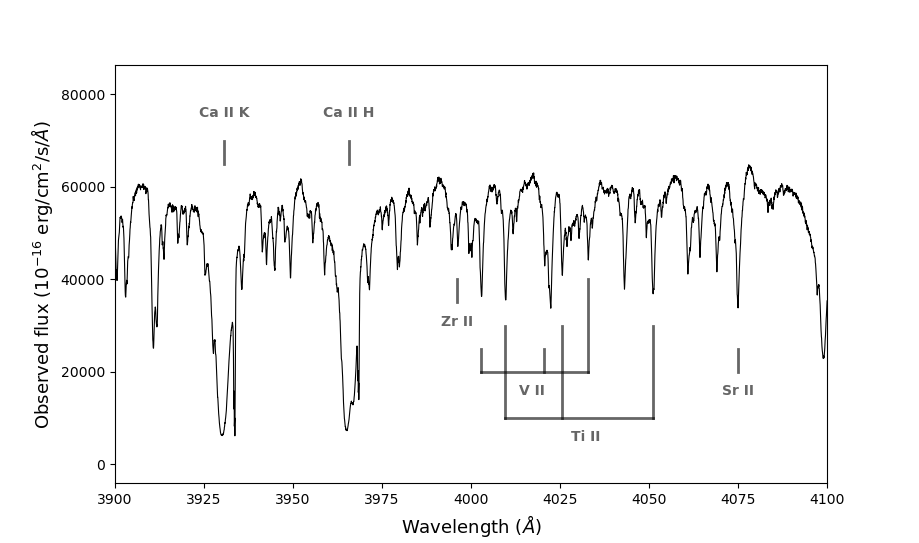}
\caption{A limited wavelength range, covering the region nearby \CaII H\&K lines, of the spectrum of V906 Car obtained on March 24, 2018, with VLT-UVES, rescaled at the resolution of ANDES, $R \sim 100,000$. A bunch of THEA lines, including Zr\,{\sc ii} and Sr\,{\sc ii},  are clearly identified thanks to the resolution available with ANDES, and reported in the plot.}
\label{fig:THEA}
\end{figure}

\subsection{Novae}
\label{subsec:novae}
Classical novae (CNe) are non-disrupting thermonuclear explosions happening on the surface of a white dwarf that is accreting matter from a companion late-type main sequence star \citep{Bode2008,DellaValle2020}. These cataclysmic events lead to the ejection of the accreted shell into the interstellar medium, giving rise to multi-wavelength emission visible for months, if not years, after the outburst. Very high-energy emission in $\gamma$-rays has recently been discovered \citep{fermi14} from the presence of shocks originating from the multi-component nature of classical novae ejecta \citep{mukai19}. Evidence for internal shock interaction, simultaneously to $\gamma$-ray and optical flares, within nova outflows has been observed in high-resolution spectra due to the appearance of new absorption systems and the evolution of P-Cygni absorption lines of the main line transitions detected in CN spectra \citep{aydi20a}, including transient heavy element absorption \citep{williams2008}. Untangling multiple components expanding at different velocities within the nova ejecta \citep{aydi20b} will help understand the still unknown origin of the shock interactions giving rise to the high-energy emission. 

The resolution provided by ANDES will also allow us to identify and measure the abundance of narrow transient heavy element absorption components (namely {\it THEA}, \citealt{williams2008}), which are generally low-ionisation transition of neutral and ionised Fe-peak elements, including heavier elements such as Sr, Y, and Ba (see Fig.~\ref{fig:THEA}. These narrow lines have been detected during the bright optically thick phases of several CNe, and they exhibit only a single, low velocity, absorption component, suggesting that they are located in the slow toroidal ejecta component, but whose origin is not clearly identified yet. Among these lines, we have also identified in few cases the presence of the resonance line of the neutral lithium (Li\,{\sc i} at 670.8 nm), which has been theoretically predicted to be synthesised in novae already in the '70s \citep{Arnould1975,Starrfield1978}, but was only discovered in CNe recently, thanks to the use of very high-resolution spectrographs \citep{izzo15}.  ANDES will enable us to measure the abundance of THEA lines, compare their inferred abundance pattern with the spectra of late-type main sequence stars that are similar to the donor secondary star, and shed light on their origin. At the same time, detection of lithium will be crucial to determine the lithium yield from a large sample of novae, possibly simultaneously to observations in the near-UV to measure the abundance of its parent element (7Be\,{\sc ii} at 313.0 nm), and finally to understand and characterise the role of novae as one of the main factories of lithium in the Galaxy \citep{Molaro2023} and in nearby systems such as the Magellanic Clouds \citep{Izzo2022}.

\subsection{Specific requirements for the transient science cases}
The majority of extra-galactic transient science investigations focus on spectroscopic analysis of individual objects, where stringent atmospheric conditions are not necessary, and can be conducted with seeing-limited observations. The ability of ANDES to simultaneously cover a wide wavelength range from 0.4 to 1.8 microns in a single exposure offers a significant advantage.
Observations of extragalactic transients need the availability of a target of opportunity (ToO) mode, while the rapidly fading GRB afterglows require a rapid response mode (RRM) similar to that available on the VLT. Particularly, catching the fading afterglow minutes from the GRB trigger will benefit from an enhanced sensitivity enabled by a lower resolution ($R\approx 50,000$) mode (see Section~\ref{sec:R50}).

\section{Additional science cases}
\label{sec:addit}

The science cases described in the previous sections were chosen by the ANDES science team as the most compelling to be carried out with this instrument in the field of galaxy formation and evolution and the intergalactic medium. However, they are far from exhausting the extragalactic science that will be enabled by ANDES. 

All the studies devoted to the investigation of the physical and chemical properties of the IGM, CGM and ISM at $z\gtrsim 2.5-3.0$ will be improved in terms of statistics (number of targets), sensitivity as a function of exposure time and details of the velocity profile of absorption lines thanks to the characteristics of ELT-ANDES (but see also Sect.~\ref{sec:Uband}). 
Furthermore, ANDES could be used to observe pairs or small groups of quasars at small angular separations ($\lesssim 1$ arcmin) on the sky  to determine the small-scale variation of chemical properties in the CGM (see also Sect.~\ref{sec:synergy} for the synergy with other ELT instruments).  

In the following, we briefly describe three example science cases further describing the capabilities of ANDES.

\subsection{\HeII reionisation}
\label{subsec:HeII}
In the cosmic reionisation theory, young stars play a key role in the reionisation of \HI\ at $z\gtrsim 6$ \citep[e.g.][]{fan2006}, but they are typically unable to further doubly ionise helium. Instead, quasars are thought to be the most likely ionising sources responsible for the \HeII reionisation, which started at $z\gtrsim5$ and completed at $z\gtrsim 3$ \citep[e.g.][]{mcquinn2016}. The \CIV\ absorption doublet at rest frame $\lambda\lambda$1548.2,1550.8 \AA\  in the spectra of bright background quasars at $z\gtrsim 4$ is the best tracer of the \HeII reionisation \citep[e.g.][]{yu2021}. This is because the \CIV\ doublet is strong and easy to identify, located at significantly longer wavelengths than the hydrogen Lyman-$\alpha$ emission line in the quasar spectrum, and the C$^{3+}$ ion has a comparable ionisation potential as He$^+$. Using high resolution optical spectra, after firmly identifying the \CIV\ absorbers, it is then easier to identify other associated metal absorbers such as \SiIV \citep[e.g.][]{dodorico2022}, as well as the Lyman-$\alpha$ lines embedded in the forest \citep[e.g.][]{simcoe2004}. The identification of these multi-ion systems will help us to study the physical and chemical properties of the IGM in the \HeII reionisation epoch (e.g., \citealt{simcoe2004,songaila05}), when the AGN become more and more important in the cosmic reionisation (e.g., \citealt{fauchergiguere2009}), and some large-scale gravitationally bound systems gradually become virialized (e.g., \citealt{wangt2016,tozzi2022}). High-resolution optical spectroscopy observations of $z\gtrsim 4$ quasars with $\sim8\rm~m$ telescopes typically allow the detection of \CIV\ absorbers in $i<19\rm~mag$ quasar spectra to an equivalent width of EW$_{\rm r}\gtrsim0.01$ \AA\ (corresponding to $\log (N($\CIV$)/{\rm cm}^2)\gtrsim 12.5$; e.g., \citealt{cooksey2013,dodorico2022}). This leads to a small sample size due to either a small number of bright enough quasars (seriously affecting the statistics at the high $N$(\CIV) end) or a small number of \CIV\ absorbers above the currently reached detection limits. With the high-resolution mode of ELT-ANDES ($R\sim100,000$), it will be much more efficient to reach a lower \CIV\ column density detection limit of $\log (N($\CIV$)/{\rm cm}^2)\lesssim 12$ (assuming $i\approx20\rm~mag$, $\rm exposure\approx2\rm~hours$, $\rm S/N\sim27$). This is critical to constrain the \CIV\ column density distribution function and its redshift evolution (e.g., \citealt{ellison2000,dodorico2010}). 
Furthermore, if a lower resolution, higher throughput mode of ANDES at $R\sim50,000$ will be implemented, an increase in sensitivity of about a factor of two is expected, which implies that the same threshold in \CIV\ column density could be reached in the same exposure time for targets $\sim1.0-1.5$ magnitudes fainter.  


\subsection{Probing the circumgalactic medium of local galaxies}
\label{subsec:localgal}
The \CaII H\&K lines near 3950 \AA\,are the only lines in the optical that can be used to study the cold/warm  CGM in absorption at $z\approx 0$. 
Previous surveys of \CaII absorption in the CGM of the Milky Way and other low-redshift galaxies using  VLT-UVES and Keck-HIRES \citep{benbekhti2008,richter2011,benbekhti2012} demonstrate that these lines are particularly useful to study the small-scale structure at pc and sub-pc levels in the neutral CGM down to neutral gas column densities of log\,$N$(H\,{\sc i}$)=18$. 

ANDES and its planned spectral capabilities in the U-band will initiate a new era in using the \CaII H\&K lines to characterize the internal density structure in $z\approx 0$ CGM clouds. In principle, every extragalactic spectrum recorded with ANDES in the U-band can be used to investigate the foreground CGM of  the Milky Way, tidal gas structures in the Local Group (e.g., the Magellanic Stream), and other circumgalactic gas features of  foreground galaxies. The systematic comparison of \CaII H\&K absorption patterns as a function of the lines of sight angular separations along thousands of them will provide crucial information on the small-scale distribution and coherence length of neutral gas structures in the (multi-phase) CGM.

Spectra of extragalactic sightlines can also be systematically  combined with ANDES U-band spectra of (in projection) adjacent, distant halos stars \citep{battaglia2017} or stars in other Local Group member galaxies to constrain the distances of the Milky Way's high-velocity clouds and the Magellanic Stream using the distance-bracketing method \citep[e.g.][]{wakker2007,wakker2008,thom2008}. This will be an essential step to reconstruct the 3D distribution and total mass of the Milky Way's CGM, to determine the Milky Way's gas-accretions rate, and to pinpoint the exact location and extent of the Magellanic Stream in the context of the on-going assembly of the Local Group.

With its capabilities to obtain high-resolution optical spectra of extragalactic background sources at extremely high S/N ($>2000$), ANDES also will potentially open a completely new window to study warm-hot gas in the intergalactic medium in the halos of foreground galaxies at low redshift by using the forbidden intersystem lines of [Fe\,{\sc x}] and [Fe\,{\sc xiv}] near 6375 and 5303 \AA. 
As originally proposed in \citet{york1983}, and later re-evaluated by other groups \citep[e.g.][]{hobbs1984,pettini1989,richter2014}, these extremely weak lines have a great potential to be used as tracers of million-degree halo gas at the viral temperatures of (massive) galaxies and hot gas in the shock-heated component of intergalactic medium (the WHIM). This requires, however, optical spectra of background AGN with a S/N of several thousand because of the extremely small transition probabilities of [Fe\,{\sc x}] and [Fe\,{\sc xiv}]. The combined use of individual ANDES high-sensitivity spectra and clever stacking strategies will enable us to search for warm-hot gas using [Fe\,{\sc x}] and [Fe\,{\sc xiv}] in different cosmological environments and to provide a census of the ``missing baryons'' in the WHIM in the local Universe \citep[][]{fresco2020}.

\subsection{Low-mass Black Holes}
\label{subsec:BH}

Supermassive black holes (BH; M$_{\rm BH} \sim 10^5- 10^{10}$ M) are key actors in the formation and evolution of galaxies. The feedback due to radiation and winds from the accretion disk surrounding the BH, can shape the host galaxies and explain the present-day empirical scaling relations between their properties and the BH mass \citep[e.g.][]{fabian2012,kormendy2013,king2015}, and possibly also the dearth of massive galaxies \citep[e.g.][]{behroozi2018}. 
However, this paradigm is ultimately  mostly based on several tens direct BH mass measurements in massive early type galaxies with $M_{\rm BH} > 10^7$ $M_{\sun}$ and little is known for the less massive galaxies which are predicted to host BH with $M_{\rm BH} < 10^7$ $M_{\sun}$ \citep[see e.g.][]{kormendy2013,greene2020}.

It is currently believed that low mass galaxies undergo a different growth than massive ones and, as a result, the distribution of their BH masses should retain an imprint of the original seed mass distribution \citep[e.g.][]{volonteri2008,reines2022}: BH growth models predict significantly different $M_{\rm BH} - \sigma$ relations at $M_{\rm BH} \sim 10^5-10^6$  $M_{\sun}$, for different initial mass functions of the seed BH, i.e. for different seed formation mechanisms. 
JWST has started to probe this BH mass regime at high redshift \citep[][]{kocevski2023,harikane2023,maiolino2023c,maiolino2024}, but in type 1 AGNs and using the so-called virial relations and single epoch observations, whose validity at high redshift (and specifically for the class of highly accreting BH identified by JWST) is uncertain. 

To properly test the BH seeding scenarios, direct measurements of BH masses in the range $\sim 10^5-10^6$  $M_{\sun}$ are needed.
This is feasible in the local universe, provided that one has high enough spectral resolution and high enough angular resolution.
A standard approach based on the measurement of gas and/or stellar kinematics would require to spatially resolve the BH sphere of influence which, for a $\sim 10^5$ $M_{\sun}$ BH, in a galaxy with central stellar velocity dispersion of about 50 \kms\  located at distance $D\sim20$ Mpc (e.g., in the Virgo cluster),
has a projected diameter of 4 mas. The diffraction limited spatial resolution of the ELT at
1.2~$\mu$m, 2.2~$\mu$m (e.g. at the location of Paschen-$\beta$, Br$\gamma$, H2) is 6 and 12 mas respectively,  meaning that even the ELT cannot reach the required spatial resolution. 

Spectroastrometry of gas emission lines is a method complementing the classical techniques and which, by exploiting high spectral resolution, allows to measure gas rotational velocities on scales as small as 1/10 of the spatial resolution. This method consists in tracing the spatial centroid of the emission line in many independent spectral channels: if the gas is circularly rotating in a disk, then the centroid will present a unique variation with wavelength which can be modelled to provide the mass of the central BH. 
The method has been tested and successfully used to measure BH masses \citep{gnerucci2010,gnerucci2011,gnerucci2013}. The same method has been successfully adopted in interferometric observations of the quasars 3C273 with VLTI-Gravity to spatially resolve, for the first time, the Broad Line Region and  revel its rotation \citep{gravitycoll2018}. 

By reaching even only 1/5 of the spatial resolution of the ELT, with spectroastrometry it will be possible to spatially resolve scales down to 1 and 2.5 mas at $\lambda \sim 1.2$~$\mu$m and 2.2~$\mu$m, respectively. Spectroastrometry with
the ELT will therefore allow us to spatially resolve the sphere of influence of $10^5$ $M_{\sun}$ BHs at a distance of 20 Mpc. 
To adequately sample the rotation curve of a $10^5$ $M_{\sun}$ BH at $D\sim20$ Mpc, at least 10 spectral channels should be sampling the nuclear rotation
curve ($30-60$ \kms), therefore the spectral resolution must be $(60 - 30)/10/2 = 3$ \kms, corresponding to R~$\sim 100,000$. To adequately map the centroid position of the central
unresolved source an IFU sampling the diffraction limited ELT PSF is needed. Complete near-IR spectral coverage will allow the use of several gas emission lines, Pa-$\beta$ in J, [FeII] in
H, and H2 and Br-$\gamma$ in K.
In summary, an IFU mode for ANDES working at the diffraction limit of the ELT and a spectral resolution of $\sim100,000$ would be able to detect BHs with masses down to $10^5$
$M_{\sun}$ up to a distance of $\sim20$ Mpc (i.e. Virgo).

\section{Extensions to ANDES design} 

\subsection{The benefits of the U-band}
\label{sec:Uband}

In Section~\ref{subsec:localgal}, we highlight how the extension of ANDES to the U-band ($0.35-0.4$ $\mu$m) would be of fundamental importance for the study of the CGM of the Milky Way and of the gas distribution in the Local Group. 
More generally, we would like to stress that almost the entire field of diffuse gas studies (IGM, CGM and ISM) over the Universe's history is based on absorption lines from electronic transitions, the vast majority of which occurring in the far UV. An extension to the U-band then corresponds directly to an extension in available redshifts. For instance, the study of gas metallicity requires access to \HI\ lines, and studying the transition to cold gas and interplay with stars (as well as AGN activity) requires covering H$_2$ lines. For systems at the raise and peak of cosmic star-formation history, these lines will be redshifted in the U-band.

\subsection{A higher sensitivity, lower resolving-power mode}
\label{sec:R50}
The majority of the science cases proposed in this work rely on the detection and measurement of spectral features which are generally resolved at a resolving power of $R\sim 50,000$. The comparison of quasar metal absorption lines  observed at $R\gtrsim 100,000$ and $R\sim 50,000$ shows how the velocity profile of complex absorption systems does not change substantially with the increase in resolution, although the different velocity components (and in particular the narrow ones) become more visible \citep[see e.g.][]{berg2022}. 


One of the aims of ELT-ANDES is to push the limits of the study of chemical enrichment to the early universe: this requires fine sensitivity. At $z\sim6$ most quasars have magnitudes fainter than $J_{\rm AB} \sim 19$ \citep{dodorico2023} and the brightest GRB afterglows at $z\sim6$ have magnitudes around $m_i\sim20$~mag that fade within hours \citep{hartoog2015,Tanvir21}. Targets at higher redshift will be fainter. An intermediate resolution mode ($R\sim50,000$) on ANDES will enable a $\sim60$\% higher sensitivity than with the full-resolution mode, in the J band. This will be crucial to be able to reach the sensitivity levels needed to study the faint high-$z$ sources. 

Simulations of GRB afterglow spectra show that the expected fraction of the total population of GRBs at $z>6$ is $\sim6$\% \citep{ghirlanda21}. Their magnitude 0.5 days after the GRB trigger in the observer frame is in the range $J_{\rm AB} \sim 17-27$, with a fraction of 20\%(35\%) of afterglows being brighter than 20(21) mag. In case of an RRM trigger just 1 hour after the explosion, 60\%(85\%) are brighter than 20(21) mag. Requiring a S/N $>10$ per spectral pixel to analyse absorption lines at the nominal resolution of $R$=100,000, a single 50 minutes integration on a mag=20 source is adequate, while a 21 mag source requires 4 hours on target\footnote{
\href{http://tirgo.arcetri.inaf.it/nicoletta/etc_andes_sn_com.html}{Based on Version 1.1 of the ANDES Exposure Time Calculator}}.

In summary, a lower spectral resolution mode ($R\sim50,000$), which additionally benefits from an improved instrument throughput, is highly desirable because it would increase the sensitivity of ANDES by a factor of $\sim$2 compared to the baseline instrument configuration.


\section{Synergy with present and future surveys and facilities}
\label{sec:synergy}
\subsection{Selection of ANDES targets}
Several upcoming telescopes and surveys will be important for optimising the selection of targets for ANDES. This includes Euclid and the Nancy Grace Roman Space Telescope, which are expected to extend the available sample of bright ($m_\mathrm{AB}<22$) $z\gtrsim 7$ quasars \citep{Barnett19,schindler2023,Tee23}, and the Legacy Survey of Space and Time (LSST) at the Vera C. Rubin Observatory \citep{ivezic2019}, which will allow the detection of super-luminous SNe up to $z\approx 5$ \citep{Villar18}.
The Dark Energy Spectroscopic Instrument (DESI) has recently added several hundred objects to the known set of $z\approx 5$--7 quasars at $m_\mathrm{AB}\lesssim 21$ in the southern hemisphere, and is projected to double their sample in coming years \citep{yang2023}. In the same redshift range, the future 4MOST survey is moreover expected to find $\sim 4\times 10^4$ quasars at $z > 5$ and $m_\mathrm{AB}\lesssim 22.5$ \citep{Merloni19}. On a longer horizon, the planned ESA THESEUS \citep{amati21} or NASA Gamow Explorer \citep{white21} missions may allow the detection of $z\gtrsim 6$ GRBs with afterglows that for a few hours will remain sufficiently bright for follow-up observations with ANDES \citep{ghirlanda21}.

\subsection{ANDES complemented by other instruments and future facilities}
The aspiration of ANDES to detect very metal-poor gas clouds and the chemical signatures of Pop~III stars in absorption offers strong synergies with other optical and infrared spectrographs that will operate on ELT and that are currently operating on JWST. Follow-up observations of quasar fields hosting relevant absorbers will be possible at, e.g., ELT with the MOSAIC spectrograph. This instrument will be vital in identifying both galaxies associated with the absorbers, including the host galaxies, and characterizing the larger-scale galaxy environment near this gas by reaching low stellar-mass systems. The prospects of ANDES for studying low-metallicity gas at high spectral resolution along the line of sight to quasars fainter than what can currently be reached will thus be well-matched to upcoming ELT capabilities to probe the faint end of the galaxy population that clusters with these gas absorption systems. 

While ANDES will be able to characterise the distribution and occurrence of absorbing gas clouds that bear the chemical imprints of the earliest generations of stars, JWST is already searching for Pop~III-powered galaxies that lack strong metal emission lines but exhibit very strong hydrogen and helium emission lines \citep[see early results by][]{Vanzella23,maiolino2023b}. In the future, it may be possible to combine the two approaches, and use JWST to search the fields around absorbers that exhibit signatures of Pop~III enrichment, to potentially identify the galaxies sites of recent Pop~III star formation. 

In the study of the high-redshift CGM, the combined use of the HARMONI/MOSAIC instruments at the ELT for the fainter, more numerous targets and of ANDES for the brighter quasars in the same field will moreover allow us to reconstruct the 3D distribution and physical properties of the gas. At lower redshifts, the tomographic mapping of CGM in Ca\,{\sc ii} from ANDES with H\,{\sc i} 21cm data from SKA will furthermore boost our understanding of the large-scale and small-scale neutral gas distribution around galaxies and its role for galaxy evolution.

In the study of cosmic reionisation, observations of the redshifted 21 cm signal from neutral hydrogen with SKA will allow new ways to connect small-scale (ANDES) and large-scale (SKA) constraints on the redshift evolution of the reionisation process. For rare high-redshift quasars or GRB afterglows that appear sufficiently bright at both near-IR and radio wavelengths, combined detections of metal (ANDES) and 21 cm neutral-hydrogen (SKA) absorption lines along the same sightlines may furthermore be possible \citep{Bhagwat22}.

While our focus is mainly on the ELT, we note that there will be two other extremely large telescopes, the Thirty Meter Telescope (TMT) and the Giant Magellan Telescope (GMT). They will be equipped with suites of instruments which will also provide complementary capabilities in terms of high-resolution (e.g. G-CLEF for GMT and MODHIS for TMT), multi-object (e.g. GMACS at the GMT, WFOS at the TMT) and integral field spectroscopy (e.g. GMTIFS at the GMT).




\section*{Declarations}

\begin{itemize}
\item Funding \\
J.S. Bolton is supported by Science and Technology Facilities Council (STFC) consolidated grant ST/X000982/1. 
A. De Cia acknowledges support by the Swiss National Science Foundation under grant 185692. 
S. Salvadori acknowledges support from the European Research Council (ERC) Starting Grant NEFERTITI H2020/808240. This project has received funding from the ERC under the European Union's Horizon 2020 research and innovation programme (grant agreement No 757535) and by Fondazione Cariplo (grant No 2018-2329). E. Zackrisson acknowledge funding from the Swedish National Space Board and project grant 2022-03804 from the Swedish Research Council (Vetenskapsr\aa{}det), and has benefited from a sabbatical at the Swedish Collegium for Advanced Study.  
\item Competing interests \\
The authors have no competing interests to declare that are relevant to the content of this article.
\item Availability of data and materials \\
Data used in the current study are available from the corresponding author on reasonable request.
\item Authors' contributions\\
V. D'Odorico: coordination, general writing and finalization of the paper; 
J. S. Bolton: coordination and writing of Sect.~\ref{sec:reioniz}; 
A. De Cia: coordination and writing of Sect.~\ref{sec:metal}; 
L. Christensen: coordination and writing of Sect.~\ref{sec:transient}; 
E. Zackrisson: writing of Sect.~\ref{sec:synergy}; 
S. Salvadori: writing of Subsect.~\ref{subsec:[X/Y]}; 
L. Izzo: writing of Subsect.~\ref{subsec:novae};
J. Li: writing of Subsect.~\ref{subsec:HeII};
P. Richter: writing of Subsect.~\ref{subsec:localgal}; 
R. Maiolino, A. Marconi: writing of Subsect.~\ref{subsec:BH}; 
A. Kordt, A. Saccardi, I. Vanni: help with simulations/forecasts, figures;
P. Di Marcantonio, R. Maiolino, A. Marconi, L. Origlia, A. Zanutta: ANDES project office support. \\
All authors provided comments to the draft. 
\end{itemize}

\bibliography{biblio_ANDES_WG3}

\begin{thebibliography}{204}
\providecommand{\natexlab}[1]{#1}
\providecommand{\url}[1]{{#1}}
\providecommand{\urlprefix}{URL }
\providecommand{\doi}[1]{\url{https://doi.org/#1}}
\providecommand{\eprint}[2][]{\url{#2}}
 \bibcommenthead

\bibitem[{{Abel} et~al(2002){Abel}, {Bryan}, and {Norman}}]{abel2002}
{Abel} T, {Bryan} GL, {Norman} ML (2002) {The Formation of the First Star in the Universe}. Science 295(5552):93--98. \doi{10.1126/science.295.5552.93}, {\href{https://arxiv.org/abs/astro-ph/0112088}{{arXiv:astro-ph/0112088}}} {[astro-ph]}

\bibitem[{{Ackermann} et~al(2014){Ackermann}, {Ajello}, {Albert}, {Baldini}, {Ballet}, {Barbiellini}, {Bastieri}, {Bellazzini}, {Bissaldi}, {Blandford}, {Bloom}, {Bottacini}, {Brandt}, {Bregeon}, {Bruel}, {Buehler}, {Buson}, {Caliandro}, {Cameron}, {Caragiulo}, {Caraveo}, {Cavazzuti}, {Charles}, {Chekhtman}, {Cheung}, {Chiang}, {Chiaro}, {Ciprini}, {Claus}, {Cohen-Tanugi}, {Conrad}, {Corbel}, {D'Ammando}, {de Angelis}, {den Hartog}, {de Palma}, {Dermer}, {Desiante}, {Digel}, {Di Venere}, {do Couto e Silva}, {Donato}, {Drell}, {Drlica-Wagner}, {Favuzzi}, {Ferrara}, {Focke}, {Franckowiak}, {Fuhrmann}, {Fukazawa}, {Fusco}, {Gargano}, {Gasparrini}, {Germani}, {Giglietto}, {Giordano}, {Giroletti}, {Glanzman}, {Godfrey}, {Grenier}, {Grove}, {Guiriec}, {Hadasch}, {Harding}, {Hayashida}, {Hays}, {Hewitt}, {Hill}, {Hou}, {Jean}, {Jogler}, {J{\'o}hannesson}, {Johnson}, {Johnson}, {Kerr}, {Kn{\"o}dlseder}, {Kuss}, {Larsson}, {Latronico}, {Lemoine-Goumard}, {Longo}, {Loparco}, {Lott}, {Lovellette}, {Lubrano}, {Manfreda},
  {Martin}, {Massaro}, {Mayer}, {Mazziotta}, {McEnery}, {Michelson}, {Mitthumsiri}, {Mizuno}, {Monzani}, {Morselli}, {Moskalenko}, {Murgia}, {Nemmen}, {Nuss}, {Ohsugi}, {Omodei}, {Orienti}, {Orlando}, {Ormes}, {Paneque}, {Panetta}, {Perkins}, {Pesce-Rollins}, {Piron}, {Pivato}, {Porter}, {Rain{\`o}}, {Rando}, {Razzano}, {Razzaque}, {Reimer}, {Reimer}, {Reposeur}, {Saz Parkinson}, {Schaal}, {Schulz}, {Sgr{\`o}}, {Siskind}, {Spandre}, {Spinelli}, {Stawarz}, {Suson}, {Takahashi}, {Tanaka}, {Thayer}, {Thayer}, {Thompson}, {Tibaldo}, {Tinivella}, {Torres}, {Tosti}, {Troja}, {Uchiyama}, {Vianello}, {Winer}, {Wolff}, {Wood}, {Wood}, {Wood}, {Charbonnel}, {Corbet}, {De Gennaro Aquino}, {Edlin}, {Mason}, {Schwarz}, {Shore}, {Starrfield}, {Teyssier}, and {Fermi-LAT Collaboration}}]{fermi14}
{Ackermann} M, {Ajello} M, {Albert} A, et~al (2014) {Fermi establishes classical novae as a distinct class of gamma-ray sources}. Science 345(6196):554--558. \doi{10.1126/science.1253947}, {\href{https://arxiv.org/abs/1408.0735}{{arXiv:1408.0735}}} {[astro-ph.HE]}

\bibitem[{{Amati} et~al(2021){Amati}, {O'Brien}, {G{\"o}tz}, {Bozzo}, {Santangelo}, {Tanvir}, {Frontera}, {Mereghetti}, {Osborne}, {Blain}, {Basa}, {Branchesi}, {Burderi}, {Caballero-Garc{\'\i}a}, {Castro-Tirado}, {Christensen}, {Ciolfi}, {De Rosa}, {Doroshenko}, {Ferrara}, {Ghirlanda}, {Hanlon}, {Heddermann}, {Hutchinson}, {Labanti}, {Le Floch}, {Lerman}, {Paltani}, {Reglero}, {Rezzolla}, {Rosati}, {Salvaterra}, {Stratta}, {Tenzer}, and {Theseus Consortium}}]{amati21}
{Amati} L, {O'Brien} PT, {G{\"o}tz} D, et~al (2021) {The THESEUS space mission: science goals, requirements and mission concept}. Experimental Astronomy 52(3):183--218. \doi{10.1007/s10686-021-09807-8}, {\href{https://arxiv.org/abs/2104.09531}{{arXiv:2104.09531}}} {[astro-ph.IM]}

\bibitem[{{Arnould} and {Norgaard}(1975)}]{Arnould1975}
{Arnould} M, {Norgaard} H (1975) {The Explosive Thermonuclear Formation of 7Li and 11B}. \aap 42:55

\bibitem[{{Asplund} et~al(2009){Asplund}, {Grevesse}, {Sauval}, and {Scott}}]{Asplund2009}
{Asplund} M, {Grevesse} N, {Sauval} AJ, et~al (2009) {The Chemical Composition of the Sun}. ARA\&A 47(1):481--522. \doi{10.1146/annurev.astro.46.060407.145222}, {\href{https://arxiv.org/abs/0909.0948}{{arXiv:0909.0948}}} {[astro-ph.SR]}

\bibitem[{{Asthana} et~al(2024){Asthana}, {Haehnelt}, {Kulkarni}, {Bolton}, {Gaikwad}, {Keating}, and {Puchwein}}]{Asthana2024}
{Asthana} S, {Haehnelt} MG, {Kulkarni} G, et~al (2024) {The impact of faint AGN discovered by JWST on reionization}. arXiv e-prints arXiv:2409.15453. \doi{10.48550/arXiv.2409.15453}, {\href{https://arxiv.org/abs/2409.15453}{{arXiv:2409.15453}}} {[astro-ph.GA]}

\bibitem[{{Aydi} et~al(2020{\natexlab{a}}){Aydi}, {Chomiuk}, {Izzo}, {Harvey}, {Leahy-McGregor}, {Strader}, {Buckley}, {Sokolovsky}, {Kawash}, {Kochanek}, {Linford}, {Metzger}, {Mukai}, {Orio}, {Shappee}, {Shishkovsky}, {Steinberg}, {Swihart}, {Sokoloski}, {Walter}, and {Woudt}}]{aydi20b}
{Aydi} E, {Chomiuk} L, {Izzo} L, et~al (2020{\natexlab{a}}) {Early Spectral Evolution of Classical Novae: Consistent Evidence for Multiple Distinct Outflows}. \apj 905(1):62. \doi{10.3847/1538-4357/abc3bb}, {\href{https://arxiv.org/abs/2010.07481}{{arXiv:2010.07481}}} {[astro-ph.HE]}

\bibitem[{{Aydi} et~al(2020{\natexlab{b}}){Aydi}, {Sokolovsky}, {Chomiuk}, {Steinberg}, {Li}, {Vurm}, {Metzger}, {Strader}, {Mukai}, {Pejcha}, {Shen}, {Wade}, {Kuschnig}, {Moffat}, {Pablo}, {Pigulski}, {Popowicz}, {Weiss}, {Zwintz}, {Izzo}, {Pollard}, {Handler}, {Ryder}, {Filipovi{\'c}}, {Alsaberi}, {Manojlovi{\'c}}, {Lopes de Oliveira}, {Walter}, {Vallely}, {Buckley}, {Brown}, {Harvey}, {Kawash}, {Kniazev}, {Kochanek}, {Linford}, {Mikolajewska}, {Molaro}, {Orio}, {Page}, {Shappee}, and {Sokoloski}}]{aydi20a}
{Aydi} E, {Sokolovsky} KV, {Chomiuk} L, et~al (2020{\natexlab{b}}) {Direct evidence for shock-powered optical emission in a nova}. Nature Astronomy 4:776--780. \doi{10.1038/s41550-020-1070-y}, {\href{https://arxiv.org/abs/2004.05562}{{arXiv:2004.05562}}} {[astro-ph.HE]}

\bibitem[{{Battaglia} et~al(2017){Battaglia}, {North}, {Jablonka}, {Shetrone}, {Minniti}, {D{\'\i}az}, {Starkenburg}, and {Savoy}}]{battaglia2017}
{Battaglia} G, {North} P, {Jablonka} P, et~al (2017) {What is the Milky Way outer halo made of?. High resolution spectroscopy of distant red giants}. \aap 608:A145. \doi{10.1051/0004-6361/201731879}, {\href{https://arxiv.org/abs/1710.01320}{{arXiv:1710.01320}}} {[astro-ph.GA]}

\bibitem[{{Becker} et~al(2015){Becker}, {Bolton}, and {Lidz}}]{becker2015}
{Becker} GD, {Bolton} JS, {Lidz} A (2015) {Reionisation and High-Redshift Galaxies: The View from Quasar Absorption Lines}. PASA 32:45. \doi{10.1017/pasa.2015.45}, {\href{https://arxiv.org/abs/1510.03368}{{arXiv:1510.03368}}} {[astro-ph.CO]}

\bibitem[{{Becker} et~al(2019){Becker}, {Pettini}, {Rafelski}, {D'Odorico}, {Boera}, {Christensen}, {Cupani}, {Ellison}, {Farina}, {Fumagalli}, {L{\'o}pez}, {Neeleman}, {Ryan-Weber}, and {Worseck}}]{becker2019}
{Becker} GD, {Pettini} M, {Rafelski} M, et~al (2019) {The Evolution of O I over $3.2 < z < 6.5$: Reionization of the Circumgalactic Medium}. \apj 883(2):163. \doi{10.3847/1538-4357/ab3eb5}, {\href{https://arxiv.org/abs/1907.02983}{{arXiv:1907.02983}}} {[astro-ph.GA]}

\bibitem[{{Becker} et~al(2021){Becker}, {D'Aloisio}, {Christenson}, {Zhu}, {Worseck}, and {Bolton}}]{becker2021}
{Becker} GD, {D'Aloisio} A, {Christenson} HM, et~al (2021) {The mean free path of ionizing photons at $5 < z < 6$: evidence for rapid evolution near reionization}. \mnras 508(2):1853--1869. \doi{10.1093/mnras/stab2696}, {\href{https://arxiv.org/abs/2103.16610}{{arXiv:2103.16610}}} {[astro-ph.CO]}

\bibitem[{{Beers} and {Christlieb}(2005)}]{Beers05}
{Beers} TC, {Christlieb} N (2005) {The Discovery and Analysis of Very Metal-Poor Stars in the Galaxy}. \araa 43(1):531--580. \doi{10.1146/annurev.astro.42.053102.134057}

\bibitem[{{Behroozi} and {Silk}(2018)}]{behroozi2018}
{Behroozi} P, {Silk} J (2018) {The most massive galaxies and black holes allowed by {\ensuremath{\Lambda}}CDM}. \mnras 477(4):5382--5387. \doi{10.1093/mnras/sty945}, {\href{https://arxiv.org/abs/1609.04402}{{arXiv:1609.04402}}} {[astro-ph.GA]}

\bibitem[{{Ben Bekhti} et~al(2008){Ben Bekhti}, {Richter}, {Westmeier}, and {Murphy}}]{benbekhti2008}
{Ben Bekhti} N, {Richter} P, {Westmeier} T, et~al (2008) {Ca II and Na I absorption signatures from extraplanar gas in the halo of the Milky Way}. \aap 487(2):583--594. \doi{10.1051/0004-6361:20079067}, {\href{https://arxiv.org/abs/0806.3204}{{arXiv:0806.3204}}} {[astro-ph]}

\bibitem[{{Ben Bekhti} et~al(2012){Ben Bekhti}, {Winkel}, {Richter}, {Kerp}, {Klein}, and {Murphy}}]{benbekhti2012}
{Ben Bekhti} N, {Winkel} B, {Richter} P, et~al (2012) {An absorption-selected survey of neutral gas in the Milky Way halo. New results based on a large sample of Ca II, Na I, and H I spectra towards QSOs}. \aap 542:A110. \doi{10.1051/0004-6361/201118673}, {\href{https://arxiv.org/abs/1203.5603}{{arXiv:1203.5603}}} {[astro-ph.GA]}

\bibitem[{{Berg} et~al(2022){Berg}, {Cupani}, {Figueira}, and {Mehner}}]{berg2022}
{Berg} TAM, {Cupani} G, {Figueira} P, et~al (2022) {Performance of ESPRESSO'S high-resolution 4 {\texttimes} 2 binning for characterising intervening absorbers towards faint quasars}. \aap 662:A35. \doi{10.1051/0004-6361/202243208}

\bibitem[{{Bhagwat} et~al(2022){Bhagwat}, {Ciardi}, {Zackrisson}, and {Schaye}}]{Bhagwat22}
{Bhagwat} A, {Ciardi} B, {Zackrisson} E, et~al (2022) {Cospatial 21 cm and metal-line absorbers in the epoch of reionization - I. Incidence and observability}. \mnras 517(2):2331--2342. \doi{10.1093/mnras/stac2663}, {\href{https://arxiv.org/abs/2209.10573}{{arXiv:2209.10573}}} {[astro-ph.CO]}

\bibitem[{{Bode} and {Evans}(2008)}]{Bode2008}
{Bode} MF, {Evans} A (2008) {Classical Novae}, vol~43

\bibitem[{{Bolmer} et~al(2019){Bolmer}, {Ledoux}, {Wiseman}, {De Cia}, {Selsing}, {Schady}, {Greiner}, {Savaglio}, {Burgess}, {D'Elia}, {Fynbo}, {Goldoni}, {Hartmann}, {Heintz}, {Jakobsson}, {Japelj}, {Kaper}, {Tanvir}, {Vreeswijk}, and {Zafar}}]{bolmer19}
{Bolmer} J, {Ledoux} C, {Wiseman} P, et~al (2019) {Evidence for diffuse molecular gas and dust in the hearts of gamma-ray burst host galaxies. Unveiling the nature of high-redshift damped Lyman-{\ensuremath{\alpha}} systems}. \aap 623:A43. \doi{10.1051/0004-6361/201834422}, {\href{https://arxiv.org/abs/1810.06403}{{arXiv:1810.06403}}} {[astro-ph.GA]}

\bibitem[{{Bolton} et~al(2012){Bolton}, {Becker}, {Raskutti}, {Wyithe}, {Haehnelt}, and {Sargent}}]{Bolton2012}
{Bolton} JS, {Becker} GD, {Raskutti} S, et~al (2012) {Improved measurements of the intergalactic medium temperature around quasars: possible evidence for the initial stages of He II reionization at z=6}. \mnras 419(4):2880--2892. \doi{10.1111/j.1365-2966.2011.19929.x}, {\href{https://arxiv.org/abs/1110.0539}{{arXiv:1110.0539}}} {[astro-ph.CO]}

\bibitem[{{Bosman} et~al(2022){Bosman}, {Davies}, {Becker}, {Keating}, {Davies}, {Zhu}, {Eilers}, {D'Odorico}, {Bian}, {Bischetti}, {Cristiani}, {Fan}, {Farina}, {Haehnelt}, {Hennawi}, {Kulkarni}, {Mesinger}, {Meyer}, {Onoue}, {Pallottini}, {Qin}, {Ryan-Weber}, {Schindler}, {Walter}, {Wang}, and {Yang}}]{bosman2022}
{Bosman} SEI, {Davies} FB, {Becker} GD, et~al (2022) {Hydrogen reionization ends by z = 5.3: Lyman-{\ensuremath{\alpha}} optical depth measured by the XQR-30 sample}. \mnras 514(1):55--76. \doi{10.1093/mnras/stac1046}, {\href{https://arxiv.org/abs/2108.03699}{{arXiv:2108.03699}}} {[astro-ph.CO]}

\bibitem[{{Bromm}(2013)}]{bromm2013}
{Bromm} V (2013) {Formation of the first stars}. Reports on Progress in Physics 76(11):112901. \doi{10.1088/0034-4885/76/11/112901}, {\href{https://arxiv.org/abs/1305.5178}{{arXiv:1305.5178}}} {[astro-ph.CO]}

\bibitem[{{Chen} and {Gnedin}(2021)}]{ChenGnedin2021}
{Chen} H, {Gnedin} NY (2021) {Recovering Density Fields inside Quasar Proximity Zones at z 6}. \apj 916(2):118. \doi{10.3847/1538-4357/ac0429}, {\href{https://arxiv.org/abs/2101.11627}{{arXiv:2101.11627}}} {[astro-ph.CO]}

\bibitem[{{Chen} et~al(2022){Chen}, {Eilers}, {Bosman}, {Gnedin}, {Fan}, {Wang}, {Yang}, {D'Odorico}, {Becker}, {Bischetti}, {Mazzucchelli}, {Mesinger}, and {Pallottini}}]{chen2022}
{Chen} H, {Eilers} AC, {Bosman} SEI, et~al (2022) {Measuring the Density Fields around Bright Quasars at $z \sim 6$ with XQR-30 Spectra}. \apj 931(1):29. \doi{10.3847/1538-4357/ac658d}, {\href{https://arxiv.org/abs/2110.13917}{{arXiv:2110.13917}}} {[astro-ph.GA]}

\bibitem[{{Chen} et~al(2023{\natexlab{a}}){Chen}, {Yan}, {Kangas}, {Lunnan}, {Schulze}, {Sollerman}, {Perley}, {Chen}, {Taggart}, {Hinds}, {Gal-Yam}, {Wang}, {Andreoni}, {Bellm}, {Bloom}, {Burdge}, {Burgos}, {Cook}, {Dahiwale}, {De}, {Dekany}, {Dugas}, {Frederik}, {Fremling}, {Graham}, {Hankins}, {Ho}, {Jencson}, {Karambelkar}, {Kasliwal}, {Kulkarni}, {Laher}, {Rusholme}, {Sharma}, {Taddia}, {Tartaglia}, {Thomas}, {Tzanidakis}, {Van Roestel}, {Walter}, {Yang}, {Yao}, and {Yaron}}]{chen2023b}
{Chen} ZH, {Yan} L, {Kangas} T, et~al (2023{\natexlab{a}}) {The Hydrogen-poor Superluminous Supernovae from the Zwicky Transient Facility Phase I Survey. I. Light Curves and Measurements}. \apj 943(1):41. \doi{10.3847/1538-4357/aca161}, {\href{https://arxiv.org/abs/2202.02059}{{arXiv:2202.02059}}} {[astro-ph.HE]}

\bibitem[{{Chen} et~al(2023{\natexlab{b}}){Chen}, {Yan}, {Kangas}, {Lunnan}, {Sollerman}, {Schulze}, {Perley}, {Chen}, {Taggart}, {Hinds}, {Gal-Yam}, {Wang}, {De}, {Bellm}, {Bloom}, {Dekany}, {Graham}, {Kasliwal}, {Kulkarni}, {Laher}, {Neill}, and {Rusholme}}]{chen2023a}
{Chen} ZH, {Yan} L, {Kangas} T, et~al (2023{\natexlab{b}}) {The Hydrogen-poor Superluminous Supernovae from the Zwicky Transient Facility Phase I Survey. II. Light-curve Modeling and Characterization of Undulations}. \apj 943(1):42. \doi{10.3847/1538-4357/aca162}, {\href{https://arxiv.org/abs/2202.02060}{{arXiv:2202.02060}}} {[astro-ph.HE]}

\bibitem[{{Christensen} et~al(2023){Christensen}, {Jakobsen}, {Willott}, {Arribas}, {Bunker}, {Charlot}, {Maiolino}, {Marshall}, {Perna}, and {{\"U}bler}}]{christensen2023}
{Christensen} L, {Jakobsen} P, {Willott} C, et~al (2023) {Metal enrichment and evolution in four z > 6.5 quasar sightlines observed with JWST/NIRSpec}. \aap 680:A82. \doi{10.1051/0004-6361/202347943}, {\href{https://arxiv.org/abs/2309.06470}{{arXiv:2309.06470}}} {[astro-ph.GA]}

\bibitem[{{Cooke} et~al(2012){Cooke}, {Sullivan}, {Gal-Yam}, {Barton}, {Carlberg}, {Ryan-Weber}, {Horst}, {Omori}, and {D{\'\i}az}}]{cooke12}
{Cooke} J, {Sullivan} M, {Gal-Yam} A, et~al (2012) {Superluminous supernovae at redshifts of 2.05 and 3.90}. \nat 491(7423):228--231. \doi{10.1038/nature11521}, {\href{https://arxiv.org/abs/1211.2003}{{arXiv:1211.2003}}} {[astro-ph.CO]}

\bibitem[{{Cooke} et~al(2011){Cooke}, {Pettini}, {Steidel}, {Rudie}, and {Nissen}}]{Cooke11}
{Cooke} R, {Pettini} M, {Steidel} CC, et~al (2011) {The most metal-poor damped Ly{\ensuremath{\alpha}} systems: insights into chemical evolution in the very metal-poor regime}. \mnras 417(2):1534--1558. \doi{10.1111/j.1365-2966.2011.19365.x}

\bibitem[{{Cooke} et~al(2017){Cooke}, {Pettini}, and {Steidel}}]{Cooke17}
{Cooke} RJ, {Pettini} M, {Steidel} CC (2017) {Discovery of the most metal-poor damped Lyman-{\ensuremath{\alpha}} system}. \mnras 467(1):802--811. \doi{10.1093/mnras/stx037}, {\href{https://arxiv.org/abs/1701.03103}{{arXiv:1701.03103}}} {[astro-ph.CO]}

\bibitem[{{Cooksey} et~al(2013){Cooksey}, {Kao}, {Simcoe}, {O'Meara}, and {Prochaska}}]{cooksey2013}
{Cooksey} KL, {Kao} MM, {Simcoe} RA, et~al (2013) {Precious Metals in SDSS Quasar Spectra. I. Tracking the Evolution of Strong, $1.5 < z < 4.5$ C IV Absorbers with Thousands of Systems}. \apj 763(1):37. \doi{10.1088/0004-637X/763/1/37}, {\href{https://arxiv.org/abs/1204.2827}{{arXiv:1204.2827}}} {[astro-ph.CO]}

\bibitem[{{Crighton} et~al(2016){Crighton}, {O'Meara}, and {Murphy}}]{Crighton16}
{Crighton} NHM, {O'Meara} JM, {Murphy} MT (2016) {Possible Population III remnants at redshift 3.5}. \mnras 457(1):L44--L48. \doi{10.1093/mnrasl/slv191}, {\href{https://arxiv.org/abs/1512.00477}{{arXiv:1512.00477}}} {[astro-ph.GA]}

\bibitem[{{Cucchiara} et~al(2011){Cucchiara}, {Levan}, {Fox}, {Tanvir}, {Ukwatta}, {Berger}, {Kr{\"u}hler}, {K{\"u}pc{\"u} Yolda{\c{s}}}, {Wu}, {Toma}, {Greiner}, {Olivares}, {Rowlinson}, {Amati}, {Sakamoto}, {Roth}, {Stephens}, {Fritz}, {Fynbo}, {Hjorth}, {Malesani}, {Jakobsson}, {Wiersema}, {O'Brien}, {Soderberg}, {Foley}, {Fruchter}, {Rhoads}, {Rutledge}, {Schmidt}, {Dopita}, {Podsiadlowski}, {Willingale}, {Wolf}, {Kulkarni}, and {D'Avanzo}}]{cucchiara11}
{Cucchiara} A, {Levan} AJ, {Fox} DB, et~al (2011) {A Photometric Redshift of z \raisebox{-0.5ex}\textasciitilde 9.4 for GRB 090429B}. \apj 736(1):7. \doi{10.1088/0004-637X/736/1/7}, {\href{https://arxiv.org/abs/1105.4915}{{arXiv:1105.4915}}} {[astro-ph.CO]}

\bibitem[{{Davies} and {Hennawi}(2023)}]{DaviesHennawi2023}
{Davies} FB, {Hennawi} JF (2023) {Signatures of Small-scale Structure of the Pre-reionization Intergalactic Medium in $z\gtrsim7$ Quasar Proximity Zones}. arXiv e-prints arXiv:2312.06763. \doi{10.48550/arXiv.2312.06763}, {\href{https://arxiv.org/abs/2312.06763}{{arXiv:2312.06763}}} {[astro-ph.CO]}

\bibitem[{{De Cia} et~al(2012){De Cia}, {Ledoux}, {Fox}, {Vreeswijk}, {Smette}, {Petitjean}, {Bj{\"o}rnsson}, {Fynbo}, {Hjorth}, and {Jakobsson}}]{decia12}
{De Cia} A, {Ledoux} C, {Fox} AJ, et~al (2012) {Rapid-response mode VLT/UVES spectroscopy of super iron-rich gas exposed to GRB 080310. Evidence of ionization in action and episodic star formation in the host}. \aap 545:A64. \doi{10.1051/0004-6361/201218884}, {\href{https://arxiv.org/abs/1207.6102}{{arXiv:1207.6102}}} {[astro-ph.CO]}

\bibitem[{{De Cia} et~al(2016){De Cia}, {Ledoux}, {Mattsson}, {Petitjean}, {Srianand}, {Gavignaud}, and {Jenkins}}]{decia16}
{De Cia} A, {Ledoux} C, {Mattsson} L, et~al (2016) {Dust-depletion sequences in damped Lyman-{\ensuremath{\alpha}} absorbers. A unified picture from low-metallicity systems to the Galaxy}. \aap 596:A97. \doi{10.1051/0004-6361/201527895}, {\href{https://arxiv.org/abs/1608.08621}{{arXiv:1608.08621}}} {[astro-ph.GA]}

\bibitem[{{De Cia} et~al(2018){De Cia}, {Gal-Yam}, {Rubin}, {Leloudas}, {Vreeswijk}, {Perley}, {Quimby}, {Yan}, {Sullivan}, {Fl{\"o}rs}, {Sollerman}, {Bersier}, {Cenko}, {Gal-Yam}, {Maguire}, {Ofek}, {Prentice}, {Schulze}, {Spyromilio}, {Valenti}, {Arcavi}, {Corsi}, {Howell}, {Mazzali}, {Kasliwal}, {Taddia}, and {Yaron}}]{decia18}
{De Cia} A, {Gal-Yam} A, {Rubin} A, et~al (2018) {Light Curves of Hydrogen-poor Superluminous Supernovae from the Palomar Transient Factory}. \apj 860(2):100. \doi{10.3847/1538-4357/aab9b6}, {\href{https://arxiv.org/abs/1708.01623}{{arXiv:1708.01623}}} {[astro-ph.HE]}

\bibitem[{{De Cia} et~al(2021){De Cia}, {Jenkins}, {Fox}, {Ledoux}, {Ramburuth-Hurt}, {Konstantopoulou}, {Petitjean}, and {Krogager}}]{decia21}
{De Cia} A, {Jenkins} EB, {Fox} AJ, et~al (2021) {Large metallicity variations in the Galactic interstellar medium}. \nat 597(7875):206--208. \doi{10.1038/s41586-021-03780-0}, {\href{https://arxiv.org/abs/2109.03249}{{arXiv:2109.03249}}} {[astro-ph.GA]}

\bibitem[{{D'Elia} et~al(2010){D'Elia}, {Fynbo}, {Covino}, {Goldoni}, {Jakobsson}, {Matteucci}, {Piranomonte}, {Sollerman}, {Th{\"o}ne}, {Vergani}, {Vreeswijk}, {Watson}, {Wiersema}, {Zafar}, {de Ugarte Postigo}, {Flores}, {Hjorth}, {Kaper}, {Levan}, {Malesani}, {Milvang-Jensen}, {Pian}, {Tagliaferri}, and {Tanvir}}]{delia10}
{D'Elia} V, {Fynbo} JPU, {Covino} S, et~al (2010) {VLT/X-shooter spectroscopy of the GRB 090926A afterglow}. \aap 523:A36. \doi{10.1051/0004-6361/201015216}, {\href{https://arxiv.org/abs/1007.5357}{{arXiv:1007.5357}}} {[astro-ph.HE]}

\bibitem[{{Della Valle} and {Izzo}(2020)}]{DellaValle2020}
{Della Valle} M, {Izzo} L (2020) {Observations of galactic and extragalactic novae}. \aapr 28(1):3. \doi{10.1007/s00159-020-0124-6}, {\href{https://arxiv.org/abs/2004.06540}{{arXiv:2004.06540}}} {[astro-ph.SR]}

\bibitem[{{D{\'\i}az} et~al(2021){D{\'\i}az}, {Ryan-Weber}, {Karman}, {Caputi}, {Salvadori}, {Crighton}, {Ouchi}, and {Vanzella}}]{diaz2021}
{D{\'\i}az} CG, {Ryan-Weber} EV, {Karman} W, et~al (2021) {Faint LAEs near $z > 4.7$ C IV absorbers revealed by MUSE}. \mnras 502(2):2645--2663. \doi{10.1093/mnras/staa3129}, {\href{https://arxiv.org/abs/2001.04453}{{arXiv:2001.04453}}} {[astro-ph.GA]}

\bibitem[{{D'Odorico} et~al(2010){D'Odorico}, {Calura}, {Cristiani}, and {Viel}}]{dodorico2010}
{D'Odorico} V, {Calura} F, {Cristiani} S, et~al (2010) {The rise of the C IV mass density at $z < 2.5$}. \mnras 401(4):2715--2721. \doi{10.1111/j.1365-2966.2009.15856.x}, {\href{https://arxiv.org/abs/0910.2126}{{arXiv:0910.2126}}} {[astro-ph.CO]}

\bibitem[{{D'Odorico} et~al(2016){D'Odorico}, {Cristiani}, {Pomante}, {Carswell}, {Viel}, {Barai}, {Becker}, {Calura}, {Cupani}, {Fontanot}, {Haehnelt}, {Kim}, {Miralda-Escud{\'e}}, {Rorai}, {Tescari}, and {Vanzella}}]{dodorico2016}
{D'Odorico} V, {Cristiani} S, {Pomante} E, et~al (2016) {Metals in the $z \sim 3$ intergalactic medium: results from an ultra-high signal-to-noise ratio UVES quasar spectrum}. \mnras 463(3):2690--2707. \doi{10.1093/mnras/stw2161}, {\href{https://arxiv.org/abs/1608.06116}{{arXiv:1608.06116}}} {[astro-ph.GA]}

\bibitem[{{D'Odorico} et~al(2022){D'Odorico}, {Finlator}, {Cristiani}, {Cupani}, {Perrotta}, {Calura}, {C{\`e}nturion}, {Becker}, {Berg}, {Lopez}, {Ellison}, and {Pomante}}]{dodorico2022}
{D'Odorico} V, {Finlator} K, {Cristiani} S, et~al (2022) {The evolution of the Si IV content in the Universe from the epoch of reionization to cosmic noon}. \mnras 512(2):2389--2401. \doi{10.1093/mnras/stac545}, {\href{https://arxiv.org/abs/2202.12206}{{arXiv:2202.12206}}} {[astro-ph.GA]}

\bibitem[{{D'Odorico} et~al(2023){D'Odorico}, {Ba{\~n}ados}, {Becker}, {Bischetti}, {Bosman}, {Cupani}, {Davies}, {Farina}, {Ferrara}, {Feruglio}, {Mazzucchelli}, {Ryan-Weber}, {Schindler}, {Sodini}, {Venemans}, {Walter}, {Chen}, {Lai}, {Zhu}, {Bian}, {Campo}, {Carniani}, {Cristiani}, {Davies}, {Decarli}, {Drake}, {Eilers}, {Fan}, {Gaikwad}, {Gallerani}, {Greig}, {Haehnelt}, {Hennawi}, {Keating}, {Kulkarni}, {Mesinger}, {Meyer}, {Neeleman}, {Onoue}, {Pallottini}, {Qin}, {Rojas-Ruiz}, {Satyavolu}, {Sebastian}, {Tripodi}, {Wang}, {Wolfson}, {Yang}, and {Zanchettin}}]{dodorico2023}
{D'Odorico} V, {Ba{\~n}ados} E, {Becker} GD, et~al (2023) {XQR-30: The ultimate XSHOOTER quasar sample at the reionization epoch}. \mnras 523(1):1399--1420. \doi{10.1093/mnras/stad1468}, {\href{https://arxiv.org/abs/2305.05053}{{arXiv:2305.05053}}} {[astro-ph.GA]}

\bibitem[{{Doughty} and {Finlator}(2019)}]{doughty19}
{Doughty} C, {Finlator} K (2019) {Evolution of neutral oxygen during the epoch of reionization and its use in estimating the neutral hydrogen fraction}. \mnras 489(2):2755--2768. \doi{10.1093/mnras/stz2331}, {\href{https://arxiv.org/abs/1908.08549}{{arXiv:1908.08549}}} {[astro-ph.CO]}

\bibitem[{{Ellison} et~al(2000){Ellison}, {Songaila}, {Schaye}, and {Pettini}}]{ellison2000}
{Ellison} SL, {Songaila} A, {Schaye} J, et~al (2000) {The Enrichment History of the Intergalactic Medium-Measuring the C IV/H I Ratio in the Ly{\ensuremath{\alpha}} Forest}. \aj 120(3):1175--1191. \doi{10.1086/301511}, {\href{https://arxiv.org/abs/astro-ph/0005448}{{arXiv:astro-ph/0005448}}} {[astro-ph]}

\bibitem[{{Euclid Collaboration} et~al(2019){Euclid Collaboration}, {Barnett}, {Warren}, {Mortlock}, {Cuby}, {Conselice}, {Hewett}, {Willott}, {Auricchio}, {Balaguera-Antol{\'\i}nez}, {Baldi}, {Bardelli}, {Bellagamba}, {Bender}, {Biviano}, {Bonino}, {Bozzo}, {Branchini}, {Brescia}, {Brinchmann}, {Burigana}, {Camera}, {Capobianco}, {Carbone}, {Carretero}, {Carvalho}, {Castander}, {Castellano}, {Cavuoti}, {Cimatti}, {Cl{\'e}dassou}, {Congedo}, {Conversi}, {Copin}, {Corcione}, {Coupon}, {Courtois}, {Cropper}, {Da Silva}, {Duncan}, {Dusini}, {Ealet}, {Farrens}, {Fosalba}, {Fotopoulou}, {Fourmanoit}, {Frailis}, {Fumana}, {Galeotta}, {Garilli}, {Gillard}, {Gillis}, {Graci{\'a}-Carpio}, {Grupp}, {Hoekstra}, {Hormuth}, {Israel}, {Jahnke}, {Kermiche}, {Kilbinger}, {Kirkpatrick}, {Kitching}, {Kohley}, {Kubik}, {Kunz}, {Kurki-Suonio}, {Laureijs}, {Ligori}, {Lilje}, {Lloro}, {Maiorano}, {Mansutti}, {Marggraf}, {Martinet}, {Marulli}, {Massey}, {Mauri}, {Medinaceli}, {Mei}, {Mellier}, {Metcalf}, {Metge}, {Meylan},
  {Moresco}, {Moscardini}, {Munari}, {Neissner}, {Niemi}, {Nutma}, {Padilla}, {Paltani}, {Pasian}, {Paykari}, {Percival}, {Pettorino}, {Polenta}, {Poncet}, {Pozzetti}, {Raison}, {Renzi}, {Rhodes}, {Rix}, {Romelli}, {Roncarelli}, {Rossetti}, {Saglia}, {Sapone}, {Scaramella}, {Schneider}, {Scottez}, {Secroun}, {Serrano}, {Sirri}, {Stanco}, {Sureau}, {Tallada-Cresp{\'\i}}, {Tavagnacco}, {Taylor}, {Tenti}, {Tereno}, {Toledo-Moreo}, {Torradeflot}, {Valenziano}, {Vassallo}, {Wang}, {Zacchei}, {Zamorani}, {Zoubian}, and {Zucca}}]{Barnett19}
{Euclid Collaboration}, {Barnett} R, {Warren} SJ, et~al (2019) {Euclid preparation. V. Predicted yield of redshift $7 < z < 9$ quasars from the wide survey}. \aap 631:A85. \doi{10.1051/0004-6361/201936427}, {\href{https://arxiv.org/abs/1908.04310}{{arXiv:1908.04310}}} {[astro-ph.GA]}

\bibitem[{{Fabian}(2012)}]{fabian2012}
{Fabian} AC (2012) {Observational Evidence of Active Galactic Nuclei Feedback}. \araa 50:455--489. \doi{10.1146/annurev-astro-081811-125521}, {\href{https://arxiv.org/abs/1204.4114}{{arXiv:1204.4114}}} {[astro-ph.CO]}

\bibitem[{{Fan} et~al(2006){Fan}, {Strauss}, {Becker}, {White}, {Gunn}, {Knapp}, {Richards}, {Schneider}, {Brinkmann}, and {Fukugita}}]{fan2006}
{Fan} X, {Strauss} MA, {Becker} RH, et~al (2006) {Constraining the Evolution of the Ionizing Background and the Epoch of Reionization with z\raisebox{-0.5ex}\textasciitilde6 Quasars. II. A Sample of 19 Quasars}. \aj 132(1):117--136. \doi{10.1086/504836}, {\href{https://arxiv.org/abs/astro-ph/0512082}{{arXiv:astro-ph/0512082}}} {[astro-ph]}

\bibitem[{{Fan} et~al(2023){Fan}, {Ba{\~n}ados}, and {Simcoe}}]{fanARAA2023}
{Fan} X, {Ba{\~n}ados} E, {Simcoe} RA (2023) {Quasars and the Intergalactic Medium at Cosmic Dawn}. \araa 61:373--426. \doi{10.1146/annurev-astro-052920-102455}, {\href{https://arxiv.org/abs/2212.06907}{{arXiv:2212.06907}}} {[astro-ph.GA]}

\bibitem[{{Faucher-Gigu{\`e}re} et~al(2009){Faucher-Gigu{\`e}re}, {Lidz}, {Zaldarriaga}, and {Hernquist}}]{fauchergiguere2009}
{Faucher-Gigu{\`e}re} CA, {Lidz} A, {Zaldarriaga} M, et~al (2009) {A New Calculation of the Ionizing Background Spectrum and the Effects of He II Reionization}. \apj 703(2):1416--1443. \doi{10.1088/0004-637X/703/2/1416}, {\href{https://arxiv.org/abs/0901.4554}{{arXiv:0901.4554}}} {[astro-ph.CO]}

\bibitem[{{Fontanot} et~al(2023){Fontanot}, {Cristiani}, {Grazian}, {Haardt}, {D'Odorico}, {Boutsia}, {Calderone}, {Cupani}, {Guarneri}, {Fiorin}, and {Rodighiero}}]{fontanot2023}
{Fontanot} F, {Cristiani} S, {Grazian} A, et~al (2023) {Eddington accreting black holes in the epoch of reionization}. \mnras 520(1):740--749. \doi{10.1093/mnras/stad189}, {\href{https://arxiv.org/abs/2301.07129}{{arXiv:2301.07129}}} {[astro-ph.CO]}

\bibitem[{{Fresco} et~al(2020){Fresco}, {P{\'e}roux}, {Merloni}, {Hamanowicz}, and {Szakacs}}]{fresco2020}
{Fresco} AY, {P{\'e}roux} C, {Merloni} A, et~al (2020) {Tracing the {}10$^{7}$ K warm-hot intergalactic medium with UV absorption lines}. \mnras 499(4):5230--5240. \doi{10.1093/mnras/staa2971}, {\href{https://arxiv.org/abs/2009.11346}{{arXiv:2009.11346}}} {[astro-ph.CO]}

\bibitem[{{Fumagalli} et~al(2011){Fumagalli}, {O'Meara}, and {Prochaska}}]{Fumagalli11}
{Fumagalli} M, {O'Meara} JM, {Prochaska} JX (2011) {Detection of Pristine Gas Two Billion Years After the Big Bang}. Science 334(6060):1245. \doi{10.1126/science.1213581}, {\href{https://arxiv.org/abs/1111.2334}{{arXiv:1111.2334}}} {[astro-ph.CO]}

\bibitem[{{Fynbo} et~al(2009){Fynbo}, {Jakobsson}, {Prochaska}, {Malesani}, {Ledoux}, {de Ugarte Postigo}, {Nardini}, {Vreeswijk}, {Wiersema}, {Hjorth}, {Sollerman}, {Chen}, {Th{\"o}ne}, {Bj{\"o}rnsson}, {Bloom}, {Castro-Tirado}, {Christensen}, {De Cia}, {Fruchter}, {Gorosabel}, {Graham}, {Jaunsen}, {Jensen}, {Kann}, {Kouveliotou}, {Levan}, {Maund}, {Masetti}, {Milvang-Jensen}, {Palazzi}, {Perley}, {Pian}, {Rol}, {Schady}, {Starling}, {Tanvir}, {Watson}, {Xu}, {Augusteijn}, {Grundahl}, {Telting}, and {Quirion}}]{fynbo09}
{Fynbo} JPU, {Jakobsson} P, {Prochaska} JX, et~al (2009) {Low-resolution Spectroscopy of Gamma-ray Burst Optical Afterglows: Biases in the Swift Sample and Characterization of the Absorbers}. \apjs 185(2):526--573. \doi{10.1088/0067-0049/185/2/526}, {\href{https://arxiv.org/abs/0907.3449}{{arXiv:0907.3449}}} {[astro-ph.CO]}

\bibitem[{{Fynbo} et~al(2014){Fynbo}, {Kr{\"u}hler}, {Leighly}, {Ledoux}, {Vreeswijk}, {Schulze}, {Noterdaeme}, {Watson}, {Wijers}, {Bolmer}, {Cano}, {Christensen}, {Covino}, {D'Elia}, {Flores}, {Friis}, {Goldoni}, {Greiner}, {Hammer}, {Hjorth}, {Jakobsson}, {Japelj}, {Kaper}, {Klose}, {Knust}, {Leloudas}, {Levan}, {Malesani}, {Milvang-Jensen}, {M{\o}ller}, {Nicuesa Guelbenzu}, {Oates}, {Pian}, {Schady}, {Sparre}, {Tagliaferri}, {Tanvir}, {Th{\"o}ne}, {de Ugarte Postigo}, {Vergani}, {Wiersema}, {Xu}, and {Zafar}}]{fynbo14}
{Fynbo} JPU, {Kr{\"u}hler} T, {Leighly} K, et~al (2014) {The mysterious optical afterglow spectrum of GRB 140506A at z = 0.889}. \aap 572:A12. \doi{10.1051/0004-6361/201424726}, {\href{https://arxiv.org/abs/1409.4975}{{arXiv:1409.4975}}} {[astro-ph.GA]}

\bibitem[{{Gaikwad} et~al(2020){Gaikwad}, {Rauch}, {Haehnelt}, {Puchwein}, {Bolton}, {Keating}, {Kulkarni}, {Ir{\v{s}}i{\v{c}}}, {Ba{\~n}ados}, {Becker}, {Boera}, {Zahedy}, {Chen}, {Carswell}, {Chardin}, and {Rorai}}]{gaikwad2020}
{Gaikwad} P, {Rauch} M, {Haehnelt} MG, et~al (2020) {Probing the thermal state of the intergalactic medium at $z > 5$ with the transmission spikes in high-resolution Ly {\ensuremath{\alpha}} forest spectra}. \mnras 494(4):5091--5109. \doi{10.1093/mnras/staa907}, {\href{https://arxiv.org/abs/2001.10018}{{arXiv:2001.10018}}} {[astro-ph.CO]}

\bibitem[{{Gaikwad} et~al(2021){Gaikwad}, {Srianand}, {Haehnelt}, and {Choudhury}}]{gaikwad2021}
{Gaikwad} P, {Srianand} R, {Haehnelt} MG, et~al (2021) {A consistent and robust measurement of the thermal state of the IGM at 2 {\ensuremath{\leq}} z {\ensuremath{\leq}} 4 from a large sample of Ly {\ensuremath{\alpha}} forest spectra: evidence for late and rapid He II reionization}. \mnras 506(3):4389--4412. \doi{10.1093/mnras/stab2017}, {\href{https://arxiv.org/abs/2009.00016}{{arXiv:2009.00016}}} {[astro-ph.CO]}

\bibitem[{{Gal-Yam}(2012)}]{galyam12}
{Gal-Yam} A (2012) {Luminous Supernovae}. Science 337(6097):927. \doi{10.1126/science.1203601}, {\href{https://arxiv.org/abs/1208.3217}{{arXiv:1208.3217}}} {[astro-ph.CO]}

\bibitem[{{Gal-Yam}(2019)}]{galyam19}
{Gal-Yam} A (2019) {The Most Luminous Supernovae}. \araa 57:305--333. \doi{10.1146/annurev-astro-081817-051819}, {\href{https://arxiv.org/abs/1812.01428}{{arXiv:1812.01428}}} {[astro-ph.HE]}

\bibitem[{{Gal-Yam} et~al(2009){Gal-Yam}, {Mazzali}, {Ofek}, {Nugent}, {Kulkarni}, {Kasliwal}, {Quimby}, {Filippenko}, {Cenko}, {Chornock}, {Waldman}, {Kasen}, {Sullivan}, {Beshore}, {Drake}, {Thomas}, {Bloom}, {Poznanski}, {Miller}, {Foley}, {Silverman}, {Arcavi}, {Ellis}, and {Deng}}]{galyam09}
{Gal-Yam} A, {Mazzali} P, {Ofek} EO, et~al (2009) {Supernova 2007bi as a pair-instability explosion}. \nat 462(7273):624--627. \doi{10.1038/nature08579}, {\href{https://arxiv.org/abs/1001.1156}{{arXiv:1001.1156}}} {[astro-ph.CO]}

\bibitem[{{Ghirlanda} et~al(2021){Ghirlanda}, {Salvaterra}, {Toffano}, {Ronchini}, {Guidorzi}, {Oganesyan}, {Ascenzi}, {Bernardini}, {Camisasca}, {Mereghetti}, {Nava}, {Ravasio}, {Branchesi}, {Castro-Tirado}, {Amati}, {Blain}, {Bozzo}, {O'Brien}, {G{\"o}tz}, {Le Floch}, {Osborne}, {Rosati}, {Stratta}, {Tanvir}, {Bogomazov}, {D'Avanzo}, {Hafizi}, {Mandhai}, {Melandri}, {Peer}, {Topinka}, {Vergani}, and {Zane}}]{ghirlanda21}
{Ghirlanda} G, {Salvaterra} R, {Toffano} M, et~al (2021) {Gamma ray burst studies with THESEUS}. Experimental Astronomy 52(3):277--308. \doi{10.1007/s10686-021-09763-3}, {\href{https://arxiv.org/abs/2104.10448}{{arXiv:2104.10448}}} {[astro-ph.IM]}

\bibitem[{{Gnerucci} et~al(2010){Gnerucci}, {Marconi}, {Capetti}, {Axon}, and {Robinson}}]{gnerucci2010}
{Gnerucci} A, {Marconi} A, {Capetti} A, et~al (2010) {Spectroastrometry of rotating gas disks for the detection of supermassive black holes in galactic nuclei. I. Method and simulations}. \aap 511:A19. \doi{10.1051/0004-6361/200912530}, {\href{https://arxiv.org/abs/1001.1072}{{arXiv:1001.1072}}} {[astro-ph.CO]}

\bibitem[{{Gnerucci} et~al(2011){Gnerucci}, {Marconi}, {Capetti}, {Axon}, {Robinson}, and {Neumayer}}]{gnerucci2011}
{Gnerucci} A, {Marconi} A, {Capetti} A, et~al (2011) {Spectroastrometry of rotating gas disks for the detection of supermassive black holes in galactic nuclei. II. Application to the galaxy Centaurus A (NGC 5128)}. \aap 536:A86. \doi{10.1051/0004-6361/201117388}, {\href{https://arxiv.org/abs/1110.0936}{{arXiv:1110.0936}}} {[astro-ph.CO]}

\bibitem[{{Gnerucci} et~al(2013){Gnerucci}, {Marconi}, {Capetti}, {Axon}, and {Robinson}}]{gnerucci2013}
{Gnerucci} A, {Marconi} A, {Capetti} A, et~al (2013) {Spectroastrometry of rotating gas disks for the detection of supermassive black holes in galactic nuclei. III. CRIRES observations of the Circinus galaxy}. \aap 549:A139. \doi{10.1051/0004-6361/201118709}, {\href{https://arxiv.org/abs/1211.0943}{{arXiv:1211.0943}}} {[astro-ph.CO]}

\bibitem[{{Graham} et~al(2019){Graham}, {Kulkarni}, {Bellm}, {Adams}, {Barbarino}, {Blagorodnova}, {Bodewits}, {Bolin}, {Brady}, {Cenko}, {Chang}, {Coughlin}, {De}, {Eadie}, {Farnham}, {Feindt}, {Franckowiak}, {Fremling}, {Gezari}, {Ghosh}, {Goldstein}, {Golkhou}, {Goobar}, {Ho}, {Huppenkothen}, {Ivezi{\'c}}, {Jones}, {Juric}, {Kaplan}, {Kasliwal}, {Kelley}, {Kupfer}, {Lee}, {Lin}, {Lunnan}, {Mahabal}, {Miller}, {Ngeow}, {Nugent}, {Ofek}, {Prince}, {Rauch}, {van Roestel}, {Schulze}, {Singer}, {Sollerman}, {Taddia}, {Yan}, {Ye}, {Yu}, {Barlow}, {Bauer}, {Beck}, {Belicki}, {Biswas}, {Brinnel}, {Brooke}, {Bue}, {Bulla}, {Burruss}, {Connolly}, {Cromer}, {Cunningham}, {Dekany}, {Delacroix}, {Desai}, {Duev}, {Feeney}, {Flynn}, {Frederick}, {Gal-Yam}, {Giomi}, {Groom}, {Hacopians}, {Hale}, {Helou}, {Henning}, {Hover}, {Hillenbrand}, {Howell}, {Hung}, {Imel}, {Ip}, {Jackson}, {Kaspi}, {Kaye}, {Kowalski}, {Kramer}, {Kuhn}, {Landry}, {Laher}, {Mao}, {Masci}, {Monkewitz}, {Murphy}, {Nordin}, {Patterson}, {Penprase},
  {Porter}, {Rebbapragada}, {Reiley}, {Riddle}, {Rigault}, {Rodriguez}, {Rusholme}, {van Santen}, {Shupe}, {Smith}, {Soumagnac}, {Stein}, {Surace}, {Szkody}, {Terek}, {Van Sistine}, {van Velzen}, {Vestrand}, {Walters}, {Ward}, {Zhang}, and {Zolkower}}]{graham19}
{Graham} MJ, {Kulkarni} SR, {Bellm} EC, et~al (2019) {The Zwicky Transient Facility: Science Objectives}. \pasp 131(1001):078001. \doi{10.1088/1538-3873/ab006c}, {\href{https://arxiv.org/abs/1902.01945}{{arXiv:1902.01945}}} {[astro-ph.IM]}

\bibitem[{{Gravity Collaboration} et~al(2018){Gravity Collaboration}, {Sturm}, {Dexter}, {Pfuhl}, {Stock}, {Davies}, {Lutz}, {Cl{\'e}net}, {Eckart}, {Eisenhauer}, {Genzel}, {Gratadour}, {H{\"o}nig}, {Kishimoto}, {Lacour}, {Millour}, {Netzer}, {Perrin}, {Peterson}, {Petrucci}, {Rouan}, {Waisberg}, {Woillez}, {Amorim}, {Brandner}, {F{\"o}rster Schreiber}, {Garcia}, {Gillessen}, {Ott}, {Paumard}, {Perraut}, {Scheithauer}, {Straubmeier}, {Tacconi}, and {Widmann}}]{gravitycoll2018}
{Gravity Collaboration}, {Sturm} E, {Dexter} J, et~al (2018) {Spatially resolved rotation of the broad-line region of a quasar at sub-parsec scale}. \nat 563(7733):657--660. \doi{10.1038/s41586-018-0731-9}, {\href{https://arxiv.org/abs/1811.11195}{{arXiv:1811.11195}}} {[astro-ph.GA]}

\bibitem[{{Grazian} et~al(2022){Grazian}, {Giallongo}, {Boutsia}, {Calderone}, {Cristiani}, {Cupani}, {Fontanot}, {Guarneri}, and {Ozdalkiran}}]{grazian2022}
{Grazian} A, {Giallongo} E, {Boutsia} K, et~al (2022) {The Space Density of Ultra-luminous QSOs at the End of Reionization Epoch by the QUBRICS Survey and the AGN Contribution to the Hydrogen Ionizing Background}. \apj 924(2):62. \doi{10.3847/1538-4357/ac33a4}, {\href{https://arxiv.org/abs/2110.13736}{{arXiv:2110.13736}}} {[astro-ph.GA]}

\bibitem[{{Greene} et~al(2020){Greene}, {Strader}, and {Ho}}]{greene2020}
{Greene} JE, {Strader} J, {Ho} LC (2020) {Intermediate-Mass Black Holes}. \araa 58:257--312. \doi{10.1146/annurev-astro-032620-021835}, {\href{https://arxiv.org/abs/1911.09678}{{arXiv:1911.09678}}} {[astro-ph.GA]}

\bibitem[{{Greif} et~al(2008){Greif}, {Johnson}, {Klessen}, and {Bromm}}]{greif2008}
{Greif} TH, {Johnson} JL, {Klessen} RS, et~al (2008) {The first galaxies: assembly, cooling and the onset of turbulence}. \mnras 387(3):1021--1036. \doi{10.1111/j.1365-2966.2008.13326.x}, {\href{https://arxiv.org/abs/0803.2237}{{arXiv:0803.2237}}} {[astro-ph]}

\bibitem[{{Gunn} and {Peterson}(1965)}]{gunnpeterson65}
{Gunn} JE, {Peterson} BA (1965) {On the Density of Neutral Hydrogen in Intergalactic Space.} \apj 142:1633--1636. \doi{10.1086/148444}

\bibitem[{{Hammer} et~al(2021){Hammer}, {Morris}, {Cuby}, {Kaper}, {Steinmetz}, {Afonso}, {Barbuy}, {Bergin}, {Finogenov}, {Gallego}, {Kassin}, {Miller}, {{\"O}stlin}, {Pentericci}, {Schaerer}, {Ziegler}, {Chemla}, {Dalton}, {De Frondat}, {Evans}, {Le Mignant}, {Puech}, {Rodrigues}, {Sanchez-Janssen}, {Taburet}, {Tasca}, {Yang}, {Zanchetta}, {Dohlen}, {Dubbeldam}, {El Hadi}, {Janssen}, {Kelz}, {Larrieu}, {Lewis}, {MacIntosh}, {Morris}, {Navarro}, and {Seifert}}]{hammer2021}
{Hammer} F, {Morris} S, {Cuby} JG, et~al (2021) {MOSAIC on the ELT: High-multiplex Spectroscopy to Unravel the Physics of Stars and Galaxies from the Dark Ages to the Present Day}. The Messenger 182:33--37. \doi{10.18727/0722-6691/5220}, {\href{https://arxiv.org/abs/2011.03549}{{arXiv:2011.03549}}} {[astro-ph.GA]}

\bibitem[{{Harikane} et~al(2023){Harikane}, {Zhang}, {Nakajima}, {Ouchi}, {Isobe}, {Ono}, {Hatano}, {Xu}, and {Umeda}}]{harikane2023}
{Harikane} Y, {Zhang} Y, {Nakajima} K, et~al (2023) {A JWST/NIRSpec First Census of Broad-line AGNs at z = 4-7: Detection of 10 Faint AGNs with M$_{\rm BH} \sim 10^{6}-10^{8}$ M $_{{\ensuremath{\odot}}}$ and Their Host Galaxy Properties}. \apj 959(1):39. \doi{10.3847/1538-4357/ad029e}, {\href{https://arxiv.org/abs/2303.11946}{{arXiv:2303.11946}}} {[astro-ph.GA]}

\bibitem[{{Hartoog} et~al(2015){Hartoog}, {Malesani}, {Fynbo}, {Goto}, {Kr{\"u}hler}, {Vreeswijk}, {De Cia}, {Xu}, {M{\o}ller}, {Covino}, {D'Elia}, {Flores}, {Goldoni}, {Hjorth}, {Jakobsson}, {Krogager}, {Kaper}, {Ledoux}, {Levan}, {Milvang-Jensen}, {Sollerman}, {Sparre}, {Tagliaferri}, {Tanvir}, {de Ugarte Postigo}, {Vergani}, {Wiersema}, {Datson}, {Salinas}, {Mikkelsen}, and {Aghanim}}]{hartoog2015}
{Hartoog} OE, {Malesani} D, {Fynbo} JPU, et~al (2015) {VLT/X-Shooter spectroscopy of the afterglow of the Swift GRB 130606A. Chemical abundances and reionisation at $z \sim 6$}. \aap 580:A139. \doi{10.1051/0004-6361/201425001}, {\href{https://arxiv.org/abs/1409.4804}{{arXiv:1409.4804}}} {[astro-ph.GA]}

\bibitem[{{Heger} and {Woosley}(2002)}]{Heger02}
{Heger} A, {Woosley} SE (2002) {The Nucleosynthetic Signature of Population III}. \apj 567(1):532--543. \doi{10.1086/338487}, {\href{https://arxiv.org/abs/astro-ph/0107037}{{arXiv:astro-ph/0107037}}} {[astro-ph]}

\bibitem[{{Heger} and {Woosley}(2010)}]{heger10}
{Heger} A, {Woosley} SE (2010) {Nucleosynthesis and Evolution of Massive Metal-free Stars}. \apj 724(1):341--373. \doi{10.1088/0004-637X/724/1/341}, {\href{https://arxiv.org/abs/0803.3161}{{arXiv:0803.3161}}} {[astro-ph]}

\bibitem[{{Heintz} et~al(2024){Heintz}, {Watson}, {Brammer}, {Vejlgaard}, {Hutter}, {Strait}, {Matthee}, {Oesch}, {Jakobsson}, {Tanvir}, {Laursen}, {Naidu}, {Mason}, {Killi}, {Jung}, {Hsiao}, {Abdurro'uf}, {Coe}, {Arrabal Haro}, {Finkelstein}, and {Toft}}]{heintz2024}
{Heintz} KE, {Watson} D, {Brammer} G, et~al (2024) {Strong damped Lyman-{\ensuremath{\alpha}} absorption in young star-forming galaxies at redshifts 9 to 11}. Science 384(6698):890--894. \doi{10.1126/science.adj0343}, {\href{https://arxiv.org/abs/2306.00647}{{arXiv:2306.00647}}} {[astro-ph.GA]}

\bibitem[{{Hennawi} et~al(2021){Hennawi}, {Davies}, {Wang}, and {O{\~n}orbe}}]{hennawi2021}
{Hennawi} JF, {Davies} FB, {Wang} F, et~al (2021) {Probing reionization and early cosmic enrichment with the Mg II forest}. \mnras 506(2):2963--2984. \doi{10.1093/mnras/stab1883}, {\href{https://arxiv.org/abs/2007.15747}{{arXiv:2007.15747}}} {[astro-ph.CO]}

\bibitem[{{Hirano} et~al(2014){Hirano}, {Hosokawa}, {Yoshida}, {Umeda}, {Omukai}, {Chiaki}, and {Yorke}}]{Hirano14}
{Hirano} S, {Hosokawa} T, {Yoshida} N, et~al (2014) {One Hundred First Stars: Protostellar Evolution and the Final Masses}. \apj 781(2):60. \doi{10.1088/0004-637X/781/2/60}, {\href{https://arxiv.org/abs/1308.4456}{{arXiv:1308.4456}}} {[astro-ph.CO]}

\bibitem[{{Hjorth} et~al(2003){Hjorth}, {Sollerman}, {M{\o}ller}, {Fynbo}, {Woosley}, {Kouveliotou}, {Tanvir}, {Greiner}, {Andersen}, {Castro-Tirado}, {Castro Cer{\'o}n}, {Fruchter}, {Gorosabel}, {Jakobsson}, {Kaper}, {Klose}, {Masetti}, {Pedersen}, {Pedersen}, {Pian}, {Palazzi}, {Rhoads}, {Rol}, {van den Heuvel}, {Vreeswijk}, {Watson}, and {Wijers}}]{hjorth03}
{Hjorth} J, {Sollerman} J, {M{\o}ller} P, et~al (2003) {A very energetic supernova associated with the {\ensuremath{\gamma}}-ray burst of 29 March 2003}. \nat 423(6942):847--850. \doi{10.1038/nature01750}, {\href{https://arxiv.org/abs/astro-ph/0306347}{{arXiv:astro-ph/0306347}}} {[astro-ph]}

\bibitem[{{Hobbs}(1984)}]{hobbs1984}
{Hobbs} LM (1984) {On absorption by hot interstellar gas. I. lambda 6375.} \apj 280:132--138. \doi{10.1086/161976}

\bibitem[{{Inserra} et~al(2013){Inserra}, {Smartt}, {Jerkstrand}, {Valenti}, {Fraser}, {Wright}, {Smith}, {Chen}, {Kotak}, {Pastorello}, {Nicholl}, {Bresolin}, {Kudritzki}, {Benetti}, {Botticella}, {Burgett}, {Chambers}, {Ergon}, {Flewelling}, {Fynbo}, {Geier}, {Hodapp}, {Howell}, {Huber}, {Kaiser}, {Leloudas}, {Magill}, {Magnier}, {McCrum}, {Metcalfe}, {Price}, {Rest}, {Sollerman}, {Sweeney}, {Taddia}, {Taubenberger}, {Tonry}, {Wainscoat}, {Waters}, and {Young}}]{inserra13}
{Inserra} C, {Smartt} SJ, {Jerkstrand} A, et~al (2013) {Super-luminous Type Ic Supernovae: Catching a Magnetar by the Tail}. \apj 770(2):128. \doi{10.1088/0004-637X/770/2/128}, {\href{https://arxiv.org/abs/1304.3320}{{arXiv:1304.3320}}} {[astro-ph.HE]}

\bibitem[{{Inserra} et~al(2017){Inserra}, {Nicholl}, {Chen}, {Jerkstrand}, {Smartt}, {Kr{\"u}hler}, {Anderson}, {Baltay}, {Della Valle}, {Fraser}, {Gal-Yam}, {Galbany}, {Kankare}, {Maguire}, {Rabinowitz}, {Smith}, {Valenti}, and {Young}}]{inserra17}
{Inserra} C, {Nicholl} M, {Chen} TW, et~al (2017) {Complexity in the light curves and spectra of slow-evolving superluminous supernovae}. \mnras 468(4):4642--4662. \doi{10.1093/mnras/stx834}, {\href{https://arxiv.org/abs/1701.00941}{{arXiv:1701.00941}}} {[astro-ph.HE]}

\bibitem[{{Ir{\v{s}}i{\v{c}}} et~al(2017){Ir{\v{s}}i{\v{c}}}, {Viel}, {Haehnelt}, {Bolton}, {Cristiani}, {Becker}, {D'Odorico}, {Cupani}, {Kim}, {Berg}, {L{\'o}pez}, {Ellison}, {Christensen}, {Denney}, and {Worseck}}]{Irsic2017}
{Ir{\v{s}}i{\v{c}}} V, {Viel} M, {Haehnelt} MG, et~al (2017) {New constraints on the free-streaming of warm dark matter from intermediate and small scale Lyman-$\alpha$ forest data}. \prd 96(2):023522. \doi{10.1103/PhysRevD.96.023522}, {\href{https://arxiv.org/abs/1702.01764}{{arXiv:1702.01764}}} {[astro-ph.CO]}

\bibitem[{{Ir{\v{s}}i{\v{c}}} et~al(2024){Ir{\v{s}}i{\v{c}}}, {Viel}, {Haehnelt}, {Bolton}, {Molaro}, {Puchwein}, {Boera}, {Becker}, {Gaikwad}, {Keating}, and {Kulkarni}}]{Irsic2024}
{Ir{\v{s}}i{\v{c}}} V, {Viel} M, {Haehnelt} MG, et~al (2024) {Unveiling dark matter free streaming at the smallest scales with the high redshift Lyman-alpha forest}. \prd 109(4):043511. \doi{10.1103/PhysRevD.109.043511}, {\href{https://arxiv.org/abs/2309.04533}{{arXiv:2309.04533}}} {[astro-ph.CO]}

\bibitem[{{Ivezi{\'c}} et~al(2019){Ivezi{\'c}}, {Kahn}, {Tyson}, {Abel}, {Acosta}, {Allsman}, {Alonso}, {AlSayyad}, {Anderson}, {Andrew}, {Angel}, {Angeli}, {Ansari}, {Antilogus}, {Araujo}, {Armstrong}, {Arndt}, {Astier}, {Aubourg}, {Auza}, {Axelrod}, {Bard}, {Barr}, {Barrau}, {Bartlett}, {Bauer}, {Bauman}, {Baumont}, {Bechtol}, {Bechtol}, {Becker}, {Becla}, {Beldica}, {Bellavia}, {Bianco}, {Biswas}, {Blanc}, {Blazek}, {Blandford}, {Bloom}, {Bogart}, {Bond}, {Booth}, {Borgland}, {Borne}, {Bosch}, {Boutigny}, {Brackett}, {Bradshaw}, {Brandt}, {Brown}, {Bullock}, {Burchat}, {Burke}, {Cagnoli}, {Calabrese}, {Callahan}, {Callen}, {Carlin}, {Carlson}, {Chandrasekharan}, {Charles-Emerson}, {Chesley}, {Cheu}, {Chiang}, {Chiang}, {Chirino}, {Chow}, {Ciardi}, {Claver}, {Cohen-Tanugi}, {Cockrum}, {Coles}, {Connolly}, {Cook}, {Cooray}, {Covey}, {Cribbs}, {Cui}, {Cutri}, {Daly}, {Daniel}, {Daruich}, {Daubard}, {Daues}, {Dawson}, {Delgado}, {Dellapenna}, {de Peyster}, {de Val-Borro}, {Digel}, {Doherty}, {Dubois},
  {Dubois-Felsmann}, {Durech}, {Economou}, {Eifler}, {Eracleous}, {Emmons}, {Fausti Neto}, {Ferguson}, {Figueroa}, {Fisher-Levine}, {Focke}, {Foss}, {Frank}, {Freemon}, {Gangler}, {Gawiser}, {Geary}, {Gee}, {Geha}, {Gessner}, {Gibson}, {Gilmore}, {Glanzman}, {Glick}, {Goldina}, {Goldstein}, {Goodenow}, {Graham}, {Gressler}, {Gris}, {Guy}, {Guyonnet}, {Haller}, {Harris}, {Hascall}, {Haupt}, {Hernandez}, {Herrmann}, {Hileman}, {Hoblitt}, {Hodgson}, {Hogan}, {Howard}, {Huang}, {Huffer}, {Ingraham}, {Innes}, {Jacoby}, {Jain}, {Jammes}, {Jee}, {Jenness}, {Jernigan}, {Jevremovi{\'c}}, {Johns}, {Johnson}, {Johnson}, {Jones}, {Juramy-Gilles}, {Juri{\'c}}, {Kalirai}, {Kallivayalil}, {Kalmbach}, {Kantor}, {Karst}, {Kasliwal}, {Kelly}, {Kessler}, {Kinnison}, {Kirkby}, {Knox}, {Kotov}, {Krabbendam}, {Krughoff}, {Kub{\'a}nek}, {Kuczewski}, {Kulkarni}, {Ku}, {Kurita}, {Lage}, {Lambert}, {Lange}, {Langton}, {Le Guillou}, {Levine}, {Liang}, {Lim}, {Lintott}, {Long}, {Lopez}, {Lotz}, {Lupton}, {Lust}, {MacArthur}, {Mahabal},
  {Mandelbaum}, {Markiewicz}, {Marsh}, {Marshall}, {Marshall}, {May}, {McKercher}, {McQueen}, {Meyers}, {Migliore}, {Miller}, {Mills}, {Miraval}, {Moeyens}, {Moolekamp}, {Monet}, {Moniez}, {Monkewitz}, {Montgomery}, {Morrison}, {Mueller}, {Muller}, {Mu{\~n}oz Arancibia}, {Neill}, {Newbry}, {Nief}, {Nomerotski}, {Nordby}, {O'Connor}, {Oliver}, {Olivier}, {Olsen}, {O'Mullane}, {Ortiz}, {Osier}, {Owen}, {Pain}, {Palecek}, {Parejko}, {Parsons}, {Pease}, {Peterson}, {Peterson}, {Petravick}, {Libby Petrick}, {Petry}, {Pierfederici}, {Pietrowicz}, {Pike}, {Pinto}, {Plante}, {Plate}, {Plutchak}, {Price}, {Prouza}, {Radeka}, {Rajagopal}, {Rasmussen}, {Regnault}, {Reil}, {Reiss}, {Reuter}, {Ridgway}, {Riot}, {Ritz}, {Robinson}, {Roby}, {Roodman}, {Rosing}, {Roucelle}, {Rumore}, {Russo}, {Saha}, {Sassolas}, {Schalk}, {Schellart}, {Schindler}, {Schmidt}, {Schneider}, {Schneider}, {Schoening}, {Schumacher}, {Schwamb}, {Sebag}, {Selvy}, {Sembroski}, {Seppala}, {Serio}, {Serrano}, {Shaw}, {Shipsey}, {Sick}, {Silvestri},
  {Slater}, {Smith}, {Smith}, {Sobhani}, {Soldahl}, {Storrie-Lombardi}, {Stover}, {Strauss}, {Street}, {Stubbs}, {Sullivan}, {Sweeney}, {Swinbank}, {Szalay}, {Takacs}, {Tether}, {Thaler}, {Thayer}, {Thomas}, {Thornton}, {Thukral}, {Tice}, {Trilling}, {Turri}, {Van Berg}, {Vanden Berk}, {Vetter}, {Virieux}, {Vucina}, {Wahl}, {Walkowicz}, {Walsh}, {Walter}, {Wang}, {Wang}, {Warner}, {Wiecha}, {Willman}, {Winters}, {Wittman}, {Wolff}, {Wood-Vasey}, {Wu}, {Xin}, {Yoachim}, and {Zhan}}]{ivezic2019}
{Ivezi{\'c}} {\v{Z}}, {Kahn} SM, {Tyson} JA, et~al (2019) {LSST: From Science Drivers to Reference Design and Anticipated Data Products}. \apj 873(2):111. \doi{10.3847/1538-4357/ab042c}, {\href{https://arxiv.org/abs/0805.2366}{{arXiv:0805.2366}}} {[astro-ph]}

\bibitem[{{Izzo} et~al(2015){Izzo}, {Della Valle}, {Mason}, {Matteucci}, {Romano}, {Pasquini}, {Vanzi}, {Jordan}, {Fernandez}, {Bluhm}, {Brahm}, {Espinoza}, and {Williams}}]{izzo15}
{Izzo} L, {Della Valle} M, {Mason} E, et~al (2015) {Early Optical Spectra of Nova V1369 Cen Show the Presence of Lithium}. \apjl 808(1):L14. \doi{10.1088/2041-8205/808/1/L14}, {\href{https://arxiv.org/abs/1506.08048}{{arXiv:1506.08048}}} {[astro-ph.SR]}

\bibitem[{{Izzo} et~al(2022){Izzo}, {Molaro}, {Cescutti}, {Aydi}, {Selvelli}, {Harvey}, {Agnello}, {Bonifacio}, {Della Valle}, {Guido}, and {Hernanz}}]{Izzo2022}
{Izzo} L, {Molaro} P, {Cescutti} G, et~al (2022) {Detection of $^{7}$Be II in the Small Magellanic Cloud}. \mnras 510(4):5302--5314. \doi{10.1093/mnras/stab3761}, {\href{https://arxiv.org/abs/2112.11859}{{arXiv:2112.11859}}} {[astro-ph.SR]}

\bibitem[{{Jaacks} et~al(2018){Jaacks}, {Thompson}, {Finkelstein}, and {Bromm}}]{jaacks2018}
{Jaacks} J, {Thompson} R, {Finkelstein} SL, et~al (2018) {Baseline metal enrichment from Population III star formation in cosmological volume simulations}. \mnras 475(4):4396--4410. \doi{10.1093/mnras/sty062}, {\href{https://arxiv.org/abs/1705.08059}{{arXiv:1705.08059}}} {[astro-ph.GA]}

\bibitem[{{Jiang} et~al(2022){Jiang}, {Ning}, {Fan}, {Ho}, {Luo}, {Wang}, {Wu}, {Wu}, {Yang}, and {Zheng}}]{jiang2022}
{Jiang} L, {Ning} Y, {Fan} X, et~al (2022) {Definitive upper bound on the negligible contribution of quasars to cosmic reionization}. Nature Astronomy 6:850--856. \doi{10.1038/s41550-022-01708-w}, {\href{https://arxiv.org/abs/2206.07825}{{arXiv:2206.07825}}} {[astro-ph.GA]}

\bibitem[{{Jin} et~al(2023){Jin}, {Yang}, {Fan}, {Wang}, {Ba{\~n}ados}, {Bian}, {Davies}, {Eilers}, {Farina}, {Hennawi}, {Pacucci}, {Venemans}, and {Walter}}]{jin2023}
{Jin} X, {Yang} J, {Fan} X, et~al (2023) {(Nearly) Model-independent Constraints on the Neutral Hydrogen Fraction in the Intergalactic Medium at $z \sim5-7$ Using Dark Pixel Fractions in Ly{\ensuremath{\alpha}} and Ly{\ensuremath{\beta}} Forests}. \apj 942(2):59. \doi{10.3847/1538-4357/aca678}, {\href{https://arxiv.org/abs/2211.12613}{{arXiv:2211.12613}}} {[astro-ph.CO]}

\bibitem[{{Kakiichi} et~al(2018){Kakiichi}, {Ellis}, {Laporte}, {Zitrin}, {Eilers}, {Ryan-Weber}, {Meyer}, {Robertson}, {Stark}, and {Bosman}}]{kakiichi2018}
{Kakiichi} K, {Ellis} RS, {Laporte} N, et~al (2018) {The role of galaxies and AGN in reionizing the IGM - I. Keck spectroscopy of $5 < z < 7$ galaxies in the QSO field J1148+5251}. \mnras 479(1):43--63. \doi{10.1093/mnras/sty1318}, {\href{https://arxiv.org/abs/1803.02981}{{arXiv:1803.02981}}} {[astro-ph.GA]}

\bibitem[{{Kann} et~al(2021){Kann}, {Oates}, {Rossi}, {Klose}, {Blazek}, {Ag{\"u}{\'\i} Fern{\'a}ndez}, {de Ugarte Postigo}, and {Th{\"o}ne}}]{kann21}
{Kann} DA, {Oates} SR, {Rossi} A, et~al (2021) {Highly luminous supernovae associated with gamma-ray bursts II. The Luminous Blue Bump in the Afterglow of GRB 140506A}. arXiv e-prints arXiv:2110.00110. \doi{10.48550/arXiv.2110.00110}, {\href{https://arxiv.org/abs/2110.00110}{{arXiv:2110.00110}}} {[astro-ph.HE]}

\bibitem[{{Kashino} et~al(2023){Kashino}, {Lilly}, {Matthee}, {Eilers}, {Mackenzie}, {Bordoloi}, and {Simcoe}}]{kashino2023}
{Kashino} D, {Lilly} SJ, {Matthee} J, et~al (2023) {EIGER. I. A Large Sample of [O III]-emitting Galaxies at $5.3 < z < 6.9$ and Direct Evidence for Local Reionization by Galaxies}. \apj 950(1):66. \doi{10.3847/1538-4357/acc588}, {\href{https://arxiv.org/abs/2211.08254}{{arXiv:2211.08254}}} {[astro-ph.GA]}

\bibitem[{{Keating} et~al(2014){Keating}, {Haehnelt}, {Becker}, and {Bolton}}]{keating2014}
{Keating} LC, {Haehnelt} MG, {Becker} GD, et~al (2014) {Probing the metallicity and ionization state of the circumgalactic medium at z {\ensuremath{\sim}} 6 and beyond with O I absorption}. \mnras 438(2):1820--1831. \doi{10.1093/mnras/stt2324}, {\href{https://arxiv.org/abs/1310.5859}{{arXiv:1310.5859}}} {[astro-ph.CO]}

\bibitem[{{King} and {Pounds}(2015)}]{king2015}
{King} A, {Pounds} K (2015) {Powerful Outflows and Feedback from Active Galactic Nuclei}. \araa 53:115--154. \doi{10.1146/annurev-astro-082214-122316}, {\href{https://arxiv.org/abs/1503.05206}{{arXiv:1503.05206}}} {[astro-ph.GA]}

\bibitem[{{Klessen} and {Glover}(2023)}]{klessen2023}
{Klessen} RS, {Glover} SCO (2023) {The First Stars: Formation, Properties, and Impact}. \araa 61:65--130. \doi{10.1146/annurev-astro-071221-053453}, {\href{https://arxiv.org/abs/2303.12500}{{arXiv:2303.12500}}} {[astro-ph.CO]}

\bibitem[{{Kobayashi} et~al(2006){Kobayashi}, {Umeda}, {Nomoto}, {Tominaga}, and {Ohkubo}}]{Kobayashi2006}
{Kobayashi} C, {Umeda} H, {Nomoto} K, et~al (2006) {Galactic Chemical Evolution: Carbon through Zinc}. \apj 653(2):1145--1171. \doi{10.1086/508914}, {\href{https://arxiv.org/abs/astro-ph/0608688}{{arXiv:astro-ph/0608688}}} {[astro-ph]}

\bibitem[{{Kocevski} et~al(2023){Kocevski}, {Onoue}, {Inayoshi}, {Trump}, {Arrabal Haro}, {Grazian}, {Dickinson}, {Finkelstein}, {Kartaltepe}, {Hirschmann}, {Aird}, {Holwerda}, {Fujimoto}, {Juneau}, {Amor{\'\i}n}, {Backhaus}, {Bagley}, {Barro}, {Bell}, {Bisigello}, {Calabr{\`o}}, {Cleri}, {Cooper}, {Ding}, {Grogin}, {Ho}, {Hutchison}, {Inoue}, {Jiang}, {Jones}, {Koekemoer}, {Li}, {Li}, {McGrath}, {Molina}, {Papovich}, {P{\'e}rez-Gonz{\'a}lez}, {Pirzkal}, {Wilkins}, {Yang}, and {Yung}}]{kocevski2023}
{Kocevski} DD, {Onoue} M, {Inayoshi} K, et~al (2023) {Hidden Little Monsters: Spectroscopic Identification of Low-mass, Broad-line AGNs at $z > 5$ with CEERS}. \apjl 954(1):L4. \doi{10.3847/2041-8213/ace5a0}, {\href{https://arxiv.org/abs/2302.00012}{{arXiv:2302.00012}}} {[astro-ph.GA]}

\bibitem[{{Koopmans} et~al(2015){Koopmans}, {Pritchard}, {Mellema}, {Aguirre}, {Ahn}, {Barkana}, {van Bemmel}, {Bernardi}, {Bonaldi}, {Briggs}, {de Bruyn}, {Chang}, {Chapman}, {Chen}, {Ciardi}, {Dayal}, {Ferrara}, {Fialkov}, {Fiore}, {Ichiki}, {Illiev}, {Inoue}, {Jelic}, {Jones}, {Lazio}, {Maio}, {Majumdar}, {Mack}, {Mesinger}, {Morales}, {Parsons}, {Pen}, {Santos}, {Schneider}, {Semelin}, {de Souza}, {Subrahmanyan}, {Takeuchi}, {Vedantham}, {Wagg}, {Webster}, {Wyithe}, {Datta}, and {Trott}}]{koopmans2015}
{Koopmans} L, {Pritchard} J, {Mellema} G, et~al (2015) {The Cosmic Dawn and Epoch of Reionisation with SKA}. In: Advancing Astrophysics with the Square Kilometre Array (AASKA14), p~1, \doi{10.22323/1.215.0001}, \eprint{1505.07568}

\bibitem[{{Kormendy} and {Ho}(2013)}]{kormendy2013}
{Kormendy} J, {Ho} LC (2013) {Coevolution (Or Not) of Supermassive Black Holes and Host Galaxies}. \araa 51(1):511--653. \doi{10.1146/annurev-astro-082708-101811}, {\href{https://arxiv.org/abs/1304.7762}{{arXiv:1304.7762}}} {[astro-ph.CO]}

\bibitem[{{Koutsouridou} et~al(2023){Koutsouridou}, {Salvadori}, {Sk{\'u}lad{\'o}ttir}, {Rossi}, {Vanni}, and {Pagnini}}]{Koutsouridou2023}
{Koutsouridou} I, {Salvadori} S, {Sk{\'u}lad{\'o}ttir} {\'A}, et~al (2023) {The energy distribution of the first supernovae}. \mnras 525(1):190--210. \doi{10.1093/mnras/stad2304}, {\href{https://arxiv.org/abs/2309.00045}{{arXiv:2309.00045}}} {[astro-ph.GA]}

\bibitem[{{Koutsouridou} et~al(2024){Koutsouridou}, {Salvadori}, and {Sk{\'u}lad{\'o}ttir}}]{Koutsouridou2024}
{Koutsouridou} I, {Salvadori} S, {Sk{\'u}lad{\'o}ttir} {\'A} (2024) {True Pair-instability Supernova Descendant: Implications for the First Stars' Mass Distribution}. \apjl 962(2):L26. \doi{10.3847/2041-8213/ad2466}, {\href{https://arxiv.org/abs/2312.05309}{{arXiv:2312.05309}}} {[astro-ph.GA]}

\bibitem[{{Kulkarni} et~al(2019){Kulkarni}, {Keating}, {Haehnelt}, {Bosman}, {Puchwein}, {Chardin}, and {Aubert}}]{kulkarni2019}
{Kulkarni} G, {Keating} LC, {Haehnelt} MG, et~al (2019) {Large Ly {\ensuremath{\alpha}} opacity fluctuations and low CMB {\ensuremath{\tau}} in models of late reionization with large islands of neutral hydrogen extending to $z < 5.5$}. \mnras 485(1):L24--L28. \doi{10.1093/mnrasl/slz025}, {\href{https://arxiv.org/abs/1809.06374}{{arXiv:1809.06374}}} {[astro-ph.CO]}

\bibitem[{{Leloudas} et~al(2015){Leloudas}, {Schulze}, {Kr{\"u}hler}, {Gorosabel}, {Christensen}, {Mehner}, {de Ugarte Postigo}, {Amor{\'\i}n}, {Th{\"o}ne}, {Anderson}, {Bauer}, {Gallazzi}, {He{\l}miniak}, {Hjorth}, {Ibar}, {Malesani}, {Morell}, {Vinko}, and {Wheeler}}]{leloudas15}
{Leloudas} G, {Schulze} S, {Kr{\"u}hler} T, et~al (2015) {Spectroscopy of superluminous supernova host galaxies. A preference of hydrogen-poor events for extreme emission line galaxies}. \mnras 449(1):917--932. \doi{10.1093/mnras/stv320}, {\href{https://arxiv.org/abs/1409.8331}{{arXiv:1409.8331}}} {[astro-ph.GA]}

\bibitem[{{Levan} et~al(2024){Levan}, {Gompertz}, {Salafia}, {Bulla}, {Burns}, {Hotokezaka}, {Izzo}, {Lamb}, {Malesani}, {Oates}, {Ravasio}, {Rouco Escorial}, {Schneider}, {Sarin}, {Schulze}, {Tanvir}, {Ackley}, {Anderson}, {Brammer}, {Christensen}, {Dhillon}, {Evans}, {Fausnaugh}, {Fong}, {Fruchter}, {Fryer}, {Fynbo}, {Gaspari}, {Heintz}, {Hjorth}, {Kennea}, {Kennedy}, {Laskar}, {Leloudas}, {Mandel}, {Martin-Carrillo}, {Metzger}, {Nicholl}, {Nugent}, {Palmerio}, {Pugliese}, {Rastinejad}, {Rhodes}, {Rossi}, {Saccardi}, {Smartt}, {Stevance}, {Tohuvavohu}, {van der Horst}, {Vergani}, {Watson}, {Barclay}, {Bhirombhakdi}, {Breedt}, {Breeveld}, {Brown}, {Campana}, {Chrimes}, {D'Avanzo}, {D'Elia}, {De Pasquale}, {Dyer}, {Galloway}, {Garbutt}, {Green}, {Hartmann}, {Jakobsson}, {Kerry}, {Kouveliotou}, {Langeroodi}, {Le Floc'h}, {Leung}, {Littlefair}, {Munday}, {O'Brien}, {Parsons}, {Pelisoli}, {Sahman}, {Salvaterra}, {Sbarufatti}, {Steeghs}, {Tagliaferri}, {Th{\"o}ne}, {de Ugarte Postigo}, and {Kann}}]{levan2024}
{Levan} AJ, {Gompertz} BP, {Salafia} OS, et~al (2024) {Heavy-element production in a compact object merger observed by JWST}. \nat 626(8000):737--741. \doi{10.1038/s41586-023-06759-1}, {\href{https://arxiv.org/abs/2307.02098}{{arXiv:2307.02098}}} {[astro-ph.HE]}

\bibitem[{{Lidz} et~al(2021){Lidz}, {Chang}, {Mas-Ribas}, and {Sun}}]{lidz2021}
{Lidz} A, {Chang} TC, {Mas-Ribas} L, et~al (2021) {Future Constraints on the Reionization History and the Ionizing Sources from Gamma-Ray Burst Afterglows}. \apj 917(2):58. \doi{10.3847/1538-4357/ac0af0}, {\href{https://arxiv.org/abs/2105.02293}{{arXiv:2105.02293}}} {[astro-ph.CO]}

\bibitem[{{Limongi} and {Chieffi}(2018)}]{LC2018}
{Limongi} M, {Chieffi} A (2018) {Presupernova Evolution and Explosive Nucleosynthesis of Rotating Massive Stars in the Metallicity Range -3 {\ensuremath{\leq}} [Fe/H] {\ensuremath{\leq}} 0}. \apjs 237(1):13. \doi{10.3847/1538-4365/aacb24}, {\href{https://arxiv.org/abs/1805.09640}{{arXiv:1805.09640}}} {[astro-ph.SR]}

\bibitem[{{Lu} et~al(2024){Lu}, {Mason}, {Hutter}, {Mesinger}, {Qin}, {Stark}, and {Endsley}}]{lu2024}
{Lu} TY, {Mason} CA, {Hutter} A, et~al (2024) {The reionizing bubble size distribution around galaxies}. \mnras 528(3):4872--4890. \doi{10.1093/mnras/stae266}, {\href{https://arxiv.org/abs/2304.11192}{{arXiv:2304.11192}}} {[astro-ph.GA]}

\bibitem[{{Lunnan} et~al(2015){Lunnan}, {Chornock}, {Berger}, {Rest}, {Fong}, {Scolnic}, {Jones}, {Soderberg}, {Challis}, {Drout}, {Foley}, {Huber}, {Kirshner}, {Leibler}, {Marion}, {McCrum}, {Milisavljevic}, {Narayan}, {Sanders}, {Smartt}, {Smith}, {Tonry}, {Burgett}, {Chambers}, {Flewelling}, {Kudritzki}, {Wainscoat}, and {Waters}}]{lunnan15}
{Lunnan} R, {Chornock} R, {Berger} E, et~al (2015) {Zooming In on the Progenitors of Superluminous Supernovae With the HST}. \apj 804(2):90. \doi{10.1088/0004-637X/804/2/90}, {\href{https://arxiv.org/abs/1411.1060}{{arXiv:1411.1060}}} {[astro-ph.HE]}

\bibitem[{{Madau} et~al(2001){Madau}, {Ferrara}, and {Rees}}]{madau2001}
{Madau} P, {Ferrara} A, {Rees} MJ (2001) {Early Metal Enrichment of the Intergalactic Medium by Pregalactic Outflows}. \apj 555(1):92--105. \doi{10.1086/321474}, {\href{https://arxiv.org/abs/astro-ph/0010158}{{arXiv:astro-ph/0010158}}} {[astro-ph]}

\bibitem[{{Madau} et~al(2024){Madau}, {Giallongo}, {Grazian}, and {Haardt}}]{Madau2024}
{Madau} P, {Giallongo} E, {Grazian} A, et~al (2024) {Cosmic Reionization in the JWST Era: Back to AGNs?} \apj 971(1):75. \doi{10.3847/1538-4357/ad5ce8}, {\href{https://arxiv.org/abs/2406.18697}{{arXiv:2406.18697}}} {[astro-ph.CO]}

\bibitem[{{Maiolino} et~al(2023){Maiolino}, {Scholtz}, {Curtis-Lake}, {Carniani}, {Baker}, {de Graaff}, {Tacchella}, {{\"U}bler}, {D'Eugenio}, {Witstok}, {Curti}, {Arribas}, {Bunker}, {Charlot}, {Chevallard}, {Eisenstein}, {Egami}, {Ji}, {Jones}, {Lyu}, {Rawle}, {Robertson}, {Rujopakarn}, {Perna}, {Sun}, {Venturi}, {Williams}, and {Willott}}]{maiolino2023c}
{Maiolino} R, {Scholtz} J, {Curtis-Lake} E, et~al (2023) {JADES. The diverse population of infant Black Holes at $4<z<11$: merging, tiny, poor, but mighty}. arXiv e-prints arXiv:2308.01230. \doi{10.48550/arXiv.2308.01230}, {\href{https://arxiv.org/abs/2308.01230}{{arXiv:2308.01230}}} {[astro-ph.GA]}

\bibitem[{{Maiolino} et~al(2024{\natexlab{a}}){Maiolino}, {Scholtz}, {Witstok}, {Carniani}, {D'Eugenio}, {de Graaff}, {{\"U}bler}, {Tacchella}, {Curtis-Lake}, {Arribas}, {Bunker}, {Charlot}, {Chevallard}, {Curti}, {Looser}, {Maseda}, {Rawle}, {Rodr{\'\i}guez del Pino}, {Willott}, {Egami}, {Eisenstein}, {Hainline}, {Robertson}, {Williams}, {Willmer}, {Baker}, {Boyett}, {DeCoursey}, {Fabian}, {Helton}, {Ji}, {Jones}, {Kumari}, {Laporte}, {Nelson}, {Perna}, {Sandles}, {Shivaei}, and {Sun}}]{maiolino2024}
{Maiolino} R, {Scholtz} J, {Witstok} J, et~al (2024{\natexlab{a}}) {A small and vigorous black hole in the early Universe}. \nat 627(8002):59--63. \doi{10.1038/s41586-024-07052-5}, {\href{https://arxiv.org/abs/2305.12492}{{arXiv:2305.12492}}} {[astro-ph.GA]}

\bibitem[{{Maiolino} et~al(2024{\natexlab{b}}){Maiolino}, {{\"U}bler}, {Perna}, {Scholtz}, {D'Eugenio}, {Witten}, {Laporte}, {Witstok}, {Carniani}, {Tacchella}, {Baker}, {Arribas}, {Nakajima}, {Eisenstein}, {Bunker}, {Charlot}, {Cresci}, {Curti}, {Curtis-Lake}, {de Graaff}, {Egami}, {Ji}, {Johnson}, {Kumari}, {Looser}, {Maseda}, {Nelson}, {Robertson}, {Rodr{\'\i}guez Del Pino}, {Sandles}, {Simmonds}, {Smit}, {Sun}, {Venturi}, {Williams}, and {Willmer}}]{maiolino2023b}
{Maiolino} R, {{\"U}bler} H, {Perna} M, et~al (2024{\natexlab{b}}) {JADES. Possible Population III signatures at $z = 10.6$ in the halo of GN-z11}. \aap 687:A67. \doi{10.1051/0004-6361/202347087}, {\href{https://arxiv.org/abs/2306.00953}{{arXiv:2306.00953}}} {[astro-ph.GA]}

\bibitem[{{Marconi} et~al(2022){Marconi}, {Abreu}, {Adibekyan}, {Alberti}, {Albrecht}, {Alcaniz}, {Aliverti}, {Allende Prieto}, {Alvarado G{\'o}mez}, {Amado}, {Amate}, {Andersen}, {Artigau}, {Baker}, {Baldini}, {Balestra}, {Barnes}, {Baron}, {Barros}, {Bauer}, {Beaulieu}, {Bellido-Tirado}, {Benneke}, {Bensby}, {Bergin}, {Biazzo}, {Bik}, {Birkby}, {Blind}, {Boisse}, {Bolmont}, {Bonaglia}, {Bonfils}, {Borsa}, {Brandeker}, {Brandner}, {Broeg}, {Brogi}, {Brousseau}, {Brucalassi}, {Brynnel}, {Buchhave}, {Buscher}, {Cabral}, {Calderone}, {Calvo-Ortega}, {Canto Martins}, {Cantalloube}, {Carbonaro}, {Chauvin}, {Chazelas}, {Cheffot}, {Cheng}, {Chiavassa}, {Christensen}, {Cirami}, {Cook}, {Cooke}, {Coretti}, {Covino}, {Cowan}, {Cresci}, {Cristiani}, {Cunha Parro}, {Cupani}, {D'Odorico}, {de Castro Le{\~a}o}, {De Cia}, {De Medeiros}, {Debras}, {Debus}, {Demangeon}, {Dessauges-Zavadsky}, {Di Marcantonio}, {Dionies}, {Doyon}, {Dunn}, {Ehrenreich}, {Faria}, {Feruglio}, {Fisher}, {Fontana}, {Fumagalli}, {Fusco}, {Fynbo},
  {Gabella}, {Gaessler}, {Gallo}, {Gao}, {Genolet}, {Genoni}, {Giacobbe}, {Giro}, {Gon{\c{c}}alves}, {Gonzalez}, {Gonz{\'a}lez Hern{\'a}ndez}, {Gracia T{\'e}mich}, {Haehnelt}, {Haniff}, {Hatzes}, {Helled}, {Hoeijmakers}, {Huke}, {J{\"a}rvinen}, {J{\"a}rvinen}, {Kaminski}, {Korn}, {Kouach}, {Kowzan}, {Kreidberg}, {Landoni}, {Lanotte}, {Lavail}, {Li}, {Liske}, {Lovis}, {Lucatello}, {Lunney}, {MacIntosh}, {Madhusudhan}, {Magrini}, {Maiolino}, {Malo}, {Man}, {Marquart}, {Marques}, {Martins}, {Martins}, {Maslowski}, {Mason}, {Mason}, {McCracken}, {Mergo}, {Micela}, {Mitchell}, {Molli{\`e}re}, {Monteiro}, {Montgomery}, {Mordasini}, {Morin}, {Mucciarelli}, {Murphy}, {N'Diaye}, {Neichel}, {Niedzielski}, {Niemczura}, {Nortmann}, {Noterdaeme}, {Nunes}, {Oggioni}, {Oliva}, {{\"O}nel}, {Origlia}, {{\"O}stlin}, {Palle}, {Papaderos}, {Pariani}, {Pe{\~n}ate Castro}, {Pepe}, {Perreault Levasseur}, {Petit}, {Pino}, {Piqueras}, {Pollo}, {Poppenhaeger}, {Quirrenbach}, {Rauscher}, {Rebolo}, {Redaelli}, {Reffert}, {Reid},
  {Reiners}, {Richter}, {Riva}, {Rivoire}, {Rodr{\'\i}guez-L{\'o}pez}, {Roederer}, {Romano}, {Rousseau}, {Rowe}, {Salvadori}, {Santos}, {Santos Diaz}, {Sanz-Forcada}, {Sarajlic}, {Sauvage}, {Sch{\"a}fer}, {Schiavon}, {Schmidt}, {Selmi}, {Sivanandam}, {Sordet}, {Sordo}, {Sortino}, {Sosnowska}, {Sousa}, {Stempels}, {Strassmeier}, {Su{\'a}rez Mascare{\~n}o}, {Sulich}, {Sun}, {Tanvir}, {Tenegi-Sangin{\'e}s}, {Thibault}, {Thompson}, {Tozzi}, {Turbet}, {Vall{\'e}e}, {Varas}, {Venn}, {V{\'e}ran}, {Verma}, {Viel}, {Wade}, {Waring}, {Weber}, {Weder}, {Wehbe}, {Weingrill}, {Woche}, {Xompero}, {Zackrisson}, {Zanutta}, {Zapatero Osorio}, {Zechmeister}, and {Zimara}}]{Marconi2022}
{Marconi} A, {Abreu} M, {Adibekyan} V, et~al (2022) {ANDES, the high resolution spectrograph for the ELT: science case, baseline design and path to construction}. In: {Evans} CJ, {Bryant} JJ, {Motohara} K (eds) Ground-based and Airborne Instrumentation for Astronomy IX, p 1218424, \doi{10.1117/12.2628689}

\bibitem[{{Matthee} et~al(2022){Matthee}, {Naidu}, {Pezzulli}, {Gronke}, {Sobral}, {Oesch}, {Hayes}, {Erb}, {Schaerer}, {Amor{\'\i}n}, {Tacchella}, {Paulino-Afonso}, {Llerena}, {Calhau}, and {R{\"o}ttgering}}]{matthee2022}
{Matthee} J, {Naidu} RP, {Pezzulli} G, et~al (2022) {(Re)Solving reionization with Ly{\ensuremath{\alpha}}: how bright Ly{\ensuremath{\alpha}} Emitters account for the $z \approx 2-8$ cosmic ionizing background}. \mnras 512(4):5960--5977. \doi{10.1093/mnras/stac801}, {\href{https://arxiv.org/abs/2110.11967}{{arXiv:2110.11967}}} {[astro-ph.GA]}

\bibitem[{{McLeod} et~al(2024){McLeod}, {Donnan}, {McLure}, {Dunlop}, {Magee}, {Begley}, {Carnall}, {Cullen}, {Ellis}, {Hamadouche}, and {Stanton}}]{mcleod2024}
{McLeod} DJ, {Donnan} CT, {McLure} RJ, et~al (2024) {The galaxy UV luminosity function at $z \simeq 11$ from a suite of public JWST ERS, ERO, and Cycle-1 programs}. \mnras 527(3):5004--5022. \doi{10.1093/mnras/stad3471}, {\href{https://arxiv.org/abs/2304.14469}{{arXiv:2304.14469}}} {[astro-ph.GA]}

\bibitem[{{McQuinn}(2016)}]{mcquinn2016}
{McQuinn} M (2016) {The Evolution of the Intergalactic Medium}. \araa 54:313--362. \doi{10.1146/annurev-astro-082214-122355}, {\href{https://arxiv.org/abs/1512.00086}{{arXiv:1512.00086}}} {[astro-ph.CO]}

\bibitem[{{Merloni} et~al(2019){Merloni}, {Alexander}, {Banerji}, {Boller}, {Comparat}, {Dwelly}, {Fotopoulou}, {McMahon}, {Nandra}, {Salvato}, {Croom}, {Finoguenov}, {Krumpe}, {Lamer}, {Rosario}, {Schwope}, {Shanks}, {Steinmetz}, {Wisotzki}, and {Worseck}}]{Merloni19}
{Merloni} A, {Alexander} DA, {Banerji} M, et~al (2019) {4MOST Consortium Survey 6: Active Galactic Nuclei}. The Messenger 175:42--45. \doi{10.18727/0722-6691/5125}, {\href{https://arxiv.org/abs/1903.02472}{{arXiv:1903.02472}}} {[astro-ph.GA]}

\bibitem[{{Meyer} et~al(2020){Meyer}, {Kakiichi}, {Bosman}, {Ellis}, {Laporte}, {Robertson}, {Ryan-Weber}, {Mawatari}, and {Zitrin}}]{meyer2020}
{Meyer} RA, {Kakiichi} K, {Bosman} SEI, et~al (2020) {The role of galaxies and AGN in reionizing the IGM - III. IGM-galaxy cross-correlations at $z \sim 6$ from eight quasar fields with DEIMOS and MUSE}. \mnras 494(2):1560--1578. \doi{10.1093/mnras/staa746}, {\href{https://arxiv.org/abs/1912.04314}{{arXiv:1912.04314}}} {[astro-ph.GA]}

\bibitem[{{Molaro} et~al(2023){Molaro}, {Izzo}, {Selvelli}, {Bonifacio}, {Aydi}, {Cescutti}, {Guido}, {Harvey}, {Hernanz}, and {Della Valle}}]{Molaro2023}
{Molaro} P, {Izzo} L, {Selvelli} P, et~al (2023) {$^{7}$Be detection in the 2021 outburst of RS Oph}. \mnras 518(2):2614--2626. \doi{10.1093/mnras/stac2708}, {\href{https://arxiv.org/abs/2209.11008}{{arXiv:2209.11008}}} {[astro-ph.SR]}

\bibitem[{{Mukai} and {Sokoloski}(2019)}]{mukai19}
{Mukai} K, {Sokoloski} JL (2019) {The new science of novae}. Physics Today 72(11):38--44. \doi{10.1063/PT.3.4341}

\bibitem[{{Naoz} et~al(2006){Naoz}, {Noter}, and {Barkana}}]{naoz2006}
{Naoz} S, {Noter} S, {Barkana} R (2006) {The first stars in the Universe}. \mnras 373(1):L98--L102. \doi{10.1111/j.1745-3933.2006.00251.x}, {\href{https://arxiv.org/abs/astro-ph/0604050}{{arXiv:astro-ph/0604050}}} {[astro-ph]}

\bibitem[{{Neill} et~al(2011){Neill}, {Sullivan}, {Gal-Yam}, {Quimby}, {Ofek}, {Wyder}, {Howell}, {Nugent}, {Seibert}, {Martin}, {Overzier}, {Barlow}, {Foster}, {Friedman}, {Morrissey}, {Neff}, {Schiminovich}, {Bianchi}, {Donas}, {Heckman}, {Lee}, {Madore}, {Milliard}, {Rich}, and {Szalay}}]{neill11}
{Neill} JD, {Sullivan} M, {Gal-Yam} A, et~al (2011) {The Extreme Hosts of Extreme Supernovae}. \apj 727(1):15. \doi{10.1088/0004-637X/727/1/15}, {\href{https://arxiv.org/abs/1011.3512}{{arXiv:1011.3512}}} {[astro-ph.CO]}

\bibitem[{{Nicholl} et~al(2013){Nicholl}, {Smartt}, {Jerkstrand}, {Inserra}, {McCrum}, {Kotak}, {Fraser}, {Wright}, {Chen}, {Smith}, {Young}, {Sim}, {Valenti}, {Howell}, {Bresolin}, {Kudritzki}, {Tonry}, {Huber}, {Rest}, {Pastorello}, {Tomasella}, {Cappellaro}, {Benetti}, {Mattila}, {Kankare}, {Kangas}, {Leloudas}, {Sollerman}, {Taddia}, {Berger}, {Chornock}, {Narayan}, {Stubbs}, {Foley}, {Lunnan}, {Soderberg}, {Sanders}, {Milisavljevic}, {Margutti}, {Kirshner}, {Elias-Rosa}, {Morales-Garoffolo}, {Taubenberger}, {Botticella}, {Gezari}, {Urata}, {Rodney}, {Riess}, {Scolnic}, {Wood-Vasey}, {Burgett}, {Chambers}, {Flewelling}, {Magnier}, {Kaiser}, {Metcalfe}, {Morgan}, {Price}, {Sweeney}, and {Waters}}]{nicholl13}
{Nicholl} M, {Smartt} SJ, {Jerkstrand} A, et~al (2013) {Slowly fading super-luminous supernovae that are not pair-instability explosions}. \nat 502(7471):346--349. \doi{10.1038/nature12569}, {\href{https://arxiv.org/abs/1310.4446}{{arXiv:1310.4446}}} {[astro-ph.CO]}

\bibitem[{{Oh}(2002)}]{oh2002}
{Oh} SP (2002) {Probing the dark ages with metal absorption lines}. \mnras 336(3):1021--1029. \doi{10.1046/j.1365-8711.2002.05859.x}, {\href{https://arxiv.org/abs/astro-ph/0201517}{{arXiv:astro-ph/0201517}}} {[astro-ph]}

\bibitem[{{Pagnini} et~al(2023){Pagnini}, {Salvadori}, {Rossi}, {Aguado}, {Koutsouridou}, and {Sk{\'u}lad{\'o}ttir}}]{Pagnini2023}
{Pagnini} G, {Salvadori} S, {Rossi} M, et~al (2023) {On the dearth of C-enhanced metal-poor stars in the galactic bulge}. \mnras 521(4):5699--5711. \doi{10.1093/mnras/stad912}, {\href{https://arxiv.org/abs/2303.14204}{{arXiv:2303.14204}}} {[astro-ph.GA]}

\bibitem[{{Pallottini} et~al(2014){Pallottini}, {Ferrara}, {Gallerani}, {Salvadori}, and {D'Odorico}}]{pallottini2014}
{Pallottini} A, {Ferrara} A, {Gallerani} S, et~al (2014) {Simulating cosmic metal enrichment by the first galaxies}. \mnras 440(3):2498--2518. \doi{10.1093/mnras/stu451}, {\href{https://arxiv.org/abs/1403.1261}{{arXiv:1403.1261}}} {[astro-ph.CO]}

\bibitem[{{Perley} et~al(2016){Perley}, {Quimby}, {Yan}, {Vreeswijk}, {De Cia}, {Lunnan}, {Gal-Yam}, {Yaron}, {Filippenko}, {Graham}, {Laher}, and {Nugent}}]{perley16}
{Perley} DA, {Quimby} RM, {Yan} L, et~al (2016) {Host-galaxy Properties of 32 Low-redshift Superluminous Supernovae from the Palomar Transient Factory}. \apj 830(1):13. \doi{10.3847/0004-637X/830/1/13}, {\href{https://arxiv.org/abs/1604.08207}{{arXiv:1604.08207}}} {[astro-ph.HE]}

\bibitem[{{Perley} et~al(2017){Perley}, {Kr{\"u}hler}, {Schady}, {Micha{\l}owski}, {Th{\"o}ne}, {Petry}, {Graham}, {Greiner}, {Klose}, {Schulze}, and {Kim}}]{perley17}
{Perley} DA, {Kr{\"u}hler} T, {Schady} P, et~al (2017) {A revised host galaxy association for GRB 020819B: a high-redshift dusty starburst, not a low-redshift gas-poor spiral}. \mnras 465(1):L89--L93. \doi{10.1093/mnrasl/slw221}, {\href{https://arxiv.org/abs/1609.04016}{{arXiv:1609.04016}}} {[astro-ph.HE]}

\bibitem[{{Pettini} et~al(1989){Pettini}, {Stathakis}, {D'Odorico}, {Molaro}, and {Vladilo}}]{pettini1989}
{Pettini} M, {Stathakis} R, {D'Odorico} S, et~al (1989) {Million Degree Gas in the Galactic Halo and the Large Magellanic Cloud. II. The Line of Sight to SN 1987A}. \apj 340:256. \doi{10.1086/167389}

\bibitem[{{Prochaska} et~al(2007){Prochaska}, {Chen}, {Dessauges-Zavadsky}, and {Bloom}}]{prochaska07}
{Prochaska} JX, {Chen} HW, {Dessauges-Zavadsky} M, et~al (2007) {Probing the Interstellar Medium near Star-forming Regions with Gamma-Ray Burst Afterglow Spectroscopy: Gas, Metals, and Dust}. \apj 666(1):267--280. \doi{10.1086/520042}, {\href{https://arxiv.org/abs/astro-ph/0703665}{{arXiv:astro-ph/0703665}}} {[astro-ph]}

\bibitem[{{Prochaska} et~al(2009){Prochaska}, {Sheffer}, {Perley}, {Bloom}, {Lopez}, {Dessauges-Zavadsky}, {Chen}, {Filippenko}, {Ganeshalingam}, {Li}, {Miller}, and {Starr}}]{prochaska09}
{Prochaska} JX, {Sheffer} Y, {Perley} DA, et~al (2009) {The First Positive Detection of Molecular Gas in a GRB Host Galaxy}. \apjl 691(1):L27--L32. \doi{10.1088/0004-637X/691/1/L27}, {\href{https://arxiv.org/abs/0901.0556}{{arXiv:0901.0556}}} {[astro-ph.HE]}

\bibitem[{{Puchwein} et~al(2023){Puchwein}, {Bolton}, {Keating}, {Molaro}, {Gaikwad}, {Kulkarni}, {Haehnelt}, {Ir{\v{s}}i{\v{c}}}, {{\v{S}}oltinsk{\'y}}, {Viel}, {Aubert}, {Becker}, and {Meiksin}}]{puchwein2023}
{Puchwein} E, {Bolton} JS, {Keating} LC, et~al (2023) {The Sherwood-Relics simulations: overview and impact of patchy reionization and pressure smoothing on the intergalactic medium}. \mnras 519(4):6162--6183. \doi{10.1093/mnras/stac3761}, {\href{https://arxiv.org/abs/2207.13098}{{arXiv:2207.13098}}} {[astro-ph.CO]}

\bibitem[{{Quimby} et~al(2011){Quimby}, {Kulkarni}, {Kasliwal}, {Gal-Yam}, {Arcavi}, {Sullivan}, {Nugent}, {Thomas}, {Howell}, {Nakar}, {Bildsten}, {Theissen}, {Law}, {Dekany}, {Rahmer}, {Hale}, {Smith}, {Ofek}, {Zolkower}, {Velur}, {Walters}, {Henning}, {Bui}, {McKenna}, {Poznanski}, {Cenko}, and {Levitan}}]{quimby11}
{Quimby} RM, {Kulkarni} SR, {Kasliwal} MM, et~al (2011) {Hydrogen-poor superluminous stellar explosions}. \nat 474(7352):487--489. \doi{10.1038/nature10095}, {\href{https://arxiv.org/abs/0910.0059}{{arXiv:0910.0059}}} {[astro-ph.CO]}

\bibitem[{{Quimby} et~al(2018){Quimby}, {De Cia}, {Gal-Yam}, {Leloudas}, {Lunnan}, {Perley}, {Vreeswijk}, {Yan}, {Bloom}, {Cenko}, {Cooke}, {Ellis}, {Filippenko}, {Kasliwal}, {Kleiser}, {Kulkarni}, {Matheson}, {Nugent}, {Pan}, {Silverman}, {Sternberg}, {Sullivan}, and {Yaron}}]{quimby18}
{Quimby} RM, {De Cia} A, {Gal-Yam} A, et~al (2018) {Spectra of Hydrogen-poor Superluminous Supernovae from the Palomar Transient Factory}. \apj 855(1):2. \doi{10.3847/1538-4357/aaac2f}, {\href{https://arxiv.org/abs/1802.07820}{{arXiv:1802.07820}}} {[astro-ph.HE]}

\bibitem[{{Reines}(2022)}]{reines2022}
{Reines} AE (2022) {Hunting for massive black holes in dwarf galaxies}. Nature Astronomy 6:26--34. \doi{10.1038/s41550-021-01556-0}, {\href{https://arxiv.org/abs/2201.10569}{{arXiv:2201.10569}}} {[astro-ph.GA]}

\bibitem[{{Richter} et~al(2011){Richter}, {Krause}, {Fechner}, {Charlton}, and {Murphy}}]{richter2011}
{Richter} P, {Krause} F, {Fechner} C, et~al (2011) {The neutral gas extent of galaxies as derived from weak intervening Ca II absorbers}. \aap 528:A12. \doi{10.1051/0004-6361/201015566}, {\href{https://arxiv.org/abs/1008.2201}{{arXiv:1008.2201}}} {[astro-ph.CO]}

\bibitem[{{Richter} et~al(2014){Richter}, {Fox}, {Ben Bekhti}, {Murphy}, {Bomans}, and {Frank}}]{richter2014}
{Richter} P, {Fox} AJ, {Ben Bekhti} N, et~al (2014) {High-resolution absorption spectroscopy of the circumgalactic medium of the Milky Way}. Astronomische Nachrichten 335(1):92. \doi{10.1002/asna.201312013}, {\href{https://arxiv.org/abs/1310.4514}{{arXiv:1310.4514}}} {[astro-ph.GA]}

\bibitem[{{Robert} et~al(2022){Robert}, {Murphy}, {O'Meara}, {Crighton}, and {Fumagalli}}]{Robert22}
{Robert} PF, {Murphy} MT, {O'Meara} JM, et~al (2022) {Discovery of three new near-pristine absorption clouds at z = 2.6-4.4}. \mnras 514(3):3559--3578. \doi{10.1093/mnras/stac1550}, {\href{https://arxiv.org/abs/2206.02947}{{arXiv:2206.02947}}} {[astro-ph.CO]}

\bibitem[{{Rossi} et~al(2021){Rossi}, {Salvadori}, and {Sk{\'u}lad{\'o}ttir}}]{Rossi2021}
{Rossi} M, {Salvadori} S, {Sk{\'u}lad{\'o}ttir} {\'A} (2021) {Ultra-faint dwarf galaxies: unveiling the minimum mass of the first stars}. \mnras 503(4):6026--6044. \doi{10.1093/mnras/stab821}, {\href{https://arxiv.org/abs/2103.09834}{{arXiv:2103.09834}}} {[astro-ph.GA]}

\bibitem[{{Saccardi} et~al(2023{\natexlab{a}}){Saccardi}, {Salvadori}, {D'Odorico}, {Cupani}, {Fumagalli}, {Berg}, {Becker}, {Ellison}, and {Lopez}}]{saccardi2023b}
{Saccardi} A, {Salvadori} S, {D'Odorico} V, et~al (2023{\natexlab{a}}) {Evidence of First Stars-enriched Gas in High-redshift Absorbers}. \apj 948(1):35. \doi{10.3847/1538-4357/acc39f}, {\href{https://arxiv.org/abs/2305.02346}{{arXiv:2305.02346}}} {[astro-ph.GA]}

\bibitem[{{Saccardi} et~al(2023{\natexlab{b}}){Saccardi}, {Vergani}, {De Cia}, {D'Elia}, {Heintz}, {Izzo}, {Palmerio}, {Petitjean}, {Rossi}, {de Ugarte Postigo}, {Christensen}, {Konstantopoulou}, {Levan}, {Malesani}, {M{\o}ller}, {Ramburuth-Hurt}, {Salvaterra}, {Tanvir}, {Th{\"o}ne}, {Vejlgaard}, {Fynbo}, {Kann}, {Schady}, {Watson}, {Wiersema}, {Campana}, {Covino}, {De Pasquale}, {Fausey}, {Hartmann}, {van der Horst}, {Jakobsson}, {Palazzi}, {Pugliese}, {Savaglio}, {Starling}, {Stratta}, and {Zafar}}]{saccardi2023a}
{Saccardi} A, {Vergani} SD, {De Cia} A, et~al (2023{\natexlab{b}}) {Dissecting the interstellar medium of a z = 6.3 galaxy. X-shooter spectroscopy and HST imaging of the afterglow and environment of the Swift GRB 210905A}. \aap 671:A84. \doi{10.1051/0004-6361/202244205}, {\href{https://arxiv.org/abs/2211.16524}{{arXiv:2211.16524}}} {[astro-ph.GA]}

\bibitem[{{Salvadori} et~al(2019){Salvadori}, {Bonifacio}, {Caffau}, {Korotin}, {Andreevsky}, {Spite}, and {Sk{\'u}lad{\'o}ttir}}]{salvadori2019}
{Salvadori} S, {Bonifacio} P, {Caffau} E, et~al (2019) {Probing the existence of very massive first stars}. \mnras 487(3):4261--4284. \doi{10.1093/mnras/stz1464}, {\href{https://arxiv.org/abs/1906.00994}{{arXiv:1906.00994}}} {[astro-ph.GA]}

\bibitem[{{Salvadori} et~al(2023){Salvadori}, {D'Odorico}, {Saccardi}, {Sk{\'u}lad{\'o}ttir}, and {Vanni}}]{salvadori2023}
{Salvadori} S, {D'Odorico} V, {Saccardi} A, et~al (2023) {First stars signatures in high-z absorbers}. In: Memorie della Societa Astronomica Italiana, p 215, \doi{10.36116/MEMSAIT_94N2.2023.215}, \eprint{2305.07706}

\bibitem[{{Salvaterra} et~al(2009){Salvaterra}, {Della Valle}, {Campana}, {Chincarini}, {Covino}, {D'Avanzo}, {Fern{\'a}ndez-Soto}, {Guidorzi}, {Mannucci}, {Margutti}, {Th{\"o}ne}, {Antonelli}, {Barthelmy}, {de Pasquale}, {D'Elia}, {Fiore}, {Fugazza}, {Hunt}, {Maiorano}, {Marinoni}, {Marshall}, {Molinari}, {Nousek}, {Pian}, {Racusin}, {Stella}, {Amati}, {Andreuzzi}, {Cusumano}, {Fenimore}, {Ferrero}, {Giommi}, {Guetta}, {Holland}, {Hurley}, {Israel}, {Mao}, {Markwardt}, {Masetti}, {Pagani}, {Palazzi}, {Palmer}, {Piranomonte}, {Tagliaferri}, and {Testa}}]{salvaterra09}
{Salvaterra} R, {Della Valle} M, {Campana} S, et~al (2009) {GRB090423 at a redshift of z\raisebox{-0.5ex}\textasciitilde8.1}. \nat 461(7268):1258--1260. \doi{10.1038/nature08445}, {\href{https://arxiv.org/abs/0906.1578}{{arXiv:0906.1578}}} {[astro-ph.CO]}

\bibitem[{{Schady} et~al(2024){Schady}, {Yates}, {Christensen}, {De Cia}, {Rossi}, {D'Elia}, {Heintz}, {Jakobsson}, {Laskar}, {Levan}, {Salvaterra}, {Starling}, {Tanvir}, {Th{\"o}ne}, {Vergani}, {Wiersema}, {Arabsalmani}, {Chen}, {De Pasquale}, {Fruchter}, {Fynbo}, {Garc{\'\i}a-Benito}, {Gompertz}, {Hartmann}, {Kouveliotou}, {Milvang-Jensen}, {Palazzi}, {Perley}, {Piranomonte}, {Pugliese}, {Savaglio}, {Sbarufatti}, {Schulze}, {Tagliaferri}, {de Ugarte Postigo}, {Watson}, and {Wiseman}}]{schady2024}
{Schady} P, {Yates} RM, {Christensen} L, et~al (2024) {Comparing emission- and absorption-based gas-phase metallicities in GRB host galaxies at z = 2-4 using JWST}. \mnras 529(3):2807--2831. \doi{10.1093/mnras/stae677}, {\href{https://arxiv.org/abs/2310.15967}{{arXiv:2310.15967}}} {[astro-ph.GA]}

\bibitem[{{Schindler} et~al(2023){Schindler}, {Ba{\~n}ados}, {Connor}, {Decarli}, {Fan}, {Farina}, {Mazzucchelli}, {Nanni}, {Rix}, {Stern}, {Venemans}, and {Walter}}]{schindler2023}
{Schindler} JT, {Ba{\~n}ados} E, {Connor} T, et~al (2023) {The Pan-STARRS1 $z > 5.6$ Quasar Survey. III. The $z \approx 6$ Quasar Luminosity Function}. \apj 943(1):67. \doi{10.3847/1538-4357/aca7ca}, {\href{https://arxiv.org/abs/2212.04179}{{arXiv:2212.04179}}} {[astro-ph.GA]}

\bibitem[{{Schulze} et~al(2018){Schulze}, {Kr{\"u}hler}, {Leloudas}, {Gorosabel}, {Mehner}, {Buchner}, {Kim}, {Ibar}, {Amor{\'\i}n}, {Herrero-Illana}, {Anderson}, {Bauer}, {Christensen}, {de Pasquale}, {de Ugarte Postigo}, {Gallazzi}, {Hjorth}, {Morrell}, {Malesani}, {Sparre}, {Stalder}, {Stark}, {Th{\"o}ne}, and {Wheeler}}]{schulze18}
{Schulze} S, {Kr{\"u}hler} T, {Leloudas} G, et~al (2018) {Cosmic evolution and metal aversion in superluminous supernova host galaxies}. \mnras 473(1):1258--1285. \doi{10.1093/mnras/stx2352}, {\href{https://arxiv.org/abs/1612.05978}{{arXiv:1612.05978}}} {[astro-ph.GA]}

\bibitem[{{Scovacricchi} et~al(2016){Scovacricchi}, {Nichol}, {Bacon}, {Sullivan}, and {Prajs}}]{scovacricchi16}
{Scovacricchi} D, {Nichol} RC, {Bacon} D, et~al (2016) {Cosmology with superluminous supernovae}. \mnras 456(2):1700--1707. \doi{10.1093/mnras/stv2752}, {\href{https://arxiv.org/abs/1511.06670}{{arXiv:1511.06670}}} {[astro-ph.CO]}

\bibitem[{{Sebastian} et~al(2024){Sebastian}, {Ryan-Weber}, {Davies}, {Becker}, {Keating}, {D'Odorico}, {Meyer}, {Bosman}, {Cupani}, {Kulkarni}, {Haehnelt}, {Lai}, {Eilers}, {Bischetti}, and {Gallerani}}]{sebastian2024}
{Sebastian} AM, {Ryan-Weber} E, {Davies} RL, et~al (2024) {E-XQR-30: The evolution of Mg II, C II, and O I across $2 < z < 6$}. \mnras 530(2):1829--1848. \doi{10.1093/mnras/stae789}, {\href{https://arxiv.org/abs/2403.10072}{{arXiv:2403.10072}}} {[astro-ph.GA]}

\bibitem[{{Simcoe} et~al(2004){Simcoe}, {Sargent}, and {Rauch}}]{simcoe2004}
{Simcoe} RA, {Sargent} WLW, {Rauch} M (2004) {The Distribution of Metallicity in the Intergalactic Medium at z\raisebox{-0.5ex}\textasciitilde2.5: O VI and C IV Absorption in the Spectra of Seven QSOs}. \apj 606(1):92--115. \doi{10.1086/382777}, {\href{https://arxiv.org/abs/astro-ph/0312467}{{arXiv:astro-ph/0312467}}} {[astro-ph]}

\bibitem[{{Sk{\'u}lad{\'o}ttir} et~al(2024){Sk{\'u}lad{\'o}ttir}, {Koutsouridou}, {Vanni}, {Amarsi}, {Lucchesi}, {Salvadori}, and {Aguado}}]{skuladottir2024b}
{Sk{\'u}lad{\'o}ttir} {\'A}, {Koutsouridou} I, {Vanni} I, et~al (2024) {On the Pair-instability Supernova Origin of J1010+2358}. \apjl 968(2):L23. \doi{10.3847/2041-8213/ad4b1a}, {\href{https://arxiv.org/abs/2404.19086}{{arXiv:2404.19086}}} {[astro-ph.SR]}

\bibitem[{{Sodini} et~al(2024){Sodini}, {D'Odorico}, {Salvadori}, {Vanni}, {Bischetti}, {Cupani}, {Davies}, {Becker}, {Ba{\~n}ados}, {Bosman}, {Davies}, {Paolo Farina}, {Ferrara}, {Keating}, {Kulkarni}, {Lai}, {Ryan-Weber}, {Maria Sebastian}, and {Walter}}]{sodini2024}
{Sodini} A, {D'Odorico} V, {Salvadori} S, et~al (2024) {Evidence of Pop III stars' chemical signature in neutral gas at z {\ensuremath{\sim}} 6. A study based on the E-XQR-30 spectroscopic sample}. \aap 687:A314. \doi{10.1051/0004-6361/202349062}, {\href{https://arxiv.org/abs/2404.10722}{{arXiv:2404.10722}}} {[astro-ph.GA]}

\bibitem[{{Songaila}(2005)}]{songaila05}
{Songaila} A (2005) {The Properties of Intergalactic C IV and Si IV Absorption. I. Optimal Analysis of an Extremely High Signal-to-Noise Quasar Sample}. \aj 130(5):1996--2005. \doi{10.1086/491704}, {\href{https://arxiv.org/abs/astro-ph/0507649}{{arXiv:astro-ph/0507649}}} {[astro-ph]}

\bibitem[{{Starrfield} et~al(1978){Starrfield}, {Truran}, {Sparks}, and {Arnould}}]{Starrfield1978}
{Starrfield} S, {Truran} JW, {Sparks} WM, et~al (1978) {On $^{7}$Li production in nova explosions.} \apj 222:600--603. \doi{10.1086/156175}

\bibitem[{{Sukhbold} et~al(2016){Sukhbold}, {Ertl}, {Woosley}, {Brown}, and {Janka}}]{Sukhbold16}
{Sukhbold} T, {Ertl} T, {Woosley} SE, et~al (2016) {Core-collapse Supernovae from 9 to 120 Solar Masses Based on Neutrino-powered Explosions}. \apj 821(1):38. \doi{10.3847/0004-637X/821/1/38}, {\href{https://arxiv.org/abs/1510.04643}{{arXiv:1510.04643}}} {[astro-ph.HE]}

\bibitem[{{Surace} et~al(2018){Surace}, {Whalen}, {Hartwig}, {Zackrisson}, {Glover}, {Patrick}, {Woods}, {Heger}, and {Haemmerl{\'e}}}]{surace18}
{Surace} M, {Whalen} DJ, {Hartwig} T, et~al (2018) {On the Detection of Supermassive Primordial Stars}. \apjl 869(2):L39. \doi{10.3847/2041-8213/aaf80d}, {\href{https://arxiv.org/abs/1811.08911}{{arXiv:1811.08911}}} {[astro-ph.GA]}

\bibitem[{{Takahashi} et~al(2018){Takahashi}, {Yoshida}, and {Umeda}}]{takahashi2018}
{Takahashi} K, {Yoshida} T, {Umeda} H (2018) {Stellar Yields of Rotating First Stars. II. Pair-instability Supernovae and Comparison with Observations}. \apj 857(2):111. \doi{10.3847/1538-4357/aab95f}, {\href{https://arxiv.org/abs/1803.06630}{{arXiv:1803.06630}}} {[astro-ph.SR]}

\bibitem[{{Tanvir} et~al(2021{\natexlab{a}}){Tanvir}, {Rossi}, {Xu}, {Zhu}, {Izzo}, {Kann}, {Levan}, and {Stargate Collaboration}}]{Tanvir21}
{Tanvir} N, {Rossi} A, {Xu} D, et~al (2021{\natexlab{a}}) {GRB 210905A: VLT/X-shooter spectroscopic redshift}. GRB Coordinates Network 30771:1

\bibitem[{{Tanvir} et~al(2009){Tanvir}, {Fox}, {Levan}, {Berger}, {Wiersema}, {Fynbo}, {Cucchiara}, {Kr{\"u}hler}, {Gehrels}, {Bloom}, {Greiner}, {Evans}, {Rol}, {Olivares}, {Hjorth}, {Jakobsson}, {Farihi}, {Willingale}, {Starling}, {Cenko}, {Perley}, {Maund}, {Duke}, {Wijers}, {Adamson}, {Allan}, {Bremer}, {Burrows}, {Castro-Tirado}, {Cavanagh}, {de Ugarte Postigo}, {Dopita}, {Fatkhullin}, {Fruchter}, {Foley}, {Gorosabel}, {Kennea}, {Kerr}, {Klose}, {Krimm}, {Komarova}, {Kulkarni}, {Moskvitin}, {Mundell}, {Naylor}, {Page}, {Penprase}, {Perri}, {Podsiadlowski}, {Roth}, {Rutledge}, {Sakamoto}, {Schady}, {Schmidt}, {Soderberg}, {Sollerman}, {Stephens}, {Stratta}, {Ukwatta}, {Watson}, {Westra}, {Wold}, and {Wolf}}]{tanvir09}
{Tanvir} NR, {Fox} DB, {Levan} AJ, et~al (2009) {A {\ensuremath{\gamma}}-ray burst at a redshift of z\raisebox{-0.5ex}\textasciitilde8.2}. \nat 461(7268):1254--1257. \doi{10.1038/nature08459}, {\href{https://arxiv.org/abs/0906.1577}{{arXiv:0906.1577}}} {[astro-ph.CO]}

\bibitem[{{Tanvir} et~al(2021{\natexlab{b}}){Tanvir}, {Le Floc'h}, {Christensen}, {Caruana}, {Salvaterra}, {Ghirlanda}, {Ciardi}, {Maio}, {D'Odorico}, {Piedipalumbo}, {Campana}, {Noterdaeme}, {Graziani}, {Amati}, {Bagoly}, {Bal{\'a}zs}, {Basa}, {Behar}, {De Cia}, {Della Valle}, {De Pasquale}, {Frontera}, {Gomboc}, {G{\"o}tz}, {Horvath}, {Hudec}, {Mereghetti}, {O'Brien}, {Osborne}, {Paltani}, {Rosati}, {Sergijenko}, {Stanway}, {Sz{\'e}csi}, {Toth}, {Urata}, {Vergani}, and {Zane}}]{tanvir21b}
{Tanvir} NR, {Le Floc'h} E, {Christensen} L, et~al (2021{\natexlab{b}}) {Exploration of the high-redshift universe enabled by THESEUS}. Experimental Astronomy 52(3):219--244. \doi{10.1007/s10686-021-09778-w}, {\href{https://arxiv.org/abs/2104.09532}{{arXiv:2104.09532}}} {[astro-ph.IM]}

\bibitem[{{Tee} et~al(2023){Tee}, {Fan}, {Wang}, {Yang}, {Malhotra}, and {Rhoads}}]{Tee23}
{Tee} WL, {Fan} X, {Wang} F, et~al (2023) {Predicting the Yields of $z > 6.5$ Quasar Surveys in the Era of Roman and Rubin}. \apj 956(1):52. \doi{10.3847/1538-4357/acf12d}, {\href{https://arxiv.org/abs/2308.12278}{{arXiv:2308.12278}}} {[astro-ph.GA]}

\bibitem[{{Thibodeaux} et~al(2024){Thibodeaux}, {Ji}, {Cerny}, {Kirby}, and {Simon}}]{thibodeaux2024}
{Thibodeaux} P, {Ji} AP, {Cerny} W, et~al (2024) {LAMOST J1010+2358 is not a Pair-Instability Supernova Relic}. The Open Journal of Astrophysics 7:66. \doi{10.33232/001c.122335}, {\href{https://arxiv.org/abs/2404.17078}{{arXiv:2404.17078}}} {[astro-ph.SR]}

\bibitem[{{Thom} et~al(2008){Thom}, {Peek}, {Putman}, {Heiles}, {Peek}, and {Wilhelm}}]{thom2008}
{Thom} C, {Peek} JEG, {Putman} ME, et~al (2008) {An Accurate Distance to High-Velocity Cloud Complex C}. \apj 684(1):364--372. \doi{10.1086/589960}, {\href{https://arxiv.org/abs/0712.0612}{{arXiv:0712.0612}}} {[astro-ph]}

\bibitem[{{Tornatore} et~al(2007){Tornatore}, {Ferrara}, and {Schneider}}]{tornatore2007}
{Tornatore} L, {Ferrara} A, {Schneider} R (2007) {Population III stars: hidden or disappeared?} \mnras 382(3):945--950. \doi{10.1111/j.1365-2966.2007.12215.x}, {\href{https://arxiv.org/abs/0707.1433}{{arXiv:0707.1433}}} {[astro-ph]}

\bibitem[{{Tozzi} et~al(2022){Tozzi}, {Gilli}, {Liu}, {Borgani}, {Lepore}, {Di Mascolo}, {Saro}, {Pentericci}, {Carilli}, {Miley}, {Mroczkowski}, {Pannella}, {Rasia}, {Rosati}, {Anderson}, {Calabr{\'o}}, {Churazov}, {Dannerbauer}, {Feruglio}, {Fiore}, {Gobat}, {Jin}, {Nonino}, {Norman}, and {R{\"o}ttgering}}]{tozzi2022}
{Tozzi} P, {Gilli} R, {Liu} A, et~al (2022) {The 700 ks Chandra Spiderweb Field. II. Evidence for inverse-Compton and thermal diffuse emission in the Spiderweb galaxy}. \aap 667:A134. \doi{10.1051/0004-6361/202244337}, {\href{https://arxiv.org/abs/2209.15467}{{arXiv:2209.15467}}} {[astro-ph.GA]}

\bibitem[{{Turner} et~al(2014){Turner}, {Schaye}, {Steidel}, {Rudie}, and {Strom}}]{turner2014}
{Turner} ML, {Schaye} J, {Steidel} CC, et~al (2014) {Metal-line absorption around z {\ensuremath{\approx}} 2.4 star-forming galaxies in the Keck Baryonic Structure Survey}. \mnras 445(1):794--822. \doi{10.1093/mnras/stu1801}, {\href{https://arxiv.org/abs/1403.0942}{{arXiv:1403.0942}}} {[astro-ph.CO]}

\bibitem[{{Umeda} et~al(2024){Umeda}, {Ouchi}, {Nakajima}, {Harikane}, {Ono}, {Xu}, {Isobe}, and {Zhang}}]{umeda2024}
{Umeda} H, {Ouchi} M, {Nakajima} K, et~al (2024) {JWST Measurements of Neutral Hydrogen Fractions and Ionized Bubble Sizes at z = 7{\textendash}12 Obtained with Ly{\ensuremath{\alpha}} Damping Wing Absorptions in 27 Bright Continuum Galaxies}. \apj 971(2):124. \doi{10.3847/1538-4357/ad554e}, {\href{https://arxiv.org/abs/2306.00487}{{arXiv:2306.00487}}} {[astro-ph.GA]}

\bibitem[{{Vanni} et~al(2023){Vanni}, {Salvadori}, {Sk{\'u}lad{\'o}ttir}, {Rossi}, and {Koutsouridou}}]{vanni2023a}
{Vanni} I, {Salvadori} S, {Sk{\'u}lad{\'o}ttir} {\'A}, et~al (2023) {Characterizing the true descendants of the first stars}. \mnras 526(2):2620--2644. \doi{10.1093/mnras/stad2910}, {\href{https://arxiv.org/abs/2309.07958}{{arXiv:2309.07958}}} {[astro-ph.GA]}

\bibitem[{{Vanni} et~al(2024){Vanni}, {Salvadori}, {D'Odorico}, {Becker}, and {Cupani}}]{vanni2024}
{Vanni} I, {Salvadori} S, {D'Odorico} V, et~al (2024) {Chemical Diagnostics to Unveil Environments Enriched by First Stars}. \apjl 967(2):L22. \doi{10.3847/2041-8213/ad46fa}, {\href{https://arxiv.org/abs/2402.18640}{{arXiv:2402.18640}}} {[astro-ph.GA]}

\bibitem[{{Vanzella} et~al(2023){Vanzella}, {Loiacono}, {Bergamini}, {Me{\v{s}}tri{\'c}}, {Castellano}, {Rosati}, {Meneghetti}, {Grillo}, {Calura}, {Mignoli}, {Brada{\v{c}}}, {Adamo}, {Rihtar{\v{s}}i{\v{c}}}, {Dickinson}, {Gronke}, {Zanella}, {Annibali}, {Willott}, {Messa}, {Sani}, {Acebron}, {Bolamperti}, {Comastri}, {Gilli}, {Caputi}, {Ricotti}, {Gruppioni}, {Ravindranath}, {Mercurio}, {Strait}, {Martis}, {Pascale}, {Caminha}, {Annunziatella}, and {Nonino}}]{Vanzella23}
{Vanzella} E, {Loiacono} F, {Bergamini} P, et~al (2023) {An extremely metal-poor star complex in the reionization era: Approaching Population III stars with JWST}. \aap 678:A173. \doi{10.1051/0004-6361/202346981}, {\href{https://arxiv.org/abs/2305.14413}{{arXiv:2305.14413}}} {[astro-ph.GA]}

\bibitem[{{Villar} et~al(2018){Villar}, {Nicholl}, and {Berger}}]{Villar18}
{Villar} VA, {Nicholl} M, {Berger} E (2018) {Superluminous Supernovae in LSST: Rates, Detection Metrics, and Light-curve Modeling}. \apj 869(2):166. \doi{10.3847/1538-4357/aaee6a}, {\href{https://arxiv.org/abs/1809.07343}{{arXiv:1809.07343}}} {[astro-ph.HE]}

\bibitem[{{Villasenor} et~al(2023){Villasenor}, {Robertson}, {Madau}, and {Schneider}}]{Villasenor2023}
{Villasenor} B, {Robertson} B, {Madau} P, et~al (2023) {New constraints on warm dark matter from the Lyman-{\ensuremath{\alpha}} forest power spectrum}. \prd 108(2):023502. \doi{10.1103/PhysRevD.108.023502}, {\href{https://arxiv.org/abs/2209.14220}{{arXiv:2209.14220}}} {[astro-ph.CO]}

\bibitem[{{Volonteri} et~al(2008){Volonteri}, {Lodato}, and {Natarajan}}]{volonteri2008}
{Volonteri} M, {Lodato} G, {Natarajan} P (2008) {The evolution of massive black hole seeds}. \mnras 383(3):1079--1088. \doi{10.1111/j.1365-2966.2007.12589.x}, {\href{https://arxiv.org/abs/0709.0529}{{arXiv:0709.0529}}} {[astro-ph]}

\bibitem[{{Vreeswijk} et~al(2004){Vreeswijk}, {Ellison}, {Ledoux}, {Wijers}, {Fynbo}, {M{\o}ller}, {Henden}, {Hjorth}, {Masi}, {Rol}, {Jensen}, {Tanvir}, {Levan}, {Castro Cer{\'o}n}, {Gorosabel}, {Castro-Tirado}, {Fruchter}, {Kouveliotou}, {Burud}, {Rhoads}, {Masetti}, {Palazzi}, {Pian}, {Pedersen}, {Kaper}, {Gilmore}, {Kilmartin}, {Buckle}, {Seigar}, {Hartmann}, {Lindsay}, and {van den Heuvel}}]{vreeswijk04}
{Vreeswijk} PM, {Ellison} SL, {Ledoux} C, et~al (2004) {The host of GRB 030323 at z=3.372: A very high column density DLA system with a low metallicity}. \aap 419:927--940. \doi{10.1051/0004-6361:20040086}, {\href{https://arxiv.org/abs/astro-ph/0403080}{{arXiv:astro-ph/0403080}}} {[astro-ph]}

\bibitem[{{Vreeswijk} et~al(2007){Vreeswijk}, {Ledoux}, {Smette}, {Ellison}, {Jaunsen}, {Andersen}, {Fruchter}, {Fynbo}, {Hjorth}, {Kaufer}, {M{\o}ller}, {Petitjean}, {Savaglio}, and {Wijers}}]{vreeswijk07}
{Vreeswijk} PM, {Ledoux} C, {Smette} A, et~al (2007) {Rapid-response mode VLT/UVES spectroscopy of GRB{\,}060418. Conclusive evidence for UV pumping from the time evolution of Fe II and Ni II excited- and metastable-level populations}. \aap 468(1):83--96. \doi{10.1051/0004-6361:20066780}, {\href{https://arxiv.org/abs/astro-ph/0611690}{{arXiv:astro-ph/0611690}}} {[astro-ph]}

\bibitem[{{Vreeswijk} et~al(2013){Vreeswijk}, {Ledoux}, {Raassen}, {Smette}, {De Cia}, {Wo{\'z}niak}, {Fox}, {Vestrand}, and {Jakobsson}}]{vreeswijk13}
{Vreeswijk} PM, {Ledoux} C, {Raassen} AJJ, et~al (2013) {Time-dependent excitation and ionization modelling of absorption-line variability due to GRB 080310}. \aap 549:A22. \doi{10.1051/0004-6361/201219652}, {\href{https://arxiv.org/abs/1209.1506}{{arXiv:1209.1506}}} {[astro-ph.CO]}

\bibitem[{{Vreeswijk} et~al(2014){Vreeswijk}, {Savaglio}, {Gal-Yam}, {De Cia}, {Quimby}, {Sullivan}, {Cenko}, {Perley}, {Filippenko}, {Clubb}, {Taddia}, {Sollerman}, {Leloudas}, {Arcavi}, {Rubin}, {Kasliwal}, {Cao}, {Yaron}, {Tal}, {Ofek}, {Capone}, {Kutyrev}, {Toy}, {Nugent}, {Laher}, {Surace}, and {Kulkarni}}]{vreeswijk14}
{Vreeswijk} PM, {Savaglio} S, {Gal-Yam} A, et~al (2014) {The Hydrogen-poor Superluminous Supernova iPTF 13ajg and its Host Galaxy in Absorption and Emission}. \apj 797(1):24. \doi{10.1088/0004-637X/797/1/24}, {\href{https://arxiv.org/abs/1409.8287}{{arXiv:1409.8287}}} {[astro-ph.HE]}

\bibitem[{{Wakker} et~al(2007){Wakker}, {York}, {Howk}, {Barentine}, {Wilhelm}, {Peletier}, {van Woerden}, {Beers}, {Ivezi{\'c}}, {Richter}, and {Schwarz}}]{wakker2007}
{Wakker} BP, {York} DG, {Howk} JC, et~al (2007) {Distances to Galactic High-Velocity Clouds: Complex C}. \apjl 670(2):L113--L116. \doi{10.1086/524222}, {\href{https://arxiv.org/abs/0710.3340}{{arXiv:0710.3340}}} {[astro-ph]}

\bibitem[{{Wakker} et~al(2008){Wakker}, {York}, {Wilhelm}, {Barentine}, {Richter}, {Beers}, {Ivezi{\'c}}, and {Howk}}]{wakker2008}
{Wakker} BP, {York} DG, {Wilhelm} R, et~al (2008) {Distances to Galactic High-Velocity Clouds. I. Cohen Stream, Complex GCP, Cloud g1}. \apj 672(1):298--319. \doi{10.1086/523845}, {\href{https://arxiv.org/abs/0709.1926}{{arXiv:0709.1926}}} {[astro-ph]}

\bibitem[{{Wang} et~al(2020){Wang}, {Davies}, {Yang}, {Hennawi}, {Fan}, {Barth}, {Jiang}, {Wu}, {Mudd}, {Ba{\~n}ados}, {Bian}, {Decarli}, {Eilers}, {Farina}, {Venemans}, {Walter}, and {Yue}}]{wang2020}
{Wang} F, {Davies} FB, {Yang} J, et~al (2020) {A Significantly Neutral Intergalactic Medium Around the Luminous z = 7 Quasar J0252-0503}. \apj 896(1):23. \doi{10.3847/1538-4357/ab8c45}, {\href{https://arxiv.org/abs/2004.10877}{{arXiv:2004.10877}}} {[astro-ph.GA]}

\bibitem[{{Wang} et~al(2016){Wang}, {Elbaz}, {Daddi}, {Finoguenov}, {Liu}, {Schreiber}, {Mart{\'\i}n}, {Strazzullo}, {Valentino}, {van der Burg}, {Zanella}, {Ciesla}, {Gobat}, {Le Brun}, {Pannella}, {Sargent}, {Shu}, {Tan}, {Cappelluti}, and {Li}}]{wangt2016}
{Wang} T, {Elbaz} D, {Daddi} E, et~al (2016) {Discovery of a Galaxy Cluster with a Violently Starbursting Core at z = 2.506}. \apj 828(1):56. \doi{10.3847/0004-637X/828/1/56}, {\href{https://arxiv.org/abs/1604.07404}{{arXiv:1604.07404}}} {[astro-ph.GA]}

\bibitem[{{Watson} et~al(2015){Watson}, {Christensen}, {Knudsen}, {Richard}, {Gallazzi}, and {Micha{\l}owski}}]{Watson15}
{Watson} D, {Christensen} L, {Knudsen} KK, et~al (2015) {A dusty, normal galaxy in the epoch of reionization}. \nat 519(7543):327--330. \doi{10.1038/nature14164}

\bibitem[{{Welsh} et~al(2019){Welsh}, {Cooke}, and {Fumagalli}}]{Welsh19}
{Welsh} L, {Cooke} R, {Fumagalli} M (2019) {Modelling the chemical enrichment of Population III supernovae: the origin of the metals in near-pristine gas clouds}. \mnras 487(3):3363--3376. \doi{10.1093/mnras/stz1526}, {\href{https://arxiv.org/abs/1906.00009}{{arXiv:1906.00009}}} {[astro-ph.SR]}

\bibitem[{{Welsh} et~al(2020){Welsh}, {Cooke}, {Fumagalli}, and {Pettini}}]{Welsh20}
{Welsh} L, {Cooke} R, {Fumagalli} M, et~al (2020) {A bound on the $^{12}$C/$^{13}$C ratio in near-pristine gas with ESPRESSO}. \mnras 494(1):1411--1423. \doi{10.1093/mnras/staa807}, {\href{https://arxiv.org/abs/2001.04983}{{arXiv:2001.04983}}} {[astro-ph.GA]}

\bibitem[{{White} et~al(2021){White}, {Bauer}, {Baumgartner}, {Bautz}, {Berger}, {Cenko}, {Chang}, {Falcone}, {Fausey}, {Feldman}, {Fox}, {Fox}, {Fruchter}, {Fryer}, {Ghirlanda}, {Gorski}, {Grant}, {Guiriec}, {Hart}, {Hartmann}, {Hennawi}, {Kann}, {Kaplan}, {Kennea}, {Kocevski}, {Kouveliotou}, {Lawrence}, {Levan}, {Lidz}, {Lien}, {Littenberg}, {Mas-Ribas}, {Moss}, {O'Brien}, {O'Meara}, {Palmer}, {Pasham}, {Racusin}, {Remillard}, {Roberts}, {Roming}, {Rud}, {Salvaterra}, {Sambruna}, {Seiffert}, {Sun}, {Tanvir}, {Terrile}, {Thomas}, {van der Horst}, {Verstrand}, {Willems}, {Wilson-Hodge}, {Young}, {Amati}, {Bozzo}, {Karczewski}, {Hernandez-Monteagudo}, {Rebolo Lopez}, {Genova-Santos}, {Martin}, {Granot}, {Bemiamini}, {Gil}, and {Burns}}]{white21}
{White} NE, {Bauer} FE, {Baumgartner} W, et~al (2021) {The Gamow Explorer: a Gamma-Ray Burst Observatory to study the high redshift universe and enable multi-messenger astrophysics}. In: {Siegmund} OH (ed) UV, X-Ray, and Gamma-Ray Space Instrumentation for Astronomy XXII, p 1182109, \doi{10.1117/12.2599293}, \eprint{2111.06497}

\bibitem[{{Williams} et~al(2023){Williams}, {Kennea}, {Dichiara}, {Kobayashi}, {Iwakiri}, {Beardmore}, {Evans}, {Heinz}, {Lien}, {Oates}, {Negoro}, {Cenko}, {Buisson}, {Hartmann}, {Jaisawal}, {Kuin}, {Lesage}, {Page}, {Parsotan}, {Pasham}, {Sbarufatti}, {Siegel}, {Sugita}, {Younes}, {Ambrosi}, {Arzoumanian}, {Bernardini}, {Campana}, {Capalbi}, {Caputo}, {D'A{\`\i}}, {D'Avanzo}, {D'Elia}, {De Pasquale}, {Eyles-Ferris}, {Ferrara}, {Gendreau}, {Gropp}, {Kawai}, {Klingler}, {Laha}, {Melandri}, {Mihara}, {Moss}, {O'Brien}, {Osborne}, {Palmer}, {Perri}, {Serino}, {Sonbas}, {Stamatikos}, {Starling}, {Tagliaferri}, {Tohuvavohu}, {Zane}, and {Ziaeepour}}]{williams23}
{Williams} MA, {Kennea} JA, {Dichiara} S, et~al (2023) {GRB 221009A: Discovery of an Exceptionally Rare Nearby and Energetic Gamma-Ray Burst}. \apjl 946(1):L24. \doi{10.3847/2041-8213/acbcd1}, {\href{https://arxiv.org/abs/2302.03642}{{arXiv:2302.03642}}} {[astro-ph.HE]}

\bibitem[{{Williams} et~al(2008){Williams}, {Mason}, {Della Valle}, and {Ederoclite}}]{williams2008}
{Williams} R, {Mason} E, {Della Valle} M, et~al (2008) {Transient Heavy Element Absorption Systems in Novae: Episodic Mass Ejection from the Secondary Star}. \apj 685(1):451--462. \doi{10.1086/590056}, {\href{https://arxiv.org/abs/0805.1372}{{arXiv:0805.1372}}} {[astro-ph]}

\bibitem[{{Windhorst} et~al(2018){Windhorst}, {Timmes}, {Wyithe}, {Alpaslan}, {Andrews}, {Coe}, {Diego}, {Dijkstra}, {Driver}, {Kelly}, and {Kim}}]{windhorst18}
{Windhorst} RA, {Timmes} FX, {Wyithe} JSB, et~al (2018) {On the Observability of Individual Population III Stars and Their Stellar-mass Black Hole Accretion Disks through Cluster Caustic Transits}. \apjs 234(2):41. \doi{10.3847/1538-4365/aaa760}, {\href{https://arxiv.org/abs/1801.03584}{{arXiv:1801.03584}}} {[astro-ph.GA]}

\bibitem[{{Wise} et~al(2012){Wise}, {Turk}, {Norman}, and {Abel}}]{wise2012}
{Wise} JH, {Turk} MJ, {Norman} ML, et~al (2012) {The Birth of a Galaxy: Primordial Metal Enrichment and Stellar Populations}. \apj 745(1):50. \doi{10.1088/0004-637X/745/1/50}, {\href{https://arxiv.org/abs/1011.2632}{{arXiv:1011.2632}}} {[astro-ph.CO]}

\bibitem[{{Woosley} and {Bloom}(2006)}]{woosley06}
{Woosley} SE, {Bloom} JS (2006) {The Supernova Gamma-Ray Burst Connection}. \araa 44(1):507--556. \doi{10.1146/annurev.astro.43.072103.150558}, {\href{https://arxiv.org/abs/astro-ph/0609142}{{arXiv:astro-ph/0609142}}} {[astro-ph]}

\bibitem[{{Woosley} and {Weaver}(1995)}]{woosley1995}
{Woosley} SE, {Weaver} TA (1995) {The Evolution and Explosion of Massive Stars. II. Explosive Hydrodynamics and Nucleosynthesis}. \apjs 101:181. \doi{10.1086/192237}

\bibitem[{{Xing} et~al(2023){Xing}, {Zhao}, {Liu}, {Heger}, {Han}, {Aoki}, {Chen}, {Ishigaki}, {Li}, and {Zhao}}]{xing2023}
{Xing} QF, {Zhao} G, {Liu} ZW, et~al (2023) {A metal-poor star with abundances from a pair-instability supernova}. \nat 618(7966):712--715. \doi{10.1038/s41586-023-06028-1}

\bibitem[{{Yan} et~al(2018){Yan}, {Perley}, {De Cia}, {Quimby}, {Lunnan}, {Rubin}, and {Brown}}]{yan18}
{Yan} L, {Perley} DA, {De Cia} A, et~al (2018) {Far-UV HST Spectroscopy of an Unusual Hydrogen-poor Superluminous Supernova: SN2017egm}. \apj 858(2):91. \doi{10.3847/1538-4357/aabad5}, {\href{https://arxiv.org/abs/1711.01534}{{arXiv:1711.01534}}} {[astro-ph.HE]}

\bibitem[{{Yang} et~al(2023){Yang}, {Fan}, {Gupta}, {Myers}, {Palanque-Delabrouille}, {Wang}, {Y{\`e}che}, {Aguilar}, {Ahlen}, {Alexander}, {Brooks}, {Dawson}, {de la Macorra}, {Dey}, {Dhungana}, {Fanning}, {Font-Ribera}, {Gontcho}, {Guy}, {Honscheid}, {Juneau}, {Kisner}, {Kremin}, {Le Guillou}, {Levi}, {Magneville}, {Martini}, {Meisner}, {Miquel}, {Moustakas}, {Nie}, {Percival}, {Poppett}, {Prada}, {Schlafly}, {Tarl{\'e}}, {Vargas Magana}, {Weaver}, {Wechsler}, {Zhou}, {Zhou}, and {Zou}}]{yang2023}
{Yang} J, {Fan} X, {Gupta} A, et~al (2023) {DESI z {\ensuremath{\gtrsim}} 5 Quasar Survey. I. A First Sample of 400 New Quasars at $z\sim 4.7-6.6$}. \apjs 269(1):27. \doi{10.3847/1538-4365/acf99b}, {\href{https://arxiv.org/abs/2302.01777}{{arXiv:2302.01777}}} {[astro-ph.GA]}

\bibitem[{{York} and {Cowie}(1983)}]{york1983}
{York} DG, {Cowie} LL (1983) {On the possibility of detecting very hot gas through absorption-line studies.} \apj 264:49--52. \doi{10.1086/160572}

\bibitem[{{Yoshida} et~al(2003){Yoshida}, {Abel}, {Hernquist}, and {Sugiyama}}]{yoshida2003}
{Yoshida} N, {Abel} T, {Hernquist} L, et~al (2003) {Simulations of Early Structure Formation: Primordial Gas Clouds}. \apj 592(2):645--663. \doi{10.1086/375810}, {\href{https://arxiv.org/abs/astro-ph/0301645}{{arXiv:astro-ph/0301645}}} {[astro-ph]}

\bibitem[{{Yu} et~al(2021){Yu}, {Li}, {Qu}, {Roederer}, {Bregman}, {Fan}, {Fang}, {Johnson}, {Wang}, and {Yang}}]{yu2021}
{Yu} X, {Li} JT, {Qu} Z, et~al (2021) {Probing the He II re-Ionization ERa via Absorbing C IV Historical Yield (HIERACHY) I: A strong outflow from a z 4.7 quasar}. \mnras 505(3):4444--4455. \doi{10.1093/mnras/stab1614}, {\href{https://arxiv.org/abs/2105.13498}{{arXiv:2105.13498}}} {[astro-ph.GA]}

\bibitem[{{Zhu} et~al(2022){Zhu}, {Becker}, {Bosman}, {Keating}, {D'Odorico}, {Davies}, {Christenson}, {Ba{\~n}ados}, {Bian}, {Bischetti}, {Chen}, {Davies}, {Eilers}, {Fan}, {Gaikwad}, {Greig}, {Haehnelt}, {Kulkarni}, {Lai}, {Pallottini}, {Qin}, {Ryan-Weber}, {Walter}, {Wang}, and {Yang}}]{zhu2022}
{Zhu} Y, {Becker} GD, {Bosman} SEI, et~al (2022) {Long Dark Gaps in the Ly{\ensuremath{\beta}} Forest at $z < 6$: Evidence of Ultra-late Reionization from XQR-30 Spectra}. \apj 932(2):76. \doi{10.3847/1538-4357/ac6e60}, {\href{https://arxiv.org/abs/2205.04569}{{arXiv:2205.04569}}} {[astro-ph.CO]}

\bibitem[{{Zhu} et~al(2023){Zhu}, {Becker}, {Christenson}, {D'Aloisio}, {Bosman}, {Bakx}, {D'Odorico}, {Bischetti}, {Cain}, {Davies}, {Davies}, {Eilers}, {Fan}, {Gaikwad}, {Haehnelt}, {Keating}, {Kulkarni}, {Lai}, {Ma}, {Mesinger}, {Qin}, {Satyavolu}, {Takeuchi}, {Umehata}, and {Yang}}]{zhu2023}
{Zhu} Y, {Becker} GD, {Christenson} HM, et~al (2023) {Probing Ultralate Reionization: Direct Measurements of the Mean Free Path over $5 < z < 6$}. \apj 955(2):115. \doi{10.3847/1538-4357/aceef4}, {\href{https://arxiv.org/abs/2308.04614}{{arXiv:2308.04614}}} {[astro-ph.CO]}

\end{thebibliography}

\end{document}